# The PV performance ratio paradox: annual data from large-scale, real-world PV systems show negligible meteorological and technical impact and points to dominant human factors

Hugo FM Milan[a*], Aline Q Alves[a], Thatiane AT Souza[b], Juliana M Galo[b], Alex SC Maia[c], Moisés AP Borges[a], Ciro J Egoavil[a]

[a]Electrical Engineering Department, Federal University of Rondônia

[b]Federal Institute of Rondônia

[c]State University of São Paulo, Jaboticabal Campus

*Corresponding author: hugo.milan@unir.br, DAEE BR 364, km 9,5, Porto Velho/RO, CEP 76.801-059

**Graphical abstract:**

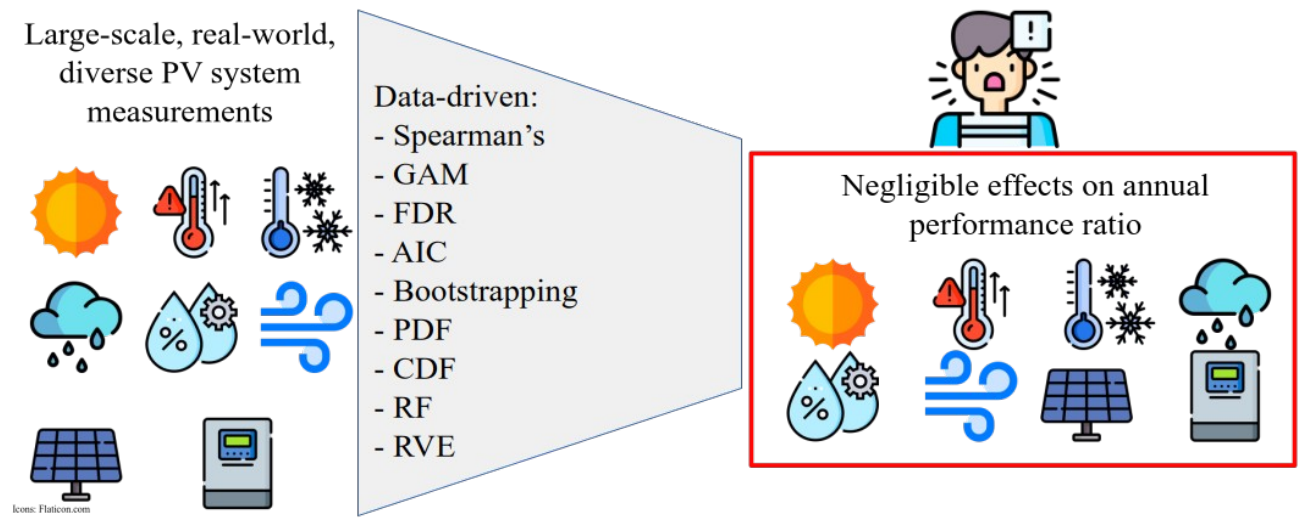

**Highlights:**

- Annual PR of large-scale, real-world PV systems from Rondônia, Brazil

- Surprisingly, effects of technical and meteorological variables were negligible

- Installation, monitoring, and maintenance quality might have stronger effects

- Policy makers could focus on educational programs to increase quality

**Abstract:** Performance ratio (PR) is a established measure of efficiency of photovoltaic (PV) systems. While previous research demonstrated the effects of meteorological and technical variables on PR, a gap persists in the literature on which variables strongly influence PR in large-scale, real-world, heterogeneous PV systems. This paper aims to fill this gap, applying data-driven models to PV systems located in Rondônia State, Brazil, to identify which variables strongly influence annual PR, and, hence, should be the target for optimization. Surprisingly, only negligible effects were found between meteorological and technical variables on annual PR, indicating that human-factors (such as installation, monitoring, and maintenance quality) might have a stronger effect. These findings indicates that, to improve performance of PV systems, policy makers could focus on creating educational programs to teach PV installers and technicians how to properly install, monitor, and maintain modern PV systems. Through estimating the probability density functions of PR, its peak value was found as 78.85% (mean 77.52%, 95% confidence interval of 76.12% to 78.84%, and 95% prediction interval of 58.83% to 92.70%). A map of annual final yield was developed for Rondônia State and can be used by entrepreneurs to quickly and cheaply estimate energy production.

**Word count:** 5,402

**Figure count:** 14

**Table count:** 1

Supplementary material available with 22 figures.

Data and code are available.

1. Introduction

Photovoltaic (PV) energy has become established world-wide, with a rapid increase over the last decade driven by decreased system costs and favorable regulatory environment (IEA. 2023; Masson et al., 2025). This expansion has led to diverse PV system applications, ranging from utility-scale solar farms, rooftops residential and commercial applications, floating (Benjamins et al., 2024; Sahin et al., 2025; Wei et al., 2025), offshore (Oyewo et al., 2025; Saincher et al., 2025; Vasuki et al., 2025), agrivoltaics (Jean and Rosentrater 2025; IEA, 2025; Vaverková et al., 2026), and animal agrivoltaics (Faria et al., 2023; Fonsêca et al., 2023; Maia et al., 2020). In South America, Brazil has seen a rapid increase in the number of PV systems (Lima et al., 2017; Nascimento et al., 2020; Rüther and Dacoregio, 2000; Silva et al., 2019; Simioni and Schaeffer, 2019), accumulating approximately 4 million systems (average size of 11,24 kW; ANEEL, 2025) throughout the country. Rondônia, a state located in the Brazilian Amazon Rainforest region, reflects this trend with more than 50 thousand PV systems (average size of 11,29 kW; ANEEL, 2025).

Performance Ratio (PR; IEC, 2021) is an standard metric to evaluate the efficiency of a PV system by project developers, investor, engineering companies, clients, and grid operators. Previous research have demonstrated that different module technologies respond differently to environmental stressors (Carr and Pryor, 2004; Minemoto et al., 2007; Rmanan et al., 2019; Wang et al., 2017), where temperature and relative humidity correlate negatively with power output (Agrawal et al., 2022; Louwen et al., 2017; Makrides et al., 2012; Rahman et al., 2015), while solar irradiation and wind speed correlate positively (Bamisile et al., 2025; Osma-Pinto and Ordóñez-Plata, 2019; Yadav and Bajpai, 2018), resulting in final yield changing with season (as an effect of meteorological changes), but PR remaining approximately constant (Malvoni et al., 2017). In addition, similar findings were reported for on-grid installations (AlSkaif et al., 2020; Bamisile et al., 2025; Ma et al., 2017; Markrides et al., 2012; Mehdi et al., 2023; Venkateswari and Sreejith, 2019). Accordingly, a rich body of literature exists that quantified PR across diverse global climates (Kumar and Kumar, 2017), with reported values as 71-95% in Algeria (Bouacha et al., 2020; Cherfa et al., 2015; Dabou et al., 2016; Necaibia et al., 2018; Sahouane et al., 2019), 73-82% in Chile (Ferrada et al., 2015), 70.4% in Ghana (Mensah et al., 2019), 70-85% in India (Bhakta and Mukherjee, 2017; Dobaria et al., 2016; Malvoni et al., 2020; Satsangi et al., 2018; Vasisht et al., 2016; Yadav and Bajpai, 2018), 67% in the Island of Crete (Kymakis et al., 2009), 84% in Italy (Malvoni et al., 2017), 79.5% in Malawi (Banda et al., 2019), 68-75% in Mauritania (Sidi et al., 2016), 73-77% in Morocco (Ameur et al., 2019), 78% in New Zealand (Emmanuel et al., 2017), 65% in Oman (Al-Badi, 2018), 84% in South Africa (Okello et al., 2015), 81% in Spain (Wang et al., 2017).

Besides meteorological and technical factors, previous research have shown that improper maintenance routines, including soiling removal, can drastically reduce PR by 50% (Bouraiou et al., 2015; Conceição et al., 2022; Liu et al., 2024; Shadid et al., 2023; Shahzad et al., 2025), resulting in a global energy loss of 4-7% (Micheli and Wilbert, 2025). While systems well maintained show PR values higher than average (Jahn et al., 1998, 2000), poor maintenance results in accumulating soiling losses with degradation losses, reported to be as high as 5.2%/year (Aboagye et al., 2022; Atsu et al., 2020; Bouraiou et al., 2015; Golive et al., 2019; Ismail et al., 2012; Mgonja and Saidi, 2017), and can result in accelerated failure (Halwachs et al., 2019; Libra et al., 2023; Nascimento et al., 2020).

Large-scale, real-world analyses have demonstrated that PR is increasing over the years, driven by improved technology (Guerrero-Lemus et al., 2019; Leloux et al., 2015), and average values of PR were reported as 73-78% in Belgium (Leloux et al., 2012a; Schardt and Heesen, 2021), 75-95% in Chile (Ascencio-Vásquez et al., 2021), 77-81% in Europe (Leloux et al., 2015; Lindig et al., 2021), 73-76% in France (Leloux et al., 2012b; Schardt and Heesen, 2021), 74% in Germany (Schardt and Heesen, 2021), 73% in Italy (Schardt and Heesen, 2021), 72-78% in Luxembourg (Heesen et al., 2019; Schardt and Heesen, 2021), 69-85% in Slovenia (Brecl et al., 2022; Seme et al., 2019), 71-79% in The Netherlands (Kausika et al., 2018; Meng et al., 2022; Schardt and Heesen, 2021; Tsafarakis et al., 2017), and 83% in UK (Taylor et al., 2015). In addition, large-scale, real-world analyses have recently shown that PR depends on race and income level of prosumers (Gherghina et al., 2025), indicating that variability of PR might be related to high-income prosumers contracting from higher quality installers (Darghouth et al., 2022; Forrester et al., 2022).

While it is established that meteorological and technical variables affect performance ratio of PV systems in controlled, small-scale experiments, it remains unclear which variables strongly influence annual PR in large-scale, real-world, heterogeneous PV systems. This paper aims to fill this gap, through the use of data-driven models (Ahmad et al., 2022) to investigate the effects of technical and meteorological variables in large-scale, real-world, heterogeneous PV systems located in Rondônia State, Brazil. Specifically, data-driven and statistical analysis were employed to identify which meteorological and technical variables strongly influence annual PR, and, hence, should be the target for optimizing PV systems.

## 2. Materials and Methods

Data and algorithms writing in R are openly available (Milan, 2025). Geo-spacial data were obtained from Brazilian official sources (FUNAI, 2019; IBGE, 2023; MMA, 2019) or using the geobr R-package (Pereira and Gonçalves, 2024).

### 2.1. Data processing

Data came from 159 PV systems in the Rondônia State, Brazil (Fig. 1), which attained the following inclusion criteria: 1) 12-months complete generation cycle, 2) less than 7 days of generation issues, and 3) only one inverter model. Data included: module brand, module model, module STC power, quantity of modules, module type (n or p), module faciality (monofacial or bifacial), module temperature coefficient for short-circuit current, module temperature coefficient for open-circuit voltage, module temperature coefficient for maximum power, inverter brand, inverter model, inverter AC power, quantity of inverters, total inverter AC power, inverter nominal efficiency, inverter European efficiency, inverter number of MPPTs and DC inputs, total DC input power (kWp), annual energy production (kWh), and city of installation. All modules were monocristalline. Outliers of annual performance ratio (Eq. 2) were removed using the interquantile range methodology (Efron and Hastie, 2011; Schardt and Heesen, 2021; Taylor et al., 2015).

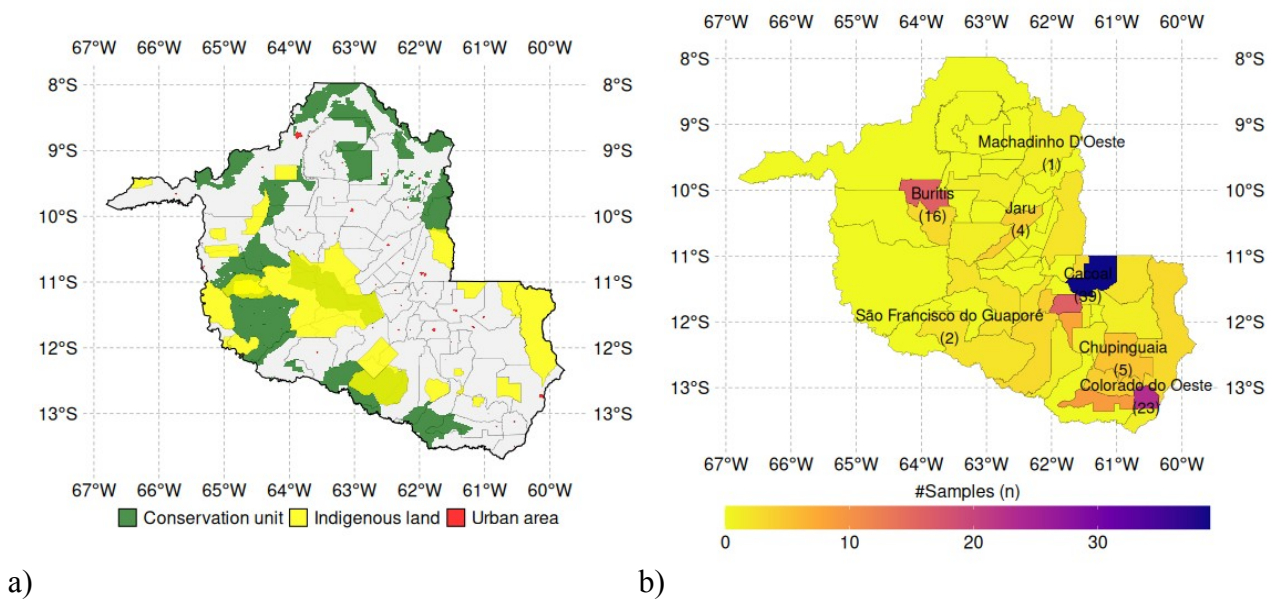

a)                                            b)

Figure 1. Map of Rondônia State showing city divisions (gray lines), conservation areas (green), indigenous lands (yellow), and urban areas (red; a). Map of Rondônia State showing that number of samples were distributed among the cities of the state, with a few clusters (b).

## 2.2. Performance metrics

Performance metrics were evaluated following IEC Standard 61724-1 (IEC, 2021). In large-scale, real-world applications, measurement and definition of PV installation tilt and azimuth angles might be challenging, specially because of multiple arrays installed on roofs with different tilt and azimuth angles and disperse location of real-world PV installations. Hence, in this paper, annual hours of peak sun were calculated using global horizontal irradiance ($GHI$, kWh/m²) as follows:

$$H_{ps@GHI}(h) = GHI(kWh/m^2) \times 365 \qquad (1)$$

where $H_{PS@GHI}$ = annual hours of peak sun at GHI.

From $H_{PS@GHI}$ and measured annual final yield ($Y_f$, kWh/kWp), annual performance ratio ($PR_{GHI}$) was calculated as previously reported (Khalid et al., 2016; Vasisht et al., 2016):

$$PR_{GHI} = \frac{Y_f(kWh/kWp)}{H_{ps@GHI}(h)} \qquad (2)$$

note that, since $PR_{GHI}$ calculated using $GHI$ can be greater than 100% for cases in which tilt and azimuth angles receive solar irradiation greater than $GHI$.

Capacity utilization factor ($CUF$) was calculated as follows:

$$CUF = \frac{Y_f(kWh/kWp) \times ISF}{365 \times 24} = \frac{PR_{GHI} \times H_{ps@GHI}(h) \times ISF}{365 \times 24} \qquad (3)$$

Data analyses were performed on $PR_{GHI}$ because of its characteristics of allowing comparison between different PV systems and because $H_{PS@GHI}$ is derived from irradiance data alone.

*2.3. Meteorological data*

Rondônia State is located in the Amazon Rainforest region, between latitudes 8º South and 14º South and longitudes 60º West and 67º West. Rondônia climate is considered tropical hot and humid all year-round (Köppen Classification: Am; Alvares et al., 2013), with two distinct seasons: dry and wet (rainy; October to April). Average annual daily solar irradiation (global horizontal, direct normal, diffuse, and tilted latitude; Fig. 2) were obtained from the Brazilian Space Research Institution (INPE; Pereira et al., 2017). As seen in Fig. 2, the State shows distinctive values near its region of highest altitude (approximately 11ºS, 64ºW), with mean and standard deviation for the whole State as 4,584 ± 92 Wh/(m². day) for global horizontal irradiation, 3,302 ± 231 Wh/(m². day) for direct normal irradiation, 2,139 ± 36 Wh/(m². day) for diffuse irradiation, and 4,651 ± 106 Wh/(m². day) for tilted latitude irradiation.

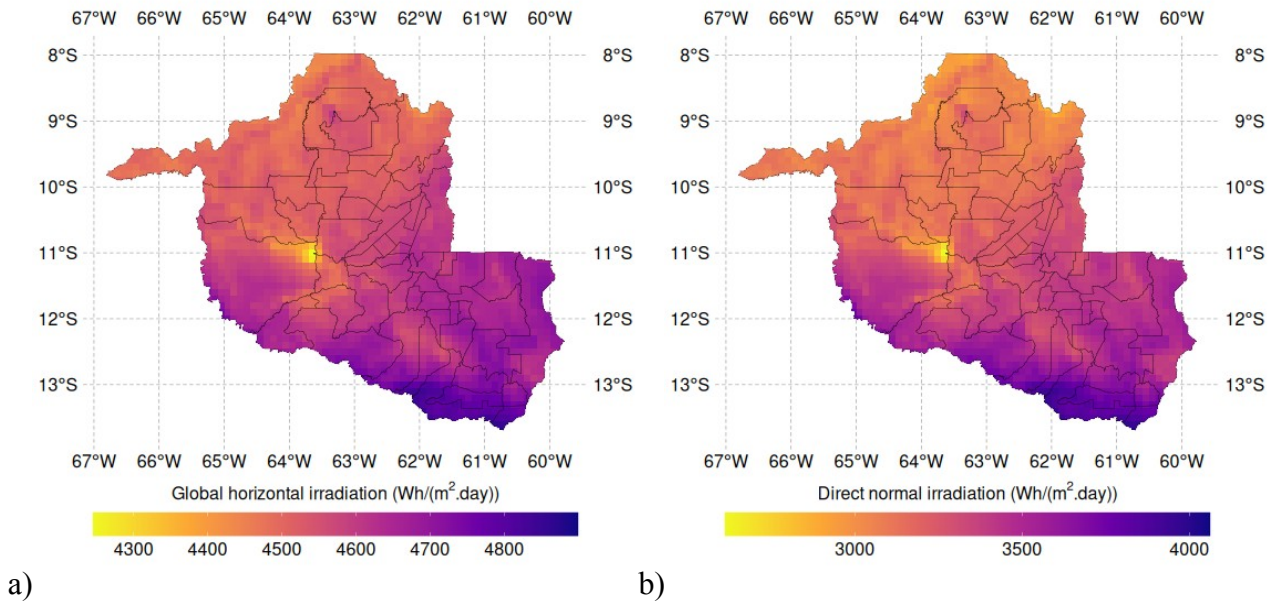

a)  b)

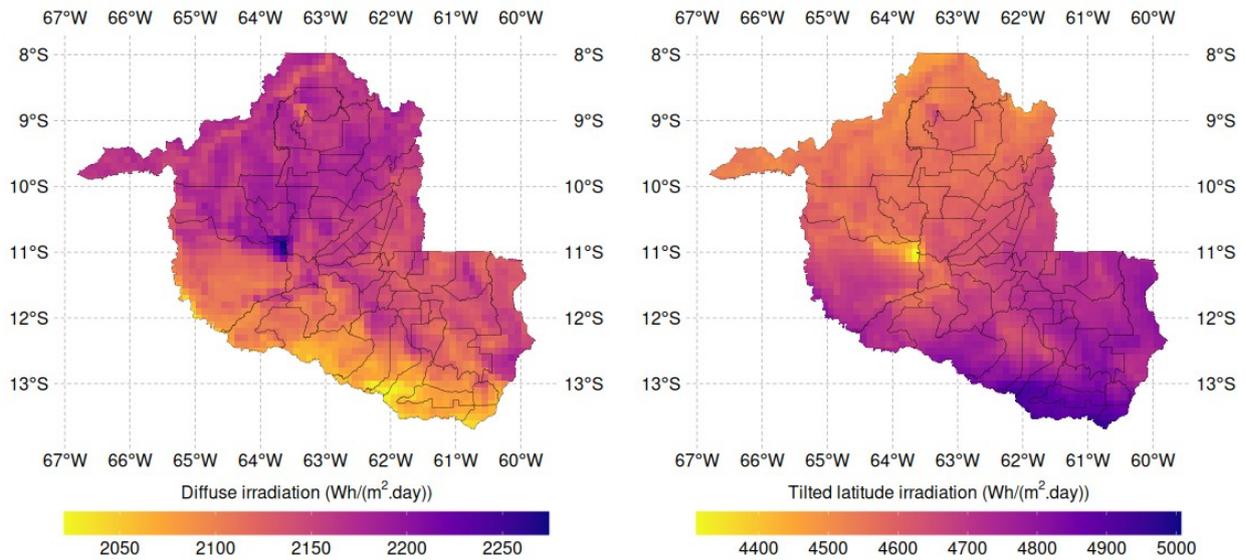

c)  d)

Figure 2. Map of Rondônia State showing city divisions and mean daily global horizontal (a; Wh/(m².day)), direct normal (b; Wh/(m².day)), diffuse (c; Wh/(m².day)), and titled latitude (d; Wh/(m².day)) irradiation data from INPE (Pereira et al., 2017). Note that map of annual hours of peak sun at global horizontal irradiance ($H_{PS@GHI}$, h) is identical to (a), only changing the scale by 1/365.

Daily meteorological data (maximum temperature, minimum temperature, precipitation, relative humidity, and wind speed at 2 m) were obtained from the Brazilian Daily Weather Gridded data (BR-DWGD; Xavier et al., 2022). Since data from PV systems are an aggregation of annual energy production, most recent annual data from BR-DWGD (year 2023) were aggregated to the PV systems dataset using the following functions: mean (Fig. 3), median (supplemental Fig. 1S), minimum (Fig. 2S), maximum (Fig. 3S), standard deviation (Fig. 4S), variance (Fig. 5S), sum (Fig. 6S), skewness (data asymmetry; Fig. 7S), and kurtosis (frequency of extreme values; Fig. 8S). As seen in Fig. 3, the State shows distinctive values near its region of highest altitude (approximately 11ºS, 64ºW), with mean and standard deviation for the whole State as 32.94 ± 0.45 ºC for maximum temperature, 22.10 ± 0.47 ºC for minimum temperature, 5.17 ± 0.88 mm for precipitation, 77.43 ± 2.43 % for relative humidity, and 0.88 ± 0.09 m/s for wind speed at 2 m.

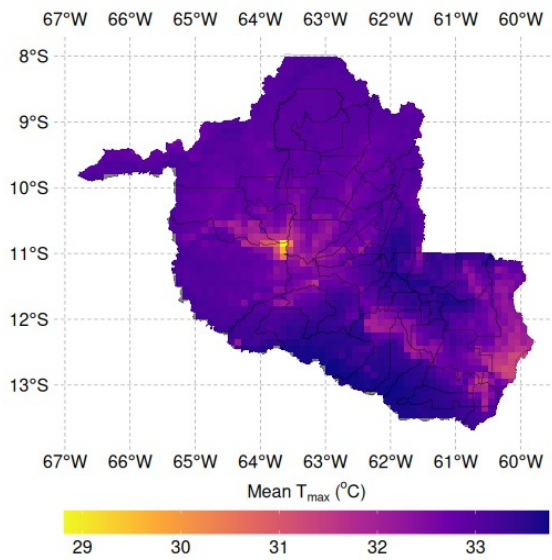
a)
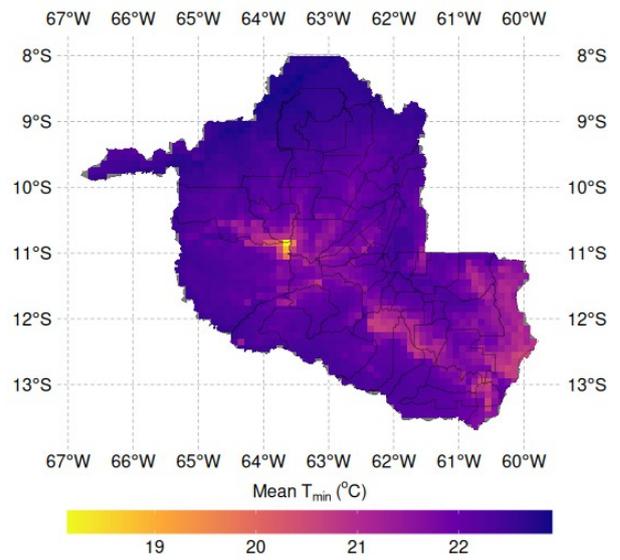
b)
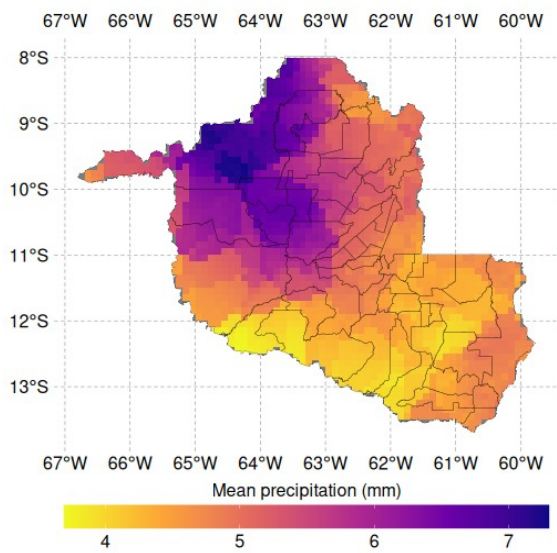
c)
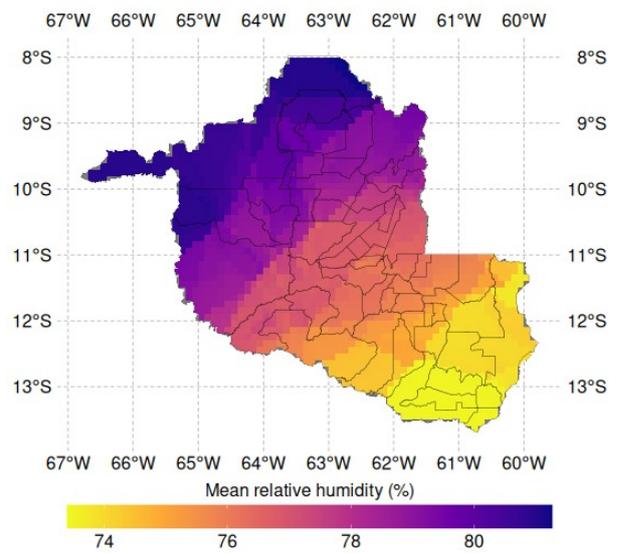
d)

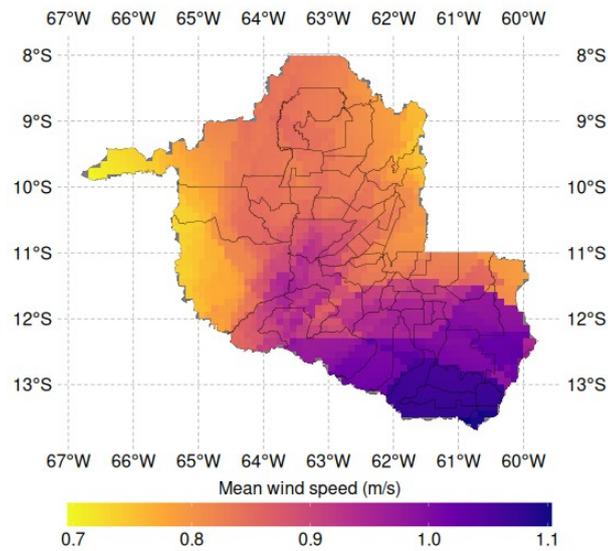

e)

Figure 3. Map of Rondônia State showing city divisions and mean of year 2023 daily maximum temperature (a; °C), minimum temperature (b; °C), precipitation (c; mm), relative humidity (d; %), and wind speed at 2 m (e; m/s) data from BR-DWGD (Xavier et al., 2022).

## 2.4. Statistical analysis

To preliminary explore monotonic relationship between continuous variables from the dataset, Spearman's rank association coefficient ($\rho$) were calculated. Spearman's coefficient is similarly interpreted as Pearson's linear correlation coefficient (r), with the difference that Spearman's coefficient measures the level of monotonic association with a nonparametric relationship. Positive values of $\rho$ indicate that when one variable increases, the other also increases, whereas negative values of $\rho$ indicate that when one variable increases, the other decreases. Interpretation from absolute values of $\rho$ is as follows: 0.00-0.19: negligible association; 0.20-0.39: weak association; 0.40-0.59: moderate association; 0.60-0.79: strong association; and 0.80-1.00: very strong association.

To investigate the effects of technical (e.g., ISF, kWp, brands, etc.) and meteorological variables on $PR_{GHI}$, first, the effects of solar irradiation on $PR_{GHI}$ were controlled. The effects of solar irradiation were controlled through fitting statistical Generalized Additive Models (GAMs; Wood, 2017) to the relationship between $PR_{GHI}$ and smooths functions for each individual solar irradiation variable (global horizontal, direct normal, diffuse, and tilted latitude) and choosing the GAM model with the best fit irradiation variable. Smooth functions were used because they allow for non-linear relationships learned from the data, without requiring explicit assumptions from researchers. To compare GAM models, p-values corrected using the False Discovery Rate (FDR; Benjamini and Hochberg, 1995; Efron and Hastie, 2021; Storey and Tibshirani, 2003) and Akaike Information Criteria (AIC) were used.

P-values correspond to the probability that a statistically more complex model does not better explains the data than a simpler model. In general, a more complex model (more parameters) is selected over a simpler model whenever a p-value less than 0.05 is found, following the principle of parsimony and to avoid overfitting to the data. Since we are training and testing different statistical models, we need to correct for the probabilities of obtaining a low p-value just by chance. To correct p-values, we adopted the False Discovery Rate (FDR; Benjamini and Hochberg, 1995;

Efron and Hastie, 2021; Storey and Tibshirani, 2003), which is a less stringent correction methodology, resulting in a probability of falsely selecting a more complex model over a simpler model (i.e., a 0.05 FDR corrected p-value means that we have a 5% chance of falsely selecting a more complex model when a simpler model should have been chosen).

In addition to the p-value, AIC values were used as a guide to select a more complex model over a simpler model. AIC is an estimator of relative quality of statistical models for a given dataset. Lower AIC values (the smaller, the better) indicate a better fit even with a more complex model. Differences up to 2 points in AIC indicate that the simpler model is as likely as the more complex model (i.e., following the principle of parsimony, the simpler model is selected), while differences greater than 7 points indicate strong preference for the more complex model. AIC values in between require further statistical investigation (such as using p-values as a guide) for a proper conclusion.

After controlling for the effects solar irradiation, the effects of technical and other meteorological variables on $PR_{GHI}$ were investigated using a forward selection method of GAM models. First, GAM models were fitted including system parameters and compared with the GAM model from the previous step. After investigating the effects of system parameters, the effects of each meteorological variable were included following the same criteria.

To investigate the approximate statistical power of our methodology to detect known physical effects of meteorological and technical variables, a statistical sensitivity analysis was performed (Champely, 2020; Cohen, 1988) using an F-test for a regression model (p-value or $\alpha = 0.05$, power = 0.80) to calculate the minimum detectable effect size ($f^2$). Interpretation from values of $f^2$ is as follows: approximately equal or lower than 0.02: small effect (i.e., can detect subtle effects on $PR_{GHI}$ that could easily be hidden by other factors); approximately equal or lower than 0.15: medium effect (i.e., can detect clear, consistent, and noticeable effects on $PR_{GHI}$); approximately equal or lower than 0.35: large effect (i.e., can only detect large, main effects on $PR_{GHI}$). From $f^2$, the minimum improvement in the coefficient of determination ($R^2$) that can be found from our procedure can be calculated. $R^2$ represents how much data variability is explained by the statistical model. For instance, $R^2 = 0\%$ means that no data variability is explained by the statistical model, where as $R^2 = 100\%$ means that all data variability is explained by the statistical model (a perfect model). In addition, multiplying the bootstrapped 95% prediction interval spread by $1 - \sqrt{(1 - R^2)}$ results in an approximation for how much improvement in the bootstrapped 95% prediction interval spread our procedure was able to detect (i.e., how much improvement in the uncertainty of $PR_{GHI}$ our statistical procedure was able to detect).

To visually evaluate homogeneity and variability of numerical data, Z-scores and Coefficients of Variation (CV) were calculated. Z-scores indicates how many sample standard deviations a sample is from the sample mean. Z-scores between -2 and +2 are usually expected. CV, calculated as a percentage of sample standard deviation over sample mean, numerically indicates data dispersion. Interpretation from absolute values of CV is as follows: <10%: highly homogeneous; 10%-20%: moderate dispersion; >20%: highly heterogeneous.

Continuous probability density function (PDF) of the variables were estimated using Gaussian kernels (density function; R Core Team, 2025). This non-parametric approach allows for smooth estimation of the data distribution without requiring any specific theoretical assumptions. From the PDF, cumulative density function (CDF) and its associated parameters were calculated. To estimate statistics (mean, 95% confidence intervals, 95% prediction intervals, percentiles, etc), a nonparametric bootstrapping technique (n = 10,000) was employed (Efron & Hastie, 2021). In essence, random samples were drawn (with replacement) from the observed dataset 10,000 times and statistics were calculated from the sampled dataset. The 95% confidence interval represents a 95% probability range for the true population mean, while the 95% prediction interval represents a 95% probability range which new samples would be observed. Percentiles represent how much of the data is expected to be below its value (e.g., a 10% percentile represents that 10% of the data is expected to be below its value).

## 3. Results and Discussion

### 3.1. Data processing

Data processing resulted in 142 samples (Fig. 4), with interquantile range outliers removal methodology resulting in the exclusion of data with $PR_{GHI}$ outside the range 53.98%-100.51%, which is a physically sounding range for $PR_{GHI}$ (Schardt and Heesen, 2021; Taylor et al., 2015). The processed data corresponded to a heterogeneous set of brands and models of modules and inverters (Fig. 5), with mean inverter AC power of 9.18 kW (DC power of 10.48 kWp), indicating that the dataset mostly contained data from residential and small commercial installations, similar to the whole State level data (11,29 kW; ANEEL, 2025).

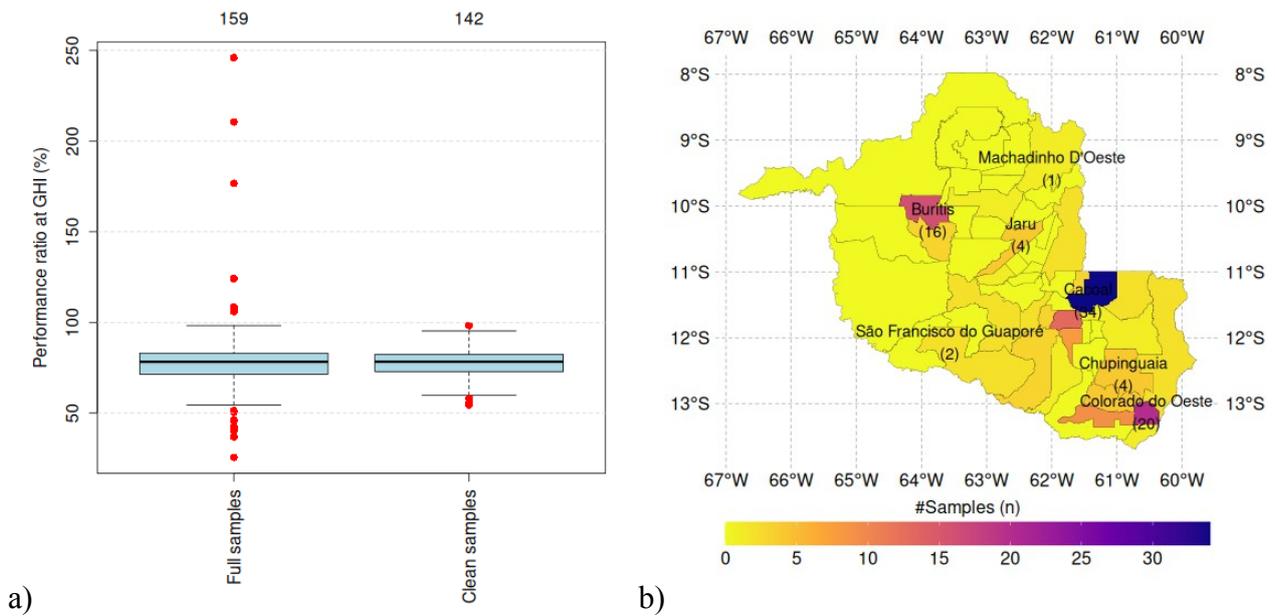

Figure 4. Boxplots for the annual performance ratio at global horizontal irradiance ($PR_{GHI}$) for the full samples and clean samples. Map of Rondônia State showing city divisions (gray lines) and the number of samples after data processing (a). Boxplots show outliers as red points, minimum value as the lower end of the dashed lines, first quartile (25% of the data falls below this value) as the lower line of the blue box, median and the black line of the blue box, third quartile (75% of the data falls below this value) as the upper line of the blue box, and maximum value as the upper of the dashed lines. Number on top of boxplots indicate number of samples.

## 3.2. Performance metrics

No clear indication of effects of brands/models of inverters or modules on $PR_{GHI}$ were seen (Fig. 5), with the exception of WEG inverter brand, which is a national top tier brand, that showed a higher $PR_{GHI}$ than the other inverter brands.

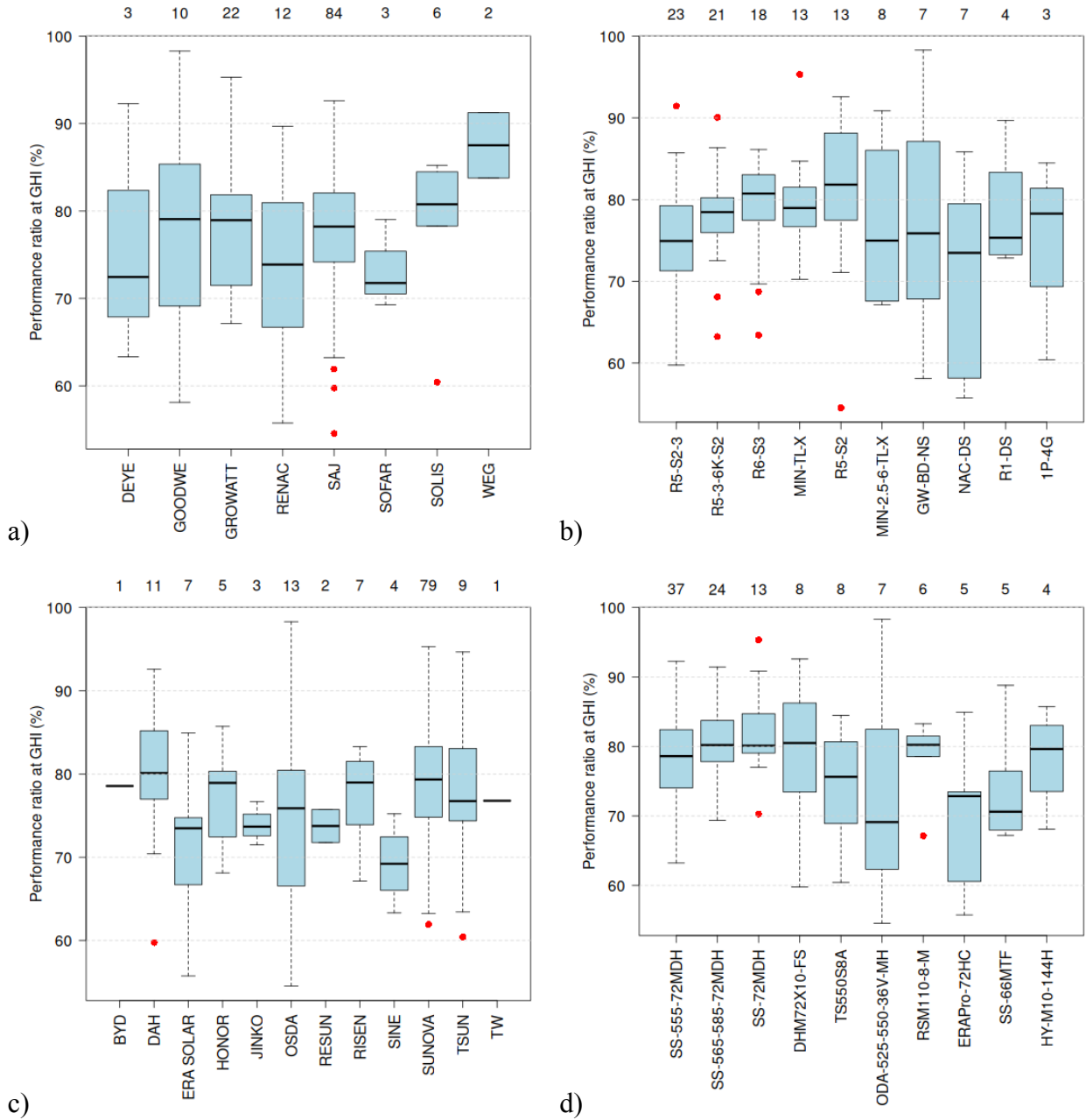

Figure 5. Boxplots for the relationship between annual performance ratio at global horizontal irradiance ($PR_{GHI}$) and inverter brand (a), top 10 inverter model with the highest number of samples (b), top 10 module brand with the highest number of samples (c), and top 10 module model with the highest number of samples (d). Boxplots show outliers as red points, minimum value as the lower end of the dashed lines, first quartile (25% of the data falls below this value) as the lower line of the blue box, median and the black line of the blue box, third quartile (75% of the data falls below this value) as the upper line of the blue box, and maximum value as the upper of the dashed lines. Number on top of boxplots indicate number of samples.

Surprisingly, no clear effect was seen on $PR_{GHI}$ from ISF (Fig. 6a), input DC (Fig. 6b), module efficiency (Fig. 6c), inverter maximum efficiency (Fig. 6d), nor the remaining technical variables from module and inverters (Fig. 9S; module temperature coefficients, inverter European efficiency, inverter number of MPPTs, inverter number of DC inputs, and inverter number of DC inputs over number of MPPTs). Similar variations were observed in The Netherlands with a large-fleet of identical PV Systems, even when comparing PV systems in the same neighborhood (Meng et al., 2022).

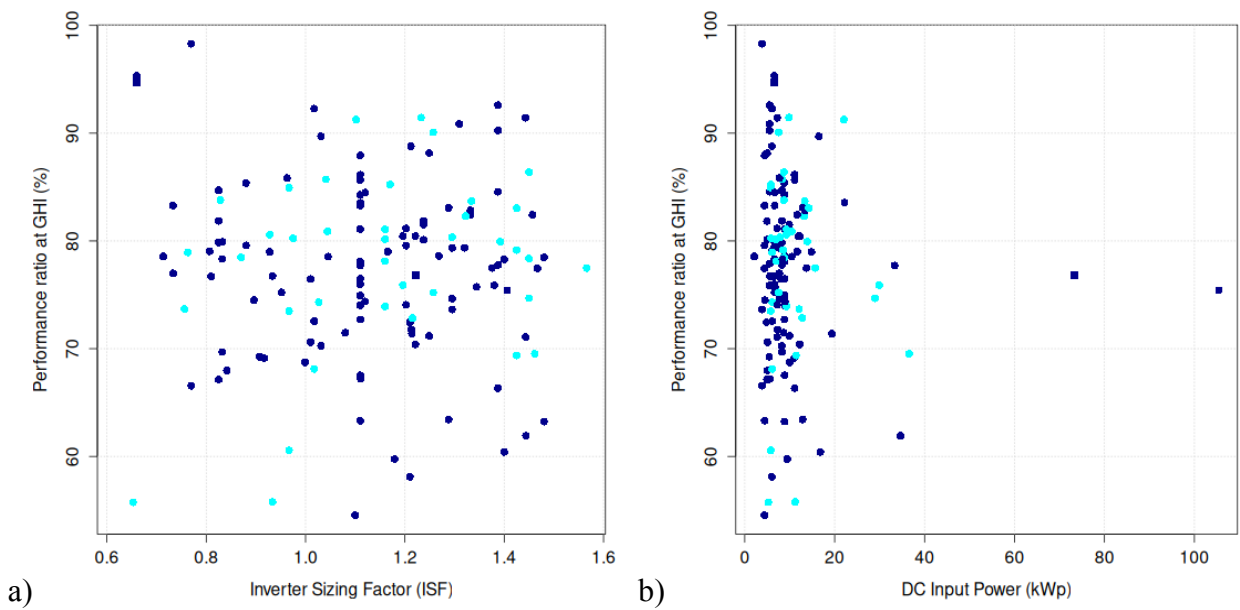

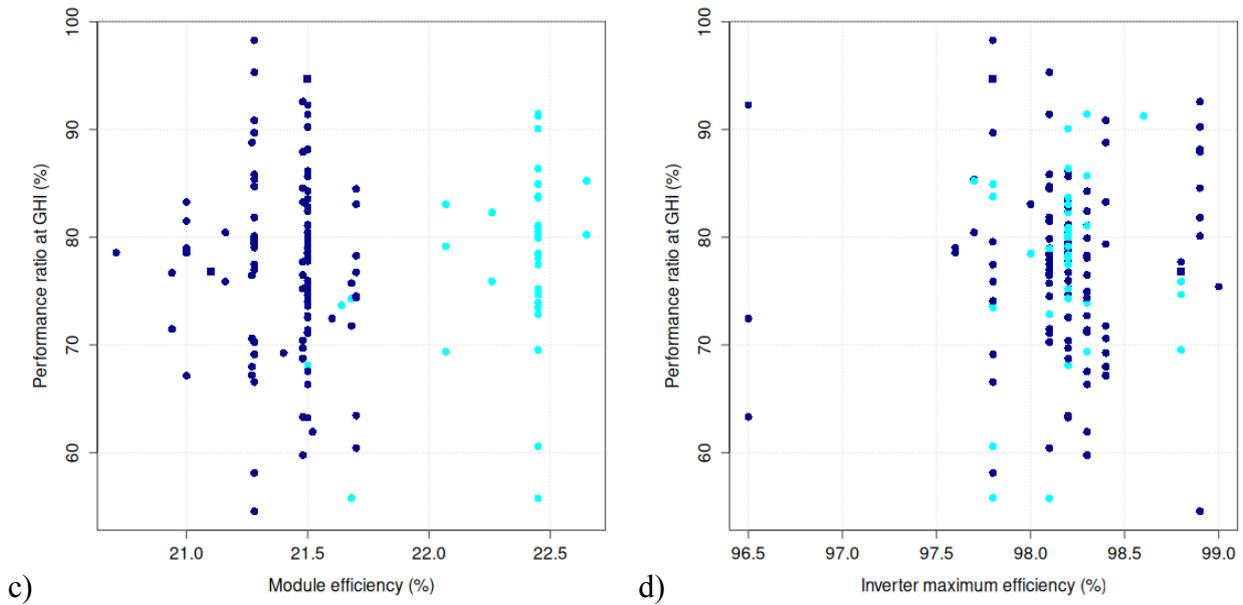

Figure 6. Relationship between annual performance ratio at global horizontal irradiance ($PR_{GHI}$) and inverter sizing factor (ISF; a), DC input power (kWp; b), module efficiency (c), and inverter maximum efficiency (d). Blue represents p-type modules (104 samples), while cyan represents n-type modules (38 samples). Circles represent monofacial modules (140 samples), while squares represent bifacial modules (2 samples).

As expected from their mathematical relationships (Eqs. 2 and 3), a positive correlation was observed between $PR_{GHI}$ and $Y_f$ (Fig. 7a) as well as between $PR_{GHI}$ and CUF (Fig. 7b).

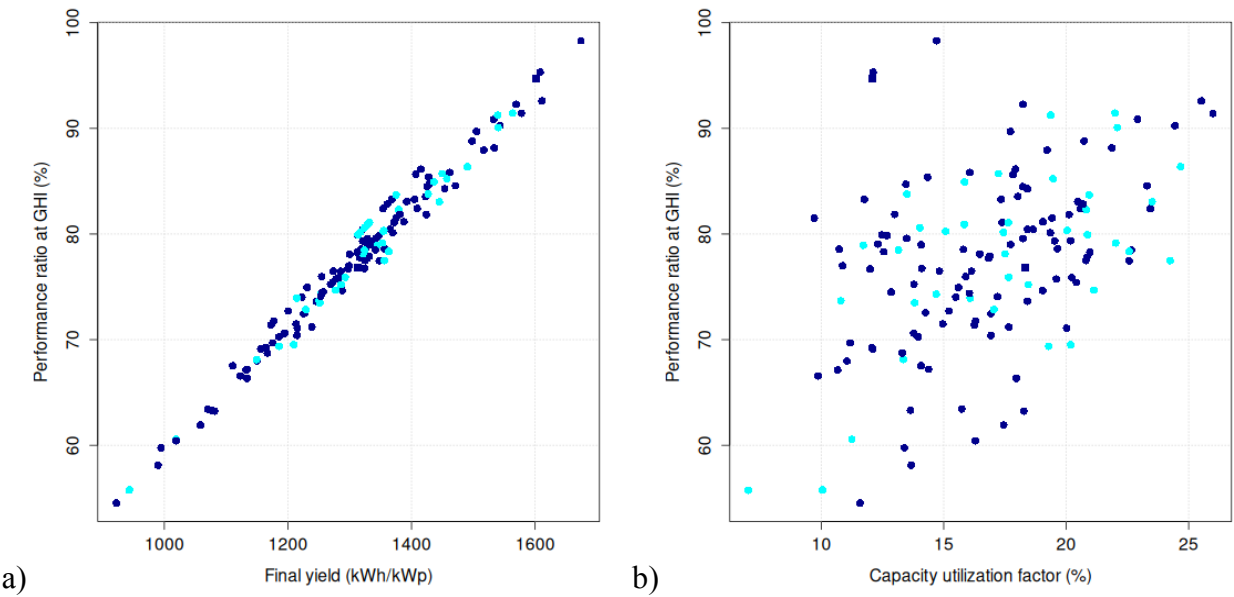

Figure 7. Relationship between annual performance ratio at global horizontal irradiance ($PR_{GHI}$) and annual final yield ($Y_f$; a) and capacity utilization factor (CUF; b). Blue represents p-type modules (104 samples), while cyan represents n-type modules (38 samples). Circles represent monofacial modules (140 samples), while squares represent bifacial modules (2 samples).

No clear effect was seen on $PR_{GHI}$ from $GHI$ (Fig. 8a), direct normal irradiation (Fig. 10Sa), diffuse irradiation (Fig. 10Sb), nor titled latitude irradiation (Fig. 10Sc), indicating that, after controlling for the effects of solar irradiation on $PR_{GHI}$, additional effects of solar irradiation on $PR_{GHI}$ are marginal. In addition, no clear effect was seen for the remaining meteorological variables (maximum temperature, minimum temperature, precipitation, relative humidity, and wind speed at 2 m) aggregated using mean (Fig. 8), median (supplemental Fig. 11S), minimum (Fig. 12S), maximum (Fig. 13S), standard deviation (Fig. 14S), variance (Fig. 15S), sum (Fig. 16S), skewness (data asymmetry; Fig. 17S), and kurtosis (frequency of extreme values; Fig. 18S).

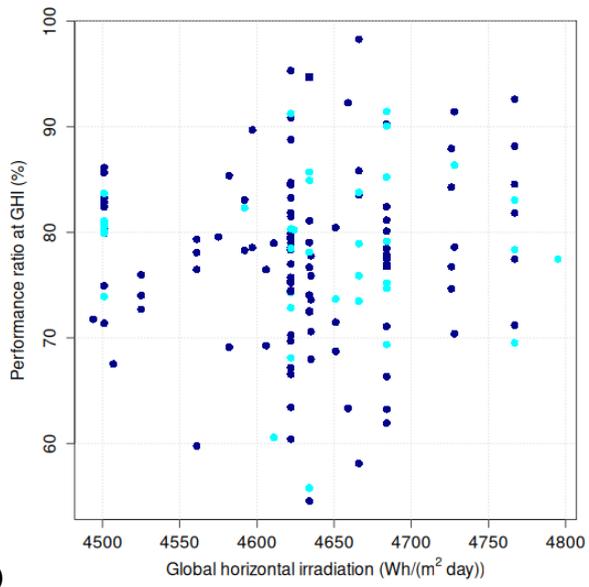
a)
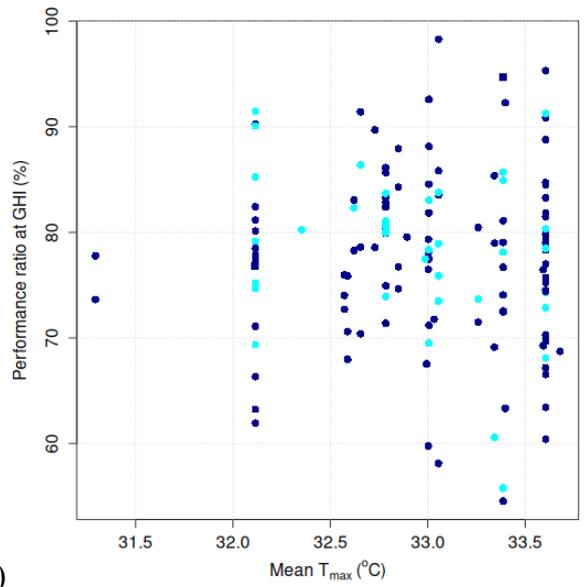
b)
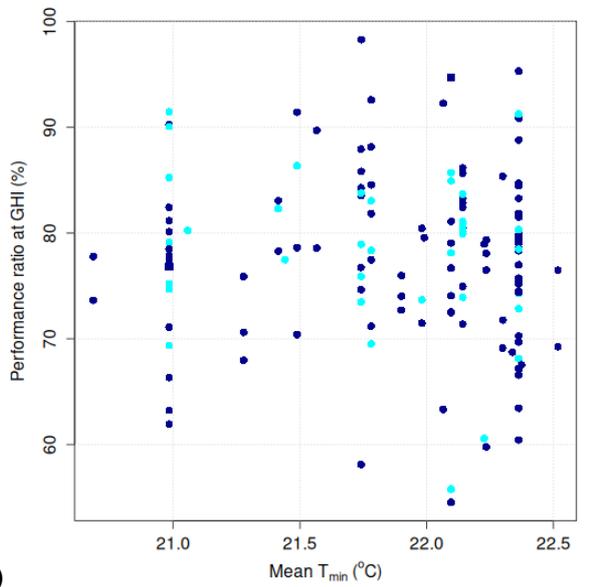
c)
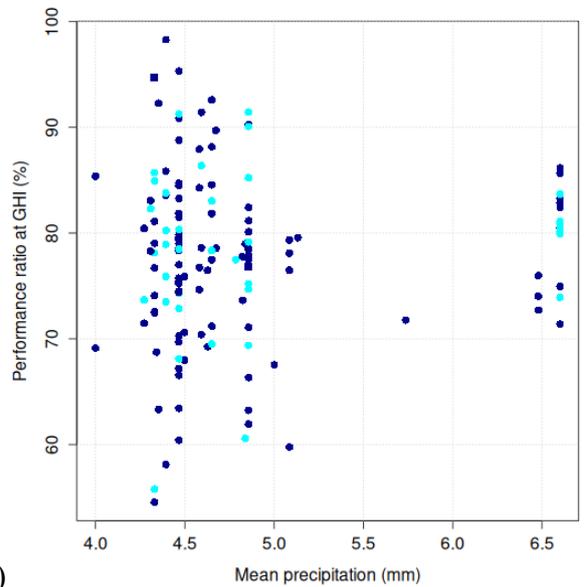
d)
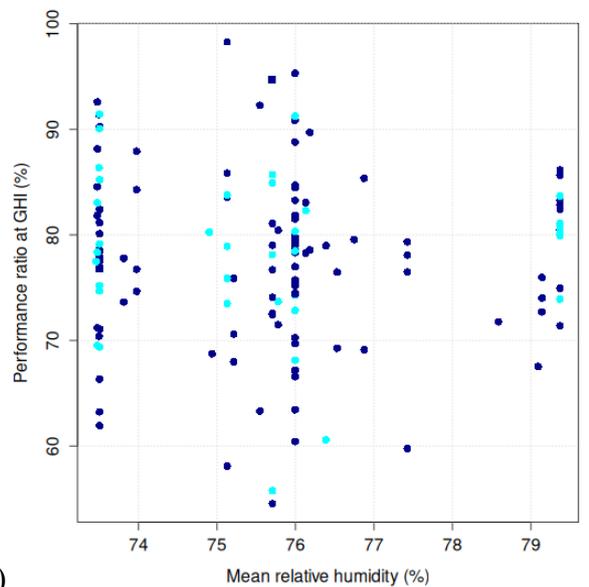
e)
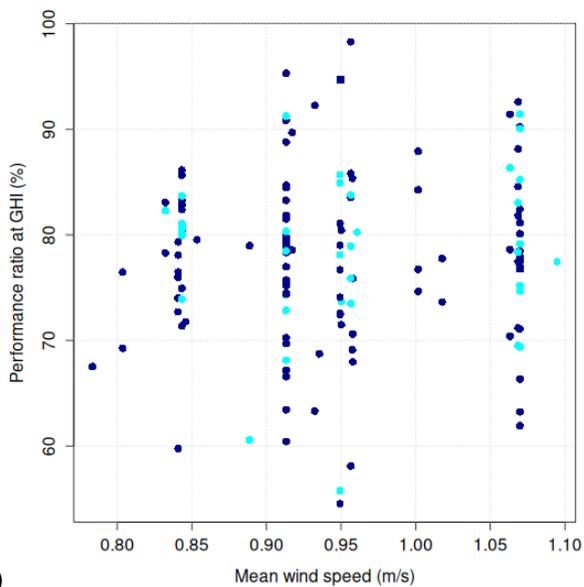
f)

Figure 8. Relationship between annual performance ratio at global horizontal irradiance ($PR_{GHI}$) and global horizontal irradiance (a; Wh/(m$^2$.day)) data from INPE (Pereira et al., 2017) and mean of year 2023 daily maximum temperature (b; °C), minimum temperature (c; °C), precipitation (d; mm), relative humidity (e; %), and wind speed at 2 m (f; m/s) data from BR-DWGD (Xavier et al., 2022). Blue represents p-type modules (104 samples), while cyan represents n-type modules (38 samples). Circles represent monofacial modules (140 samples), while squares represent bifacial modules (2 samples).

## 3.3. Statistical analysis

As expected (Fig. 9a), strong associations were found between *GHI* and remaining meteorological data, with weak associations between technical parameters. Strong correlations between *GHI* and meteorological variables indicate that solar irradiation is strongly associated with local climate. On the other hand, surprisingly, only negligible associations were found between $PR_{GHI}$ and meteorological or technical variables (Fig. 9b), with the exception of $Y_f$ and *CUF*, which are mathematically correlated to $PR_{GHI}$ (Eqs. 2 and 3).

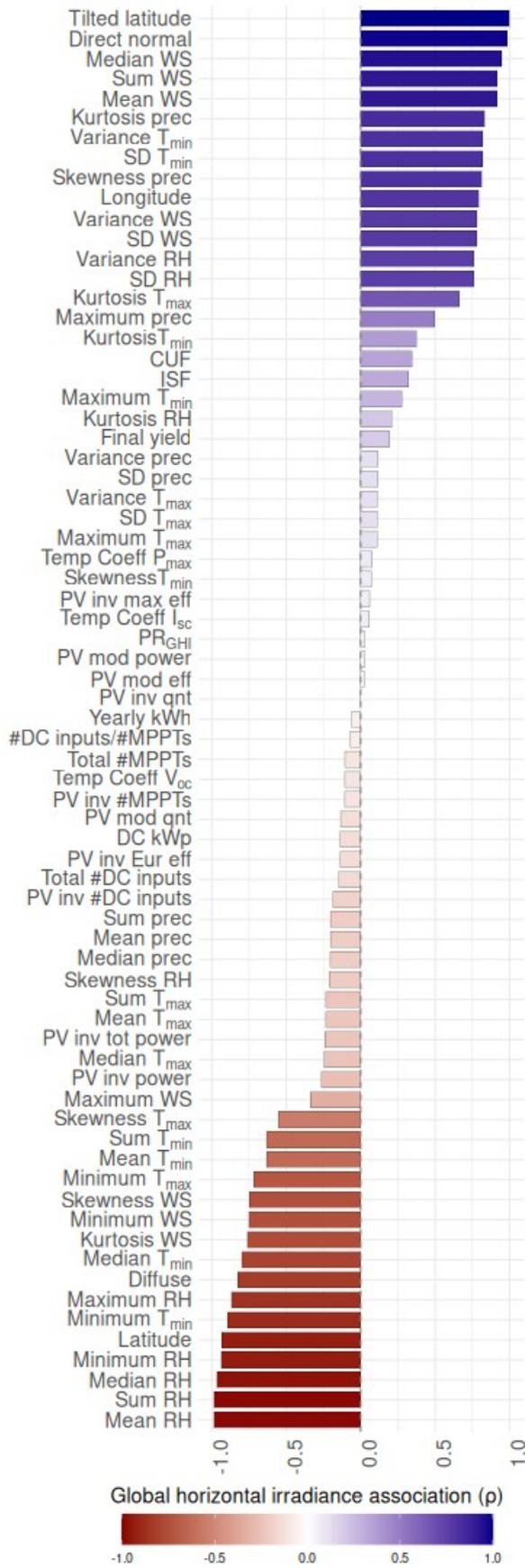
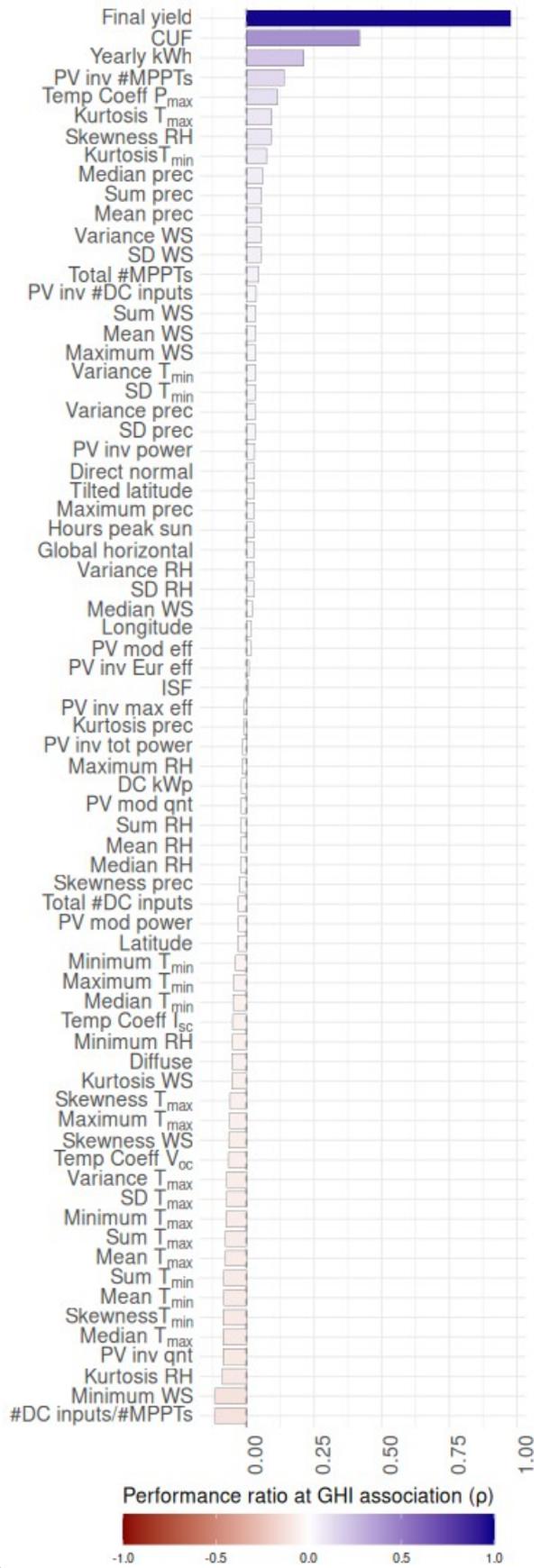

a) b)

Figure 9. Spearman's rank association coefficient ($\rho$) between global horizontal irradiance (*GHI*; a) and annual performance ratio calculated using *GHI* ($PR_{GHI}$; b) and the remaining variables in the dataset. SD: standard deviation; CUF: capacity utilization factor; ISF: inverter sizing factor; $PR_{GHI}$: annual performance ratio calculated at the global horizontal irradiance; RH: relative humidity; WS: wind speed at 2 m; Skewness: measure of data asymmetry; Kurtosis: measure of frequency of extreme values.

Similar to the finds in Fig. 9, solar irradiation and technical parameters were not statistically significant to improve GAM models for annual $PR_{GHI}$ (Fig. 10). Moreover, as seen in the relationship between the best estimation and measured annual $Y_f$ (Fig. 11), a significant portion of the data is not explained by meteorological and/or technical variables, since for a perfect model all points would lie in the red line. In addition to the methodology described in this paper, further data exploration was performed (Milan, 2025) using all meteorological and technical variables to fit a penalized GAM model, to train a Random Forest (RF) optimized via 5-fold repeated cross-validation, and to perform a Recursive Variable Elimination (RVE) strategy. From these analyses, a maximum of 4% $R^2$ was found, reinforcing that meteorological and technical variables do not explain the variance in annual $PR_{GHI}$.

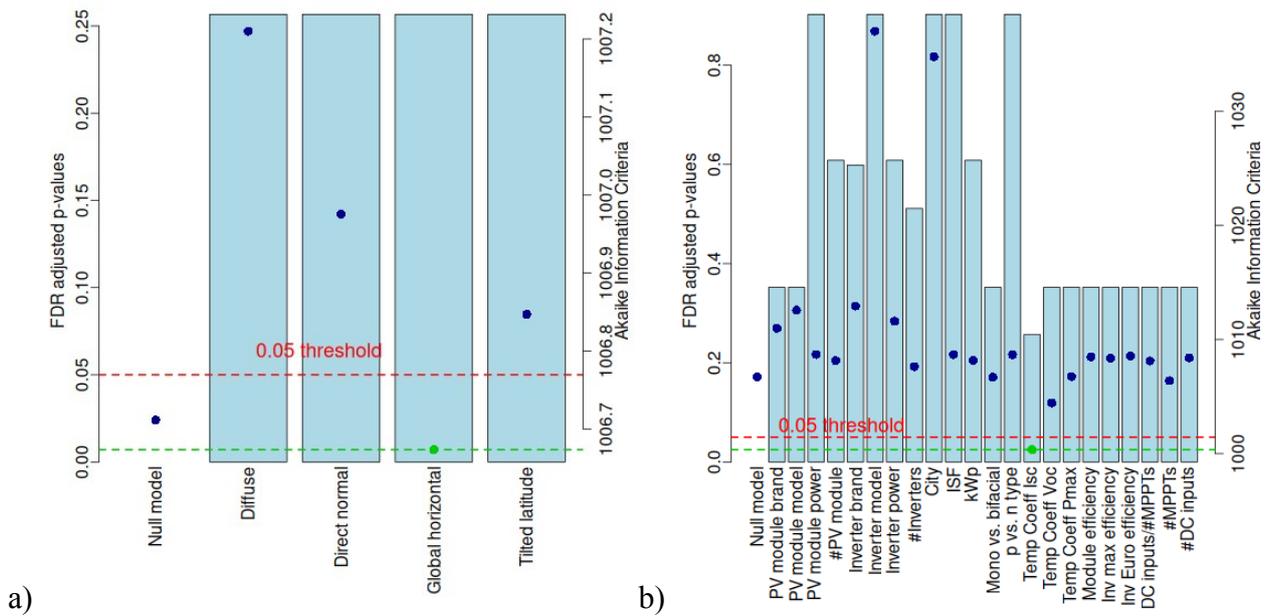

Figure 10. P-values adjusted using the False Discovery Rate (FDR) methodology and Akaike information criteria (AIC) calculated to statistically compare a Generalize Additive Models (GAM) null model (only the intercept) for predicting annual performance ratio at global horizontal irradiance ($PR_{GHI}$) vs. GAM models with solar irradiation (a) and for technical variables (b). Bars represent FDR adjusted p-values. Red dashed line indicates 0.05 p-value threshold for statistically significant more complex models. Blue circles represent AIC values. Green dashed line indicates lowest AIC value and green circle represents lowest AIC value.

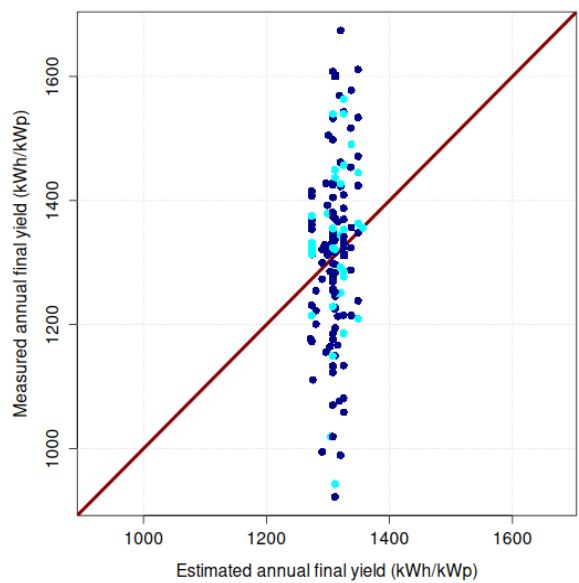

Figure 11. Best estimation vs. measured annual final yield ($Y_f$, kWh/kWp). Blue represents p-type modules, while cyan represents n-type modules. Circles represent monofacial modules, while squares represent bifacial modules. Red line represents y = x line.

When comparing the null model (only the intercept) with GAM models with solar irradiation (Fig. 10a), FDR adjusted p-values were greater than the significance threshold of 0.05 and the AIC values were approximately equal. Even though solar irradiation is the strongest factor affecting energy output (Bamisile et al., 2025), our finding indicate that, after accounting for solar irradiation in annual $PR_{GHI}$, additional effects of solar irradiation under real operational conditions, reported as only ~2% (Louwen et al., 2017), are smaller than other larger effects to be statistically detectable.

When comparing the null model with GAM models for the technical parameters (Fig. 10b), FDR adjusted p-values were, again, greater than 0.05 and AIC values were approximately equal. While the exception was for the GAM model for temperature coefficient for short-circuit current, which showed the smallest AIC value and FDR adjusted p-value slightly greater than 0.05, further statistical investigation (Fig. 19S) concluded that it was a case of overfitting, since non-linear relationships learned from the data were not physically adequate (Fig. 19S). While early studies found that PR tends to increase with kWp and some inverter and module brands perform better, these findings were likely a result of analyzing evolving technologies (Leloux et al., 2015), as recent studies indicate otherwise (Sánchez-Jiménez et al., 2025; Schardt and Heesen, 2021). In addition, commoditization of PV systems and its consequential homogenization (IEA. 2023; Masson et al., 2025) might have reduced advantages of specific brands and technology differences for inverters and modules, as recent studies have not seen significant differences between them (Tsafarakis et al., 2017).

Analogous results were found for maximum temperature, minimum temperature, precipitation, relative humidity, and wind speed at 2 m (Fig. 20S). These results indicate that, although effects of meteorological variables are real (e.g., ~-0.5%/°C; Aarich et al., 2018; Bamisile et al., 2025; ~+1.4%/m/s (Osma-Pinto and Ordóñez-Plata, 2019; ~-0.7%/RH% Rahman et al., 2015), a combination between inverters oversizing (ISF>1), which can compensate for module losses from meteorological conditions, and smaller effects of meteorological variables, might have emphasized that other non-measured variables have a stronger statistical impact on annual $PR_{GHI}$.

### *3.4. Statistical power and data heterogeneity*

Using data-driven models, we surprisingly found that for large-scale, real-world, heterogeneous PV systems, annual $PR_{GHI}$ was not statistically significantly influenced by meteorological variables or technical parameters of the PV system. This contradictory finding can be addressed by the following hypotheses:

1) The statistical procedure was not powerful enough to detect the effects of technical and meteorological variables on annual $PR_{GHI}$

2) The data showed no considerable variability in the technical and meteorological variables (i.e., input data were approximately homogeneous) to detect any effect on annual $PR_{GHI}$

3) Variance of unmeasured variables was higher than the known effects of technical and meteorological variables on annual $PR_{GHI}$

To investigate if hypothesis 1 is valid, the statistical power of this data-driven methodology was investigated. The minimum detectable effect size was $f^2 = 0.056$, which is close to the threshold of detecting small effects (0.02), indicating that this procedure was sufficiently sensitive. In addition, the minimum $R^2$ this data-driven methodology was able to detect was 5.3%, which indicates that this data-driven procedure could detect any single meteorological or technical variable that could explain at least 5.3% (a very small value) of the total variance of annual $PR_{GHI}$. This indicates that this data-driven procedure was powerful enough to identify any variable that could reduce annual $PR_{GHI}$ bootstrapped 95% prediction interval spread by 0.91% (i.e., from approximately 33.77% to 32.86%). Hence, the fact that the data-driven analysis failed to find any significant effects suggests that the true effect size of meteorological and technical variables in large-scale, real-world, heterogeneous PV systems is below this detectable threshold (i.e., explains less than 5.3% of the variance, or reduces the bootstrapped 95% prediction interval spread by less than 0.91%), indicating that hypothesis 1 is unlikely.

To investigate hypothesis 2, standardized values (Z-scores) of the numerical variables as well as Coefficients of Variation (CV) are shown in Fig. 8. Although Fig. 8 clearly shows that data were disperse (i.e., not concentrated around Z = 0), the dispersion (represented by CV) indicates that only technical variables are highly heterogeneous (CV > 20%), whereas most meteorological variables are highly homogeneous (CV < 10%). Hence, the effects of highly heterogeneous technical variables, while physically present, were uncorrelated with annual $PR_{GHI}$ (Fig. 9b) and, consequently, rendered statistically negligible (i.e., explained less than 5.3% of the variance, or reduced the bootstrapped 95% prediction interval spread by less than 0.91%), indicating that hypothesis 2 is unlikely for technical variables.

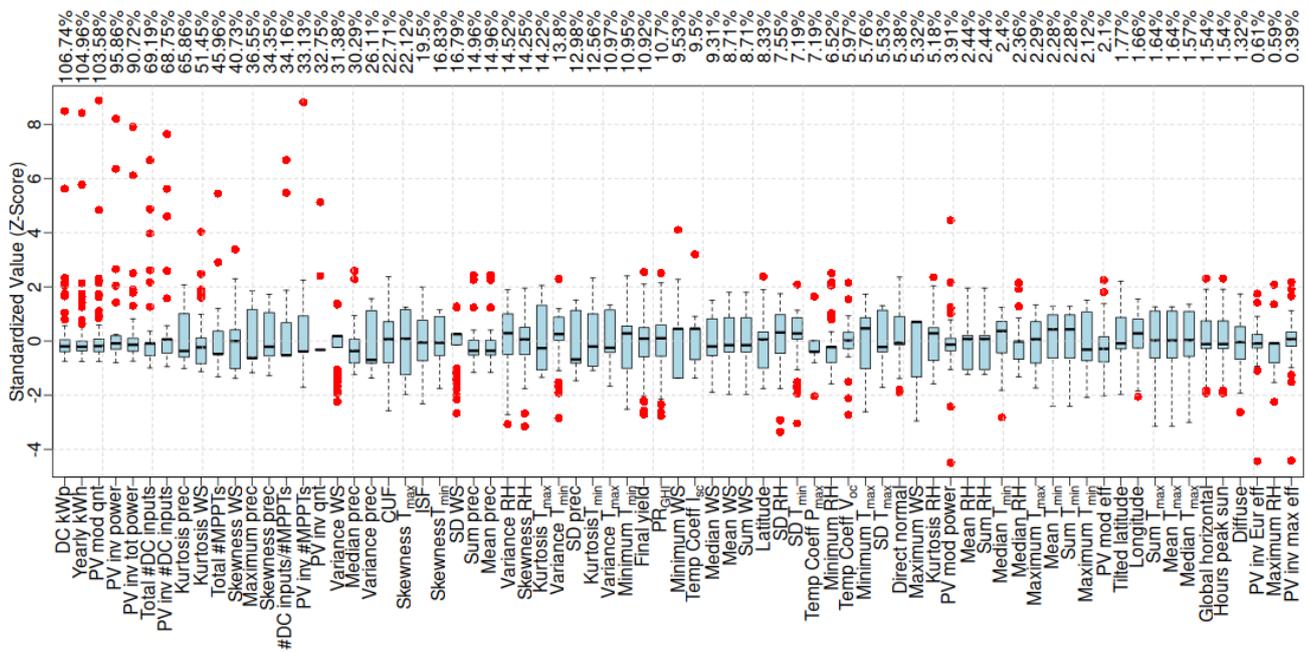

Figure 12. Standardized value (Z-Score) of all numerical variables of the dataset of this paper sorted by highest coefficient of variation (CV; percentage of sample standard deviation over sample mean). SD: standard deviation; CUF: capacity utilization factor; ISF: inverter sizing factor; $PR_{GHI}$: annual performance ratio calculated at the global horizontal irradiance; RH: relative humidity; WS: wind speed at 2 m; Skewness: measure of data asymmetry; Kurtosis: measure of frequency of extreme values.

In addition, statistical effects of meteorological variables on annual $PR_{GHI}$ were not found likely because effects of these variables in large-scale, real-world, heterogeneous, uncontrolled studies are small (Aarich et al., 2018; Bamisile et al., 2025; Louwen et al., 2017; Osma-Pinto and Ordóñez-Plata, 2019; Rahman et al., 2015). As Fig. 8 shows, meteorological data are disperse but concentrated (small CV), indicating that if effects of meteorological variables reduced the bootstrapped 95% prediction interval spread by at least 0.91% (or explained at least 5.3% of the variance), they would have been detected. In addition, current installation practices of using large ISF (>1) might have contributed to reduce the effects of meteorological variables through capping AC output at the inverter power independently of meteorological variables driving DC input power up or down. Consequently, the data indicates that hypotheses 2 can also be ruled out for meteorological variables and can be concluded that the physical effect of meteorological variable on annual $PR_{GHI}$ were small in the context of large-scale, real-world, heterogeneous PV systems.

Considering that hypotheses 1 (data-driven procedure was not statistically powerful) and 2 (data concentration and dispersion were not adequate) were found unlikely, we have a strong indication that the variance of annual $PR_{GHI}$ was significantly affected by more dominant, unmeasured variables, rendering hypothesis 3 likely. Consequently, four main unmeasured variables might have a dominant effect: 1) plane of array irradiation of the installation (which was substituted by GHI to calculate annual $PR_{GHI}$), 2) component degradation, 3) installation quality, 4) monitoring quality, and 5) maintenance quality. Although it can be argued that using GHI to calculate PR introduced bias, it can be counter-argued that this bias is non-systematic, tending to zero for large samples, rendering its effect negligible (Brecl et al., 2022; Killinger et al., 2018; Leloux et al., 2012ab, 2015). Considering that, in real-world PV systems, the tilt/azimuth installation angles follow the random roof orientation, with a large sample, positive (favorable tilt/azimuth angles) and negative (from unfavorable ones) biases are expected to cancel each other out, causing the bias to tend to zero, rendering the effects of this approximation likely small. Moreover, component degradation, while posing real effects on annual $PR_{GHI}$ (Atsu et al., 2020; Bouraiou et al., 2015), are unlike to have considerable effects on data variance because the data came from new installations (most are 2-3 years-old). Hence, unexpectedly, we tend to conclude that human-factors (installation, monitoring, and maintenance) become the more prominent source of variance, as recently proposed (Darghouth et al., 2022; Forrester et al., 2022; Gherghina et al., 2025).

Furthermore, when examined in the context of a rapidly expanding market (approximately 70 installations per day in 2024 alone; ANEEL, 2025), which implies a high number of new, perhaps less-experienced, PV installers and technicians (as personally observed by the authors), quality of human-factors can be expected to be highly variable, as previously mentioned (IEA. 2023; Leloux et al., 2012; Masson et al., 2025). For example, low installation quality can be described as non-optimal tilt/azimuth angles, strings connected to modules in varied tilt/azimuth angles, DC wiring losses (mismatch, wire gauge), poor AC connections, not respecting modules and inverters ventilation distances, installation with permanent shading, improper grounding, and among others. Low monitoring quality can hugely affect performance (Orosz et al., 2024), through not quickly detecting fault, rearming a tripped circuit breaker (Jundi et al., 2026; Nordmann et al., 2014; Tsafarakis et al., 2017), following performance metrics to maintain the system (i.e., removing soiling), and among others. Finally, maintenance quality also have a strong effect (Abdulla et al., 2024; Hernández-Callejo et al., 2019; Lindig et al., 2024), through not proper removing soiling, not removing shading from vegetation growth, not thoroughly inspecting AC/DC connections, and among others.

### 3.5. Probability density function of annual $PR_{GHI}$ and final yield map

Fig. 13 shows the probability and inverse cumulative density functions of annual $PR_{GHI}$. The peak value of annual $PR_{GHI}$ was found as 78.85% (mean 77.52%; 95% confidence interval of 76.12% to 78.84%, 2.72% spread; 95% prediction interval of 58.93% to 92.70%, 33.77% spread). Previous research reported probability density functions with similar form across Europe (Schardt and Heesen, 2021), in Slovenia (Seme et al., 2019), in The Netherlands (Tsafarakis et al., 2017), and in UK (Taylor et al., 2015).

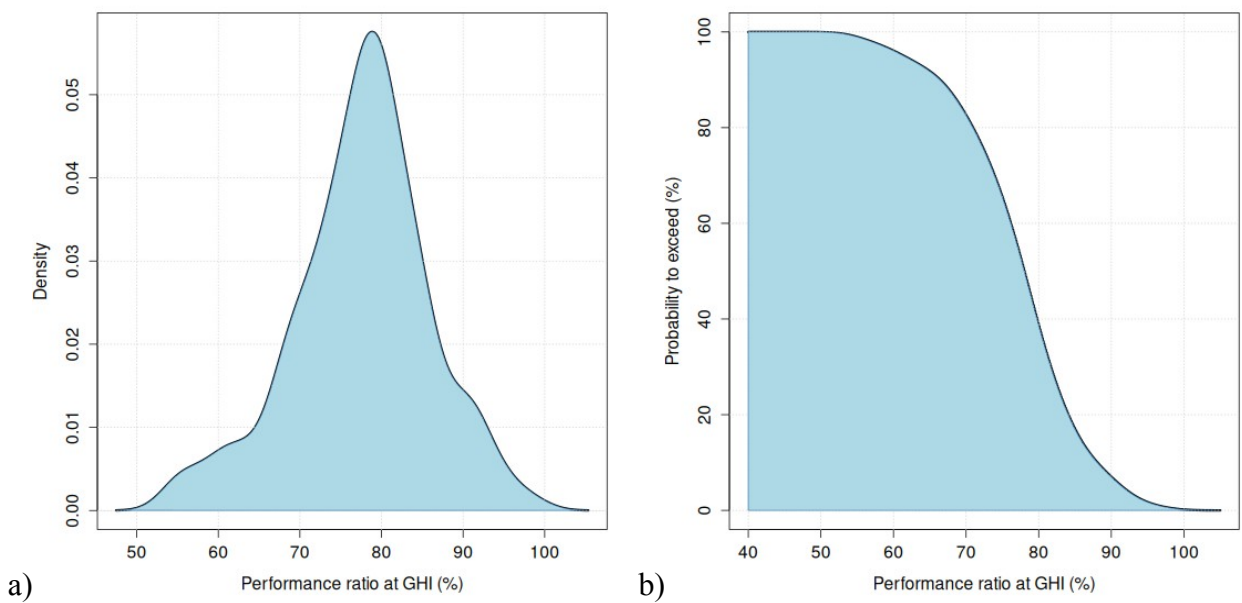

Figure 13. Estimated probability density function (PDF; a) and inverse cumulative density function (survival function; b) for annual performance ratio calculated at the global horizontal irradiance ($PR_{GHI}$).

Since our results indicate that annual $Y_f$ can be estimated solely based annual $PR_{GHI}$ for large-scale, real-world, heterogeneous, in-state PV systems, the best estimation of annual $PR_{GHI}$ can be used to estimate annual $Y_f$, leaving the large, more complex PV systems for advanced simulations, which required specialized skills, powerful computational resources and are time-consuming (Humada et al., 2020; Ríos-Ledesma et al., 2026). For instance, using the bootstrapped mean value of annual $PR_{GHI}$, an entrepreneur can quickly and cheaply estimate the energy production of a given PV system in Rondônia State (Fig. 8a; note that Fig. 8a is a scaled version of *GHI* shown in Fig. 2a; see Fig. 21S for confidence and prediction intervals) or in urban concentrations (Fig. 8b; see Fig. 22S for confidence and prediction intervals). In addition, for risk-controlling, instead of using the bootstrapped mean value of $PR_{GHI}$, an entrepreneur can use one of the bootstrapped percentiles (Tab. 1). For example, using the 10% percentile indicates that only 10% of the PV systems will have a $PR_{GHI}$ lower than 67.10%.

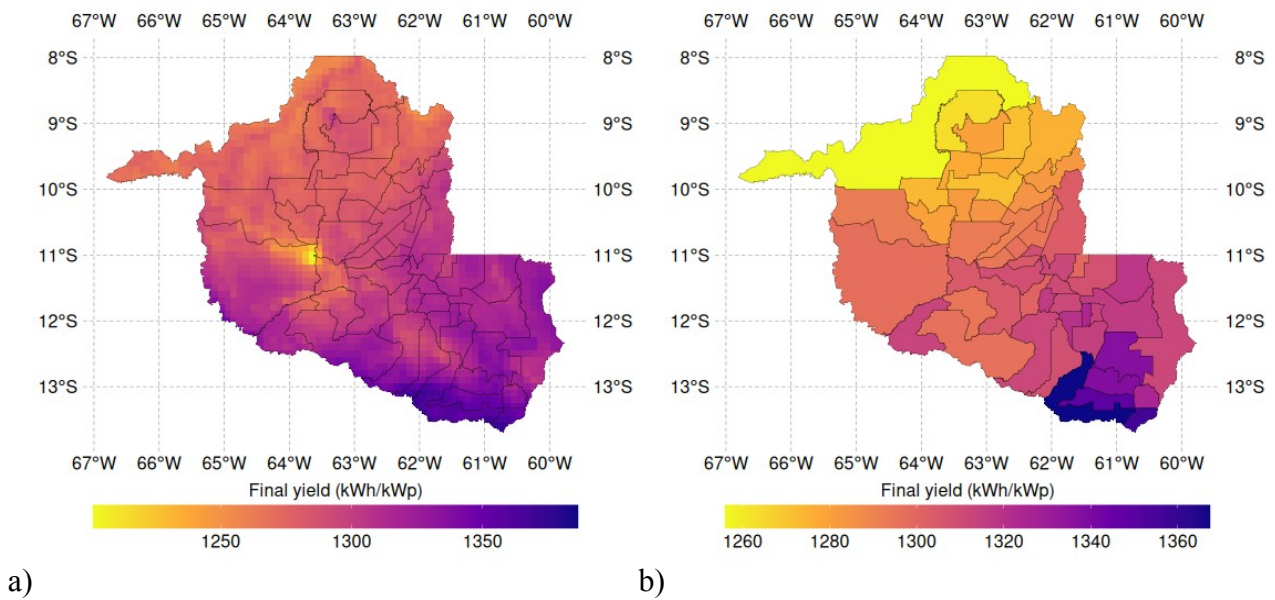

Figure 14. Map of Rondônia State showing city divisions and best estimation of annual final yield ($Y_f$, kWh/kWp) for every geographic position (a) and for the urban concentrations in each city (b).

Table 1: Bootstrapped percentiles for annual performance ratio at global horizontal irradiance ($PR_{GHI}$;%).

| Percentile (%) | Annual performance ratio at global horizontal irradiance ($PR_{GHI}$; %) |
|---:|---:|
| 10 | 67.10 |
| 20 | 71.29 |
| 30 | 74.29 |
| 40 | 76.50 |

| Percentile (%) | Annual performance ratio at global horizontal irradiance ($PR_{GHI}$; %) |
|---|---|
| 50 | 78.27 |
| 60 | 79.59 |
| 70 | 81.29 |
| 80 | 83.73 |
| 90 | 87.53 |

*3.6.   Concluding remarks*

Surprisingly, our data suggest that effects of meteorological and technical variables on annual $PR_{GHI}$ are negligible on large-scale, real-world, heterogeneous PV systems, rendering a substantial variance for unmeasured variables. In addition, our findings suggest that in the rapidly expanding PV market, the focus on optimizing component technology (e.g., n-type vs. p-type modules) may be less impactful than ensuring basic installation (Aboagye et al., 2022; Ismail et al., 2012; Leloux et al., 2015; Mgonja and Saidi, 2017 ), monitoring (Jundi et al., 2026; Nordmann et al., 2014; Orosz et al., 2024; Tsafarakis et al., 2017), and maintenance quality (Abdulla et al., 2024; Hernández-Callejo et al., 2019; Lindig et al., 2024). Hence, our results contribute to recent findings that indicate that, for improved performance, policy makers should focus on creating educational programs to teach PV installers and technicians how to properly install, monitor, and maintain modern PV systems (Darghouth et al., 2022; Forrester et al., 2022; Gherghina et al., 2025).

Ultimately, our results suggests that more research is needed from large-scale, real-world, heterogeneous PV systems, specially studies focused on cooperation between academy and inverter industry, through analyzing their big data from inverters installed worldwide.

## 4. Conclusions

Using data-driven models, we surprisingly found that for large-scale, real-world, heterogeneous PV systems, annual performance ratio calculated using the global horizontal irradiance ($PR_{GHI}$) was not statistically significant influenced by meteorological nor technical parameter of the PV system. Our findings indicate that unmeasured variables human factors (such as installation, monitoring, and maintenance quality) likely have a stronger influence on annual $PR_{GHI}$. These findings indicate that, for improved performance, policy makers should focus on educational programs to teach PV installers and technicians how to properly install, monitor, and maintain modern PV systems. Using the best data-driven model, bootstrapped mean (77.52%), 95% confidence interval (76.12% to 78.84%), and 95% prediction interval (58.83% to 92.70%) of annual $PR_{GHI}$ were calculated. From bootstrapped mean annual $PR_{GHI}$, a map of the best estimated annual final yield ($Y_f$) for Rondônia State was developed. Using the best estimated $Y_f$ allows for quick estimation of annual energy production for simpler cases, leaving more complex cases for simulation software.

**Acknowledgments**

Funding from National Council for Scientific and Technological Development (CNPq; Proc. 446340/2024-3, 314878/2025-4 e 442650/2025-6) and Federal Institute of Rondônia.

**Data availability**

All data and code used in this research are available online (Milan, 2025).

**Competing interests statement**

Nothing to declare.

**Declaration of generative AI in scientific writing**

Nothing to declare.

**Author contributions**

Conceptualization: HFMM & AQA; Data curation: AQA; Formal analysis: HFMM; Funding acquisition: HFMM, JMG, ASCM; Supervision: HFMM; Writing-original draft: AQA; Writing-review & editing: all authors.

# References

AlSkaif T, Dev S, Visser L, Hossari M, van Sark W. A systematic analysis of meteorological variables for PV output power estimation. *Renewable Energy* 153:12-22, 2020. https://doi.org/10.1016/j.renene.2020.01.150

Aarich N, Erraissi N, Akhsassi M, Bennouna A, Asselman A, Barhdadi A, Boukhattem L, Cherkaoui A, Darmane Y, Doudou A, el Fanaoui A, el Omari H, Fahoume M, Hadrami M, Moussaid D, Hartiti B, Ihlal A, Khaidar M, Lfakir A, Loudiyi K, Mabrouki M, Raoufi M, Ridah A, Saadani R, Zorkani I, Aboufirass M. Photovoltaic DC yield maps for all Morocco validated with ground measurements. *Energy for Sustainable Development* 47:158-169, 2018. https://doi.org/10.1016/j.esd.2018.10.003

Abdulla H, Sleptchenko A, Nayfeh A. Photovoltaic systems operation and maintenance: A review and future directions. *Renewable and Sustainable Energy Reviews* 195:114342, 2024. https://doi.org/10.1016/j.rser.2024.114342

Aboagye B, Gyamfi S, Ofosu EA, Djordjevic S. Investigation into the impacts of design, installation, operation and maintenance issues on performance and degradation of installed solar photovoltaic (PV) systems. *Energy for Sustainable Development* 66:165-176, 2022. https://doi.org/10.1016/j.esd.2021.12.003

Agrawl M, Chhajed P, Chowdhury A. Performance analysis of photovoltaic module with reflector: Optimizing orientation with different tilt scenarios. *Renewable Energy* 186:10-25, 2022. https://doi.org/10.1016/j.renene.2021.12.149

Ahmad T, Madonski R, Zhang D, Huang C, Mujeeb A. Data-driven probabilistic machine learning in sustainable smart energy/smart energy systems: Key developments, challenges, and future research opportunities in the context of smart grid paradigm. *Renewable and Sustainable Energy Reviews* 160:112128, 2022.

Al-Badi AH. Measured performance evaluation of a 1.4 kW grid connected desert type PV in Oman. *Energy for Sustainable Development* 47:107-113, 2018. https://doi.org/10.1016/j.esd.2018.09.007

Alvares CA, Stape JL, Sentelhas PC, Gonçalves JLM, Sparovek G. Köppen's climate classification map for Brazil. *Meteorologische Zeitschrift* 2013; 22(6): 711-728. https://dx.doi.org/10.1127/0941-2948/2013/0507

Ameur A, Sekkat A, Loudiyi K, Aggour M. Performance evaluation of different photovoltaic technologies in the region of Ifrane, Morocco. *Energy for Sustainable Development* 52:96-103, 2019. https://doi.org/10.1016/j.esd.2019.07.007

Ascencio-Vásquez J, Osorio-Aravena JC, Brecl K, Muñoz-Cerón E, Topic M. Typical Daily Profiles, a novel approach for photovoltaics performance assessment: Case study on large-scale systems in Chile. *Solar Energy* 225:357-374, 2021. https://doi.org/10.1016/j.solener.2021.07.007

Atsu D, Seres I, Aghaei M, Farkas I. Analysis of long-term performance and reliability of PV modules under tropical climatic conditions in sub-Saharan. *Renewable Energy* 162:258-295, 2020. https://doi.org/10.1016/j.renene.2020.08.021

ANEEL. *Brazilian National Agency of Electric Energy [Agência Nacional de Energia Elétrica]: Dashboard for distributed generation.* https://app.powerbi.com/view?r=eyJrIjoiY2VmMmUwN2QtYWFiOS00ZDE3LWI3NDMtZDk0NGI4MGU2NTkxIiwidCI6IjQwZDZmOWI4LWVjYTctNDZhMi05MmQ0LWVhNGU5YzAxNzBlMSIsImMiOjR9 Acessed on Oct. 17, 2025.

Bamisile O, Acen C, Cai D, Huang Q, Staffell I. The environmental factors affecting solar photovoltaic output. *Renewable and Sustainable Energy Reviews* 208:115073, 2025. https://doi.org/10.1016/j.rser.2024.115073

Banda MH, Nyeinga K, Okello D. Performance evaluation of 830 kWp grid-connected photovoltaic power plant at Kamuzu International Airport-Malawi. *Energy for Sustainable Development* 51:50-55, 2019. https://doi.org/10.1016/j.esd.2019.05.005

Benjamini Y, Hochberg Y. Controlling the false discovery rate: a practical and powerful approach to multiple testing. *Journal of the Royal Statistical Society. Series B (Methodological)* 57(1):289-300, 1995.

Benjamins S, Williamson B, Billing SL, Yuan Z, Collu M, Fox C, Hobbs L, Masden EA, Cottier-Cook EJ, Wilson B. Potential environmental impacts of floating solar photovoltaic systems. *Renewable and sustainable Energy Reviews* 199:114463, 2024. https://doi.org/10.1016/j.rser.2024.114463

Bhakta S, Mukherjee V. Performance indices evaluation and techno economic analysis of photovoltaic power plant for the application of isolated India' island. *Sustainable Energy Technologies and Assessments* 20:9-24, 2017. http://dx.doi.org/10.1016/j.seta.2017.02.002

Bouraiou A, Hamouda M, Chaker A, Mostefaoui M, Lachtar S, Sadok M, Boutasseta N, Othmani M, Issam A. Analysis and evaluation of the impact of climatic conditions on the photovoltaic modules performance in the desert environment. *Energy Conversion and Management* 106:1345-1355, 2015. http://dx.doi.org/10.1016/j.enconman.2015.10.073

Bouacha S, Malek A, Benkraouda O, Hadj Arab A, Razagui A, Boulahchiche S, Semaoui S. Performance analysis of the first photovoltaic grid-connected system in Algeria. *Energy for Sustainable Development* 57:1-11, 2020. https://doi.org/10.1016/j.esd.2020.04.002

Brecl K, Ascencio-Vásquez J, Topic M. Performance of PV systems in Slovenia with the help of typical daily profiles and automatic detection of orientation and inclination angles. *Solar Energy* 236:870-878, 2022. https://doi.org/10.1016/j.solener.2022.03.059

Carr AJ, Pryor TL. A comparison of the performance of different PV module types in temperate climates. *Solar Energy* 76:285-294, 2004. https://doi.org/10.1016/j.solener.2003.07.026

Champely S. *pwr: Basic Functions for Power Analysis*. R package version 1.3-0, 2020. https://doi.org/10.32614/CRAN.package.pwr

Cherfa F, Hadj Arab A, Oussaid R, Abdeladim K, Bouchakour S. Performance analysis of the mini-grid connected photovoltaic system at Algiers. *Energya Procedia* 83:226-236, 2015. https://doi.org/10.1016/j.egypro.2015.12.177

Cohen J. *Statistical power analysis for the behavioral sciences.* 2nd ed. Hillsdale, NJ: Lawrence Erlbaum, 1988.

Conceição R, González-Aguilar J, Merrouni AA, Romero M. Soiling effect in solar energy conversion systems: A review. *Renewable and sustainable Energy Reviews* 162:112434, 2022. https://doi.org/10.1016/j.rser.2022.112434

Dabou R, Bouchafaa F, Arab AH, Bouraiou A, Draou MD, Neçaibia A, Mostefaoui M. Monitoring and performance analysis of grid connected photovoltaic under different climatic conditions in south Algeria. *Energy Conversion and Management* 130:200-206, 2016. http://dx.doi.org/10.1016/j.enconman.2016.10.058

Darghouth NR, O'Shaughnessy E, Forrester S, Barbose G. Characterizing local rooftop solar adoption inequity in the USA. *Environmental Research Letters* 17:034028, 2022.

Dobaria B, Pandya M, Aware M. Analytical assessment of 5.05 kWp grid tied photovoltaic plant performance on the system level in a composite climate of western India. *Energy* 111:47-51, 2016. http://dx.doi.org/10.1016/j.energy.2016.05.082

Efron B, Hastie T. *Computer age statistical inference: algorithms, evidence, and data science*. Cambrige University Press, Cambridge, England, 2021.

Emmanuel M, Akinyele D, Rayudu R. Techno-economic analysis of a 10 kWp utility interactive photovoltaic system at Maungaraki school, Wellington, New Zealand. *Energy* 120:573-583, 2017. http://dx.doi.org/10.1016/j.energy.2016.11.107

Faria AFPA, Maia ASC, Moura GAB, Fonsêca VFC, Nascimento ST, Milan HFM, Gebremedhin KG. Use of solar panels for shade for Holstein heifers. *Animals* 13(3):329, 2023. https://doi.org/10.3390/ani13030329

Ferrada P, Araya F, Marzo A, Fuentealba E. Performance analysis of photovoltaic systems of two different technologies in a coastal desert climate zone of Chile. *Solar Energy* 114:356-363, 2015. http://dx.doi.org/10.1016/j.solener.2015.02.009

Fonsêca VFC, Culhari EZ, Moura GAB, Nascimento ST, Milan HFM, Neto MC, Maia ASC. Shade of solar panels relieves heat load of sheep. *Applied Animal Behaviour Science* 265:105998, 2023. https://dx.doi.org/10.1016/j.applanim.2023.105998

Forrester S, Barbose GL, O'Shaughnessy E, Darghouth NR, Montañés CC. *Residential solar-adopter income and demographic trends: November 2022 update*. Energy Technologies Area, Lawrence Berkeley National Laboratory, 2022.

FUNAI. *Brazilian National Indian Foundation [Fundação Nacional do Índio]: Indigenous conservation lands*. Loaded using function read_indigenous_land() of geobr R-package, 2019.

Gherghina M, Dokshin FA, Leffel B. Unequal solar photovoltaic performance by race and income partly reflects financing models and installer choices. *Nature Energy* 10:697-706, 2025.

Golive YR, Zachariah S, Bhaduri S, Dubey R, Chattopadhyay S, Joshi P, Kusher N, Nandi A, Chindarkar A, Vasudevan DP, Mundle P, Rambabu S, Ajeesh MV, Ingle R, Khattab AA, Singh HK, Arora BM, Narasimhan KL, Kottantharayil A, Vasi J, Shiradkar N, Singh YK, Singh R, Raghava S, Bangar M, Ganesh G, Haldkar AK, Bora B, Kumar S, Kumar R, Tripathi AK. All-India Survey of Photovoltaic Module Reliability: 2018. National Centre for Photovoltaic Research and Education (NCPRE), Indian Institute of Technology Bombay & National Institute of Solar Energy (NISE); Mumbai, India, e Gurugram, Haryana, Índia, 2019.

Guerrero-Lemus R, Cañadillas-Ramallo D, Reindl T, Valle-Feijóo JM. A simple big data methodology and analysis of the specific yield of all PV power plants in a power system over a long time period. *Renewable and Sustainable Energy Reviews* 107:123-132, 2019. https://doi.org/10.1016/j.rser.2019.02.033a

Halwachs M, Neumaier L, Vollert N, Maul L, dimitriadis S, Voronko Y, Eder GC, Omazic A, Mühleisen W, Hirschl C, Schwark M, Berger KA, Ebner R. Statistical evaluation of PV system performance and failure data among different climatic zones. *Renewable Energy* 139:1040-1060, 2019. https://doi.org/10.1016/j.renene.2019.02.135

Heesen H, Herbort V, Rumpler M. Performance of roof-top PV systems in Germany from 2012 to 2018. *Solar Energy* 194:128-135, 2019. https://doi.org/10.1016/j.solener.2019.10.019

Hernández-Callejo L, Gallardo-Saavedra S, Alonso-Gómez V. A review of photovoltaica systems: Design, operation and maintenance. *Solar Energy* 188:426-440, 2019. https://doi.org/10.1016/j.solener.2019.06.017

Humada AM, Darweesh SY, Mohammed KG, Kamil M, Mohammed SF, Kasim NK, Tahseen TA, Awad OI, Mekhilef S. Modeling of PV system and parameter extraction based on experimental data: Review and investigation. *Solar Energy* 199:742-760, 2020. https://doi.org/10.1016/j.solener.2020.02.068

IBGE. *Brazilian Institute of Geography and Statistics [Instituto Brasileiro de Geografia e Estatística]: Urbanized areas in Brazil*, 2023.

IEA. *International Energy Agency: Energy Technology Perspective 2023*. OECD Publishing, Paris, 2023. https://doi.org/10.1787/7c6b23db-en.

IEA. *International Energy Agency: Dual Land Use for Agriculture and Solar Power Production: Overview and Performance of Agrivoltaic Systems 2025*. OECD Publishing, Paris, 2025. https://doi.org/10.69766/XAEU5008

IEC. *IEC Standard 61724-1: Photovoltaic system performance – Part 1: Monitoring*. Geneva, Switzerland; 2021.

Ismail OS, Ajide OO, Akingbesote F. Performance assessment of installed solar PV system: a case study of Oke-Angula in Nigeria. *Engineering* 4:453-458, 2012. http://dx.doi.org/10.4236/eng.2012.48059

Jahn U, Niemann M, Blaesser G, Dahl R, Castello S, Clavadetscher L, Faiman D, Mayer D, van Otterdijk K, Sachau J, Sakuta K, Yamaguchi M, Zoglauer M. *International Energy Agency Task II database on photovoltaic power systems: statistical and analytical evaluation of PV operational data*. 15th European Photovoltaic Solar Energy Conference, Vienna, Austria, 1998.

Jahn U, Mayer D, Heidenreich M, Dahl R, Castello S, Clavadetscher L, Frölich A, Grimming B, Nasse W, Sakuta K, Sugiura T, van der Borg N, van Otterdijk K. *International Energy Agency PVPS Task 2: analysis of the operational performance of the IEA database PV Systems*. 16th European Photovoltaic Solar Energy Conference and Exhibition, Glasgow, United Kingdom, 2000.

Jean A, Rosentrater KA. Economic and Environmental Outlook on Agrivoltaics: Review and Perspectives. *Energies* 18:5836, 2025. https://doi.org/10.3390/en18215836

Leloux J, Narvarte L, Trebosc D. Review of the performance of residential PV systems in Belgium. *Renewable and Sustainable Energy Reviews* 16:178-184, 2012a. dx.doi.org/10.1016/j.rser.2011.07.145

Leloux J, Narvarte L, Trebosc D. Review of the performance of residential PV systems in France. *Renewable and Sustainable Energy Reviews* 16:1369-1376, 2012b. https://dx.doi.org/10.1016/j.rser.2011.10.018

Leloux J, Taylor J, Moretón R, Narvarte L, Trebosc D, Desportes A. *Monitoring 30,000 PV systems in Europe: Performance, faults, and state of the art.* 31st European Photovoltaic Solar Energy Conference and Exhibition, Hamburg, 2015.

Jundi ZS, Al-Waeli AHA, Switzner NT. A systematic review of low-cost photovoltaic monitoring systems: Technologies, challenges, and opportunities. *Renewable and Sustainable Energy Reviews* 226:116417, 2026. https://doi.org/10.1016/j.rser.2025.116417

Kausika BB, Moraitis P, van Sark WGJHM. Visualization of Operational Performance of Grid-Connected PV Systems in Selected European Countries. *Energies* 11:1330, 2018. http://dx.doi.org/10.3390/en11061330

Khalid AM, Mitra I, Warmuth W, Schacht V. Performance ratio – Crucial parameter for grid connected PV plants. *Renewable and Sustainable Energy Reviews* 65:1139-1158, 2016. http://dx.doi.org/10.1016/j.rser.2016.07.066

Killinger S, Lingfors D, Saint-Drenan YM, Moraitis P, van Sarks W, Taylor J, Engerer NA, Bright JM. On the search for representative characteristics of PV systems: Data collection and analysis of PV system azimuth, tilt, capacity, yield and shading. *Solar Energy* 173:1087-1106, 2018. https://doi.org/10.1016/j.solener.2018.08.051

Kumar M, Kumar A. Performance assessment and degradation analysis of solar photovoltaic technologies: A review. *Renewable and Sustainable Energy Reviews* 78:554-587, 2017. http://dx.doi.org/10.1016/j.rser.2017.04.083

Kymakis E, Kalykakis S, Papazoglou TM. Performance analysis of a grid connected photovoltaic park on the island of Crete. *Energy Conversion and Management* 50:433-438, 2009.

Libra M, Mrázek D, Tyukhov I, Severová L, Poulek V, Mach J, Subrt T, Beránek V, Svoboda R, Sedlácek J. Reduced real lifetime of PV panel – Economic consequences. *Solar Energy* 259:229-234, 2023. https://doi.org/10.1016/j.solener.2023.04.063

Lima LC, Ferreira LA, Morais FHBL. Performance analysis of a grid connected photovoltaic system in northeastern Brazil. *Energy for Sustainable Development* 37:79-85, 2017. http://dx.doi.org/10.1016/j.esd.2017.01.004

Lindig S, Ascencio-Vásquez J, Leloux J, Moser D, Reinder A. Performance analysis and degradation of a large fleet of PV systems. *IEEE Journal of Photovoltaics* 11(5):1312-1318, 2021. https://doi.org/10.1109/JPHOTOV.2021.3093049

Lindig S, Herz M, Ascencio-Vásquez K, Theristis M, Herteleer B, Deckx J, Anderson K. Review of Technical Photovoltaic Key Performance Indicators and the Importance of Data Quality Routines. *Solar RRL* 8:2400634, 2024. https://doi.org/10.1002/solr.202400634

Liu Y, Li H, Li L, Yin X, Wu X, Su Z, Gao F, Cai B, Tang L, Zhou S. Solar photovoltaic panel soiling accumulation and removalmethods: A review. *IET Renewable Power Generation* 18(6):4097-4118, 2024. https://doi.org/10.1049/rpg2.12940

Louwen A, de Wall AC, Schropp REI, Faaij APC, van Sark WGJHM. Comprehensive characterisation and analysis of PV module performance under real operating conditions. *Progress in Photovoltaics: Research and Applications* 25:218-232, 2017. https://doi.org/10.1002/pip.2848

Ma T, Yang H, Lu L. Long term performance analysis of a standalone photovoltaic system under real conditions. *Applied Energy* 201:320-331, 2017. http://dx.doi.org/10.1016/j.apenergy.2016.08.126

Maia ASC, Culhari EA, Fonsêcia VFC, Milan HFM, Gebremedhin KG. Photovoltaic panels as shading resources for livestock. *Journal of Cleaner Production* 258 :120551, 2020. https://doi.org/10.1016/j.jclepro.2020.120551

Makrides G, Zinsser B, Phinikarides A, Schubert M, Georghiou GE. Temperature and thermal annealing effects on different photovoltaic technologies. *Renewable Energy* 43:407-417, 2012. https://dx.doi.org/10.1016/j.renene.2011.11.046

Malvoni M, Leggieri A, Maggiotto G, Congedo PM, Giorgi MGC. Long term performance, losses and efficiency analysis of a 960 kWp photovoltaic system in the Mediterranean climate. *Energy Conversion and Management* 145:169-181, 2017. http://dx.doi.org/10.1016/j.enconman.2017.04.075

Malvoni M, Kumar NM, Chopra SS, Hatziargyriou N. Performance and degradation assessment of large-scale grid-connected solar photovoltaic power plant in tropical semi-arid environment of India. *Solar Energy* 203:101-113, 2020. https://doi.org/10.1016/j.solener.2020.04.011

Masson G, L'Epine M, Kaizuka I, Okazaki J. Trends in PV applications 2025. *International Energy Agency – Photovoltaic Power Systems Programme*, 2025. https://doi.org/10.69766/NCNN2417

Mehdi M, Ammari N, Merrouni AA, Gallasi HE, Dahmani M, Ghennioui A. An experimental comparative analysis of different PV technologies performance including the influence of hot-arid climatic parameters: Toward a realistic yield assessment for desert locations. *Renewable Energy* 205:695-716, 2023. https://doi.org/10.1016/j.renene.2023.01.082

Meng B, Loonen RCGM, Hensen JLM. Performance variability and implications for field prediction of rooftop PV systems – Analysis of 246 identical systems. *Applied Energy* 233:119550, 2022. https://doi.org/10.1016/j.apenergy.2022.119550

Mensah LD, Yamoah JO, Adaramola MS. Performance evaluation of a utility-scale grid-tied solar photovoltaic (PV) installation in Ghana. *Energy for Sustainable Development* 48:82-87, 2019. https://doi.org/10.1016/j.esd.2018.11.003

Mgonja CT, Saidi H. Effectiveness on implementation of maintenance management system for off-grid solar PV systems in public facilities – a case study of SSMP1 project in Tanzania. *International Journal of Mechanical Engineering and Technology* 8(7):869-880, 2017.

Micheli L, Wilbert S. Understanding, measuring and mitigating soiling losses in PV power systems. *International Energy Agency – Photovoltaic Power Systems Programme*, 2025.

Milan HFM. Performance_Ration_PV_Systems: Data and analysis for performance ratio of solar energy PV systems in Rondônia, Brazil. Available at https://github.com/hugomilan/performance_ratio_PV_systems, 2025. https://doi.org/10.5281/zenodo.266350

Minemoto T, Nagae S, Takakura H. Impact of spectral irradiance distribution and temperature on the outdoor performance of amorphous Si photovoltaic modules. *Solar Energy Materials and Solar Cells* 91:919-923, 2007. https://dx.doi.org/10.1016/j.solmat.2007.02.012

MMA. *Brazilian Ministry of Environment and Climate Change [Ministério do Meio Ambiente e Mudança do Clima]: Environmental conservation units*, 2019.

Necaibia A, Bouraiou A, Ziane A, Sahouane N, Hassani S, Mostefaoui M, Dabou R, Mouhadjer S. Analytical assessment of the outdoor performance and efficiency of grid-tied photovoltaic system under hot dry climate in the south of Algeria. *Energy Conversion and Management* 171:778-786, 2018. https://doi.org/10.1016/j.enconman.2018.06.020

Nordmann T, Clavadetscher L, van Sark WGJHM, Green M. Analysis of long-term performance of PV Systems: Different data resolution for different purposes. *International Energy Agency – Photovoltaic Power Systems Programme*, 2014.

Ramanan P, Murugavel KK, Karthick A. Performance analysis and energy metrics of grid-connected photovoltaic systems. *Energy for Sustainable Development* 52:104-115, 2019. https://doi.org/10.1016/j.esd.2019.08.001

Nascimento LR, Braga M, Campos RA, Naspolini HF, Rüther R. Performance assessment of solar photovoltaic technologies under different climatic conditions in Brazil. *Renewable Energy* 146:1070-1082, 2020.

Okello D, van Dyk EE, Vorster FJ. Analysis of measured and simulated performance data of a 3.2 kWp grid-connected PV system in Port Elizabeth, South Africa. *Energy Conversion and Management* 100:10-15, 2015. http://dx.doi.org/10.1016/j.enconman.2015.04.064

Orosz T, Rassõlkin A, Arsénio P, Poór P, Valme D, Sleisz A. Current Challenges in Operation, Performance, and Maintenance of Photovoltaic Panels. *Energies* 17:1306, 2024. https://doi.org/10.3390/en17061306

Osma-Pinto G, Ordóñez-Planta G. Measuring factors influencing performance of rooftop PV panels in warm tropical climates. *Solar Energy* 185:112-123, 2019. https://doi.org/10.1016/j.solener.2019.04.053

Oyewo AS, Kunkar A, Satymov R, Breyer C. A multi-sector, multi-node, and multi-scenario energy system analysis for the Caribbean with focus on the role of offshore floating photovoltaics. *Renewable and Sustainable Energy Reviews* 210:115189, 2025. https://doi.org/10.1016/j.rser.2024.115189

Pereira EB, Martins FR, Gonçalves AR, Costa RS, Lima FL, Rüther R, Abreu SL, Tiepolo GM, Pereira SV, Souza JG. *Brazilian Atlas of Solar Energy [Atlas brasileiro de energia solar]*. 2nd.ed. São José dos Campos: INPE, 2017. https://doi.org/10.34024/978851700089

Pereira HMF, Gonçalves CN. *Geobr: download official spatial data sets of Brazil*. R package version 1.9.1, 2024. https://doi.org/10.32614/CRAN.package.geobr

R Core Team. *R: A language and environment for statistical computing*. R Foundation for Statistical Computing, Vienna, Austria, 2025.

Rahman MM, Hasanuzzaman M, Rahim. Effects of various parameters on PV-module power and efficiency. *Energy Conversion and Management* 103:348-358, 2015. http://dx.doi.org/10.1016/j.enconman.2015.06.067

Ríos-Ledesma F. Barrutia L, Ramírez-Ledesma J, Narvarte L, Lorenzo E. Angular losses in photovoltaic energy estimation: A review of models and a comparative analysis using the enhanced SISIFO simulation tool. *Renewable Sustainable Energy Reviews* 226:116377, 2026. https://doi.org/10.1016/j.rser.2025.116377

Rüther R, Dacoregio MM. Performance assessment of a 2 kWp grid-connected, building-integrated, amorphous silicon photovoltaic installation in Brazil. *Progress in Photovoltaics* 8(2):257-266, 2000. https://doi.org/10.1002/(SICI)1099-159X(200003/04)8:2%3C257::AID-PIP292%3E3.0.CO;2-P

Sahin AD, Kara T, Mut AO, Korkmaz MS, Kaymak MK, Güloglu B. A comprehensive review of economic and policy considerations for floating photovoltaic systems: insight on national policies and sustainable development goals. *Renewable ans Sustainable Energy Reviews* 223:116024, 2025. https://doi.org/10.1016/j.rser.2025.116024

Sahouane N, Dabou R, Ziane A, Neçaibia A, Bouraiou A, Rouabhia A, Mohammed B. Energy and economic efficiency performance assessment of a 28 kWp photovoltaic grid-connected system under desertic weather conditions in Algeria Sahara. *Renewable Energy* 143:1318-1330, 2019. https://doi.org/10.1016/j.renene.2019.05.086

Saincher S, Sriram V, Stoesser T. Shared moorings for floating offshore renewable energy technologies: a review. *Renewable ans Sustainable Energy Reviews* 224:116064, 2025. https://doi.org/10.1016/j.rser.2025.116064

Sánchez-Jiménez JL, Jiménez-Castillo G, Rus-Casas C, Martínez-Calahorro AJ, Muñoz-Rodriguez FJ. Performance evaluation of photovoltaic self-consumption systems on industrial rooftops under continental Mediterranean climate conditions with multi-string inverter topology. *Energy Reports* 14:1020-1042, 2025. https://doi.org/10.1016/j.egyr.2025.06.047

Satsangi KP, Das DB, Babu GSS, Saxena AK. Performance analysis of grid interactive solar photovoltaic plant in India. *Energy for Sustainable Development* 47:9-16, 2018. https://doi.org/10.1016/j.esd.2018.08.003

Schardt J, Heesen H. Performance of roof-top PV systems in selected European countries from 2012 to 2019. *Solar Energy* 217:235-244, 2021. https://doi.org/10.1016/j.solener.2021.02.001

Seme S, Sredensek K, Stumberger B, Hadziselimovic M. Analysis of the performance of photovoltaaic systems in Slovenia. *Solar Energy* 180:550-558, 2019. https://doi.org/10.1016/j.solener.2019.01.062

Shadid R, Khawaja Y, Bani-Abdullah A, Akho-Zahieh M, Allahham A. Investigation of weather conditions on the output power of various photovoltaic systems. *Renewable Energy* 217:119202, 2023. https://doi.org/10.1016/j.renene.2023.119202

Shahzad N, Hussain N, Umar S, Azam MF, Yousaf T, Waqas A. Impacts of soiling on solar panel performance and state-of-the-art effective cleaning methods: A recent review. Journal of Cleaner Production 497:145119, 2025. https://doi.org/10.1016/j.jclepro.2025.145119

Sidi CEBE, Ndiaye ML, Bah ME, Mbodji A, Ndiaye A, Ndiaye PA. Performance analysis of the first large-scale (15 MWp) grid-connected photovoltaic plant in Mauritania. *Energy Conversion and Management* 119:411-421, 2016. http://dx.doi.org/10.1016/j.enconman.2016.04.070

Silva AM, Melo FC, Reis JH, Freitas LCG. The study and evaluation methods for photovoltaic modules under real operational conditions, in a region of the Brazilian Southeast. *Renewable Energy* 138:1189-1204, 2019.

Simioni T, Schaeffer R. Georeferenced operating-efficiency solar potential maps with local weather conditions – An application to Brazil. *Solar Energy* 184:345-355, 2019. https://doi.org/10.1016/j.solener.2019.04.006

Storey JD, Tibshirani R. Statistical significance for genomewide studies. *Proceedings of the National Academy of Sciences (PNAS)* 100(16):9440-9445, 2003. https://doi.org/10.1073/pnas.1530509100

Taylor J, Leloux J, Hall LMH, Everard AM, Briggs J, Buckley A. Performance of distributed PV in the UK: a statistical analysis of over 7000 systems. *31st European Photovoltaic Solar Energy Conference and Exhibition* 2263-2268, 2015.

Tsafarakis O, Moraitis P, Kausika BB, van der Velde H, 't Hart S, de Vries A, de Rijk P, de Jong MM, van Leeuwen HP, van Sark W. Three years experience in a Dutch public awareness campaign on photovoltaic system performance. *IET Renewable Power Generation* 11(10):1229-1233, 2017. https://doi.org/10.1049/iet-rpg.2016.1037

Vasisht MS, Srinivasan J, Ramasesha SK. Performance of photovoltaic installations: Effect of seasonal variations. *Solar Energy* 131:39-46, 2016. http://dx.doi.org/10.1016/j.solener.2016.02.013

Vasuki SS, Levell J, Santbergen R, Isabella O. A technical review on the energy yield estimation of offshore floating photovoltaic systems. *Renewable and Sustainable Energy Reviews* 216:115596, 2025. https://doi.org/10.1016/j.rser.2025.115596

Vaverková MD, Kousal M, Kosakiewicz M, Krysińska K, Winkler J. Agrivoltaics for sustainable land use: A critical review of synergistic and antagonistic effects. *Renewable and Sustainable Energy Reviews* 226(E):116482, 2026. https://doi.org/10.1016/j.rser.2025.116482

Venkateswari R, Sreejith. Factor influencing the efficiency of photovoltaic system. *Renewable and Sustainable Energy Reviews* 101:376-394, 2019. https://doi.org/10.1016/j.rser.2018.11.012

Wang H, Muñoz-García MA, Moreda GP, Alonso-García MC. Seasonal performance comparison of three grid connected photovoltaic systems based on different technologies operating under the same condition. *Solar Energy* 144:789-807, 2017. http://dx.doi.org/10.1016/j.solener.2017.02.006

Wei Y, Khojasteh D, Windt C, Huang L. An interdisciplinary literature review of floating solar power plants. *Renewable and Sustainable Energy Reviews* 209:115094, 2025. https://doi.org/10.1016/j.rser.2024.115094

Wood SN. *Generalized Additive Models: An introduction with R*. 2nd. ed. New York: Chapman and Hall/CRC, 2017. https://doi.org/10.1201/9781315370279

Xavier AC, Scanlon BR, King CW, Alves AI. New improved Brazilian daily weather gridded data (1961–2020). *International Journal of Climatology* 42(16):8390–8404, 2022. https://doi.org/10.1002/joc.7731

Yadav SK, Bajpai U. Performance evaluation of a rooftop solar photovoltaic power plant in Northern India. *Energy for Sustainable Development* 43:130-138, 2018. https://doi.org/10.1016/j.esd.2018.01.006

Supplementary Material

# The PV performance ratio paradox: annual data from large-scale, real-world PV systems show negligible meteorological and technical impact and points to dominant human factors

Hugo FM Milan[a*], Aline Q Alves[a], Thatiane AT Souza[b], Juliana M Galo[b], Alex SC Maia[c], Moisés AP Borges[a], Ciro J Egoavil[a]

[a]Electrical Engineering Department, Federal University of Rondônia

[b]Federal Institute of Rondônia

[c]State University of São Paulo, Jaboticabal Campus

*Corresponding author: hugo.milan@unir.br, DAEE BR 364, km 9,5, Porto Velho/RO, CEP 76.801-059

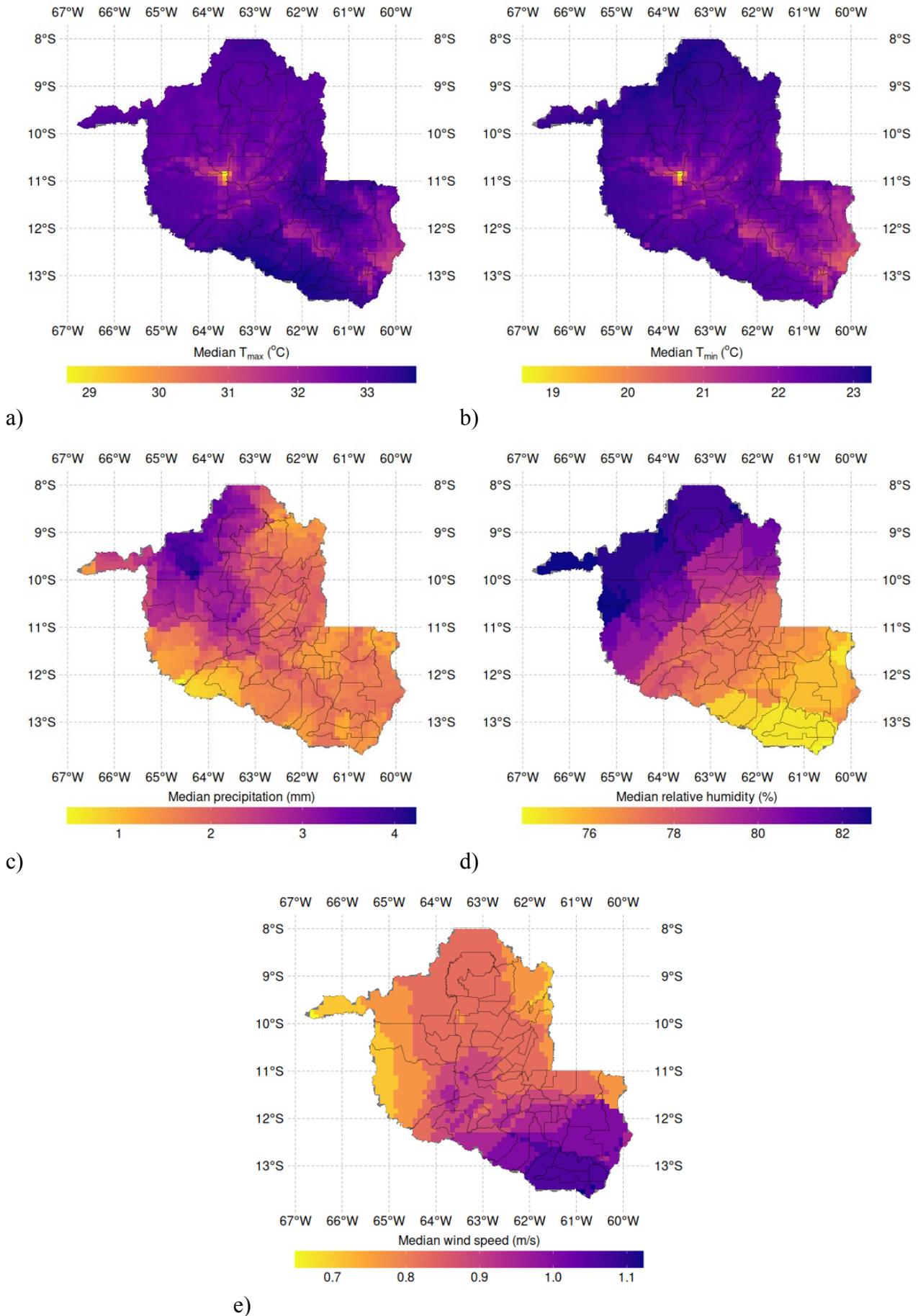

Figure 1S. Map of Rondônia State, showing city divisions and median of year 2023 daily maximum temperature (a; ºC), minimum temperature (b; ºC), precipitation (c; mm), relative humidity (d; %), and wind speed at 2 m (e; m/s) data from BR-DWGD (Xavier et al., 2022).

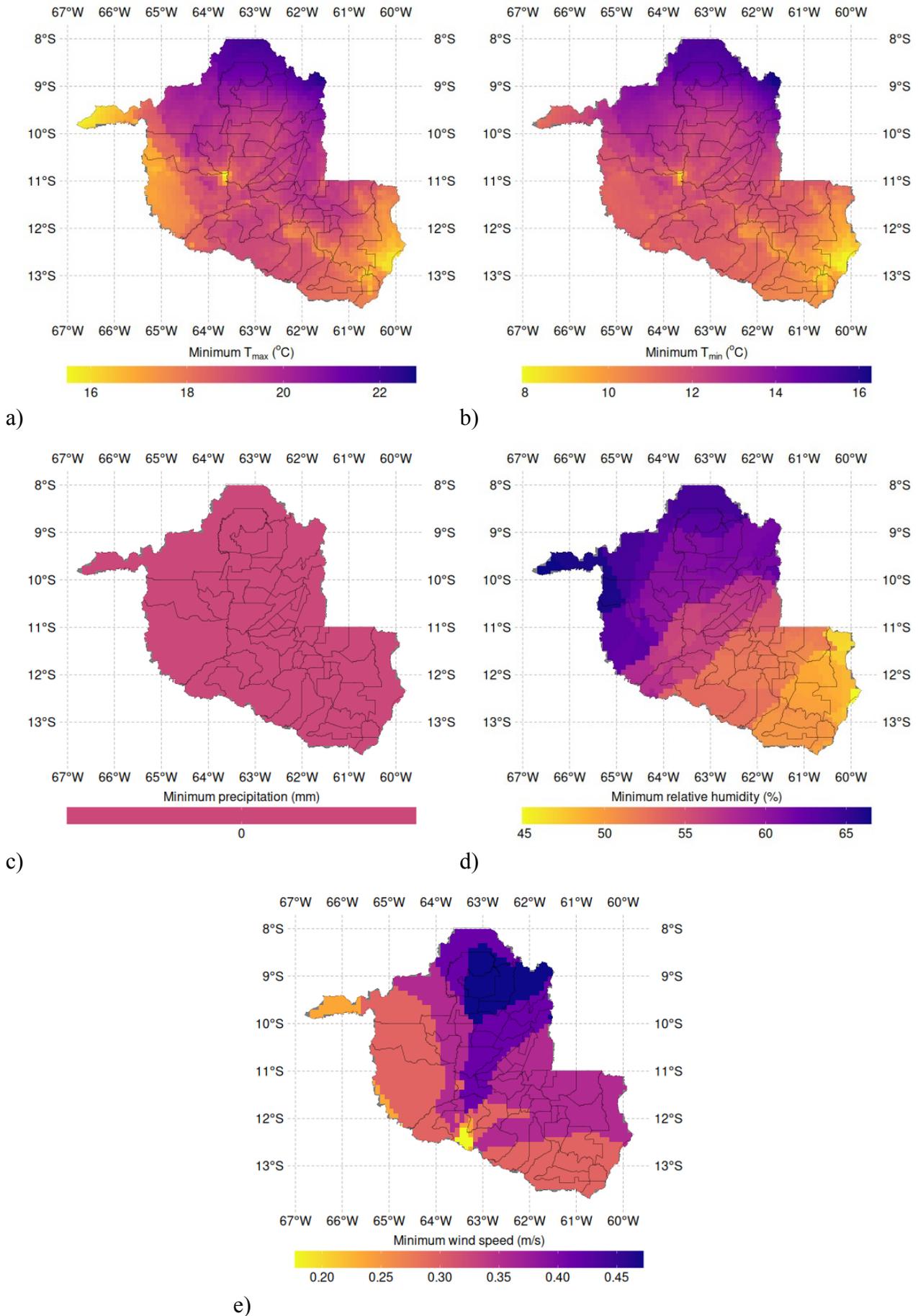

Figure 2S. Map of Rondônia State, showing city divisions and minimum of year 2023 daily maximum temperature (a; ºC), minimum temperature (b; ºC), precipitation (c; mm), relative humidity (d; %), and wind speed at 2 m (e; m/s) data from BR-DWGD (Xavier et al., 2022).

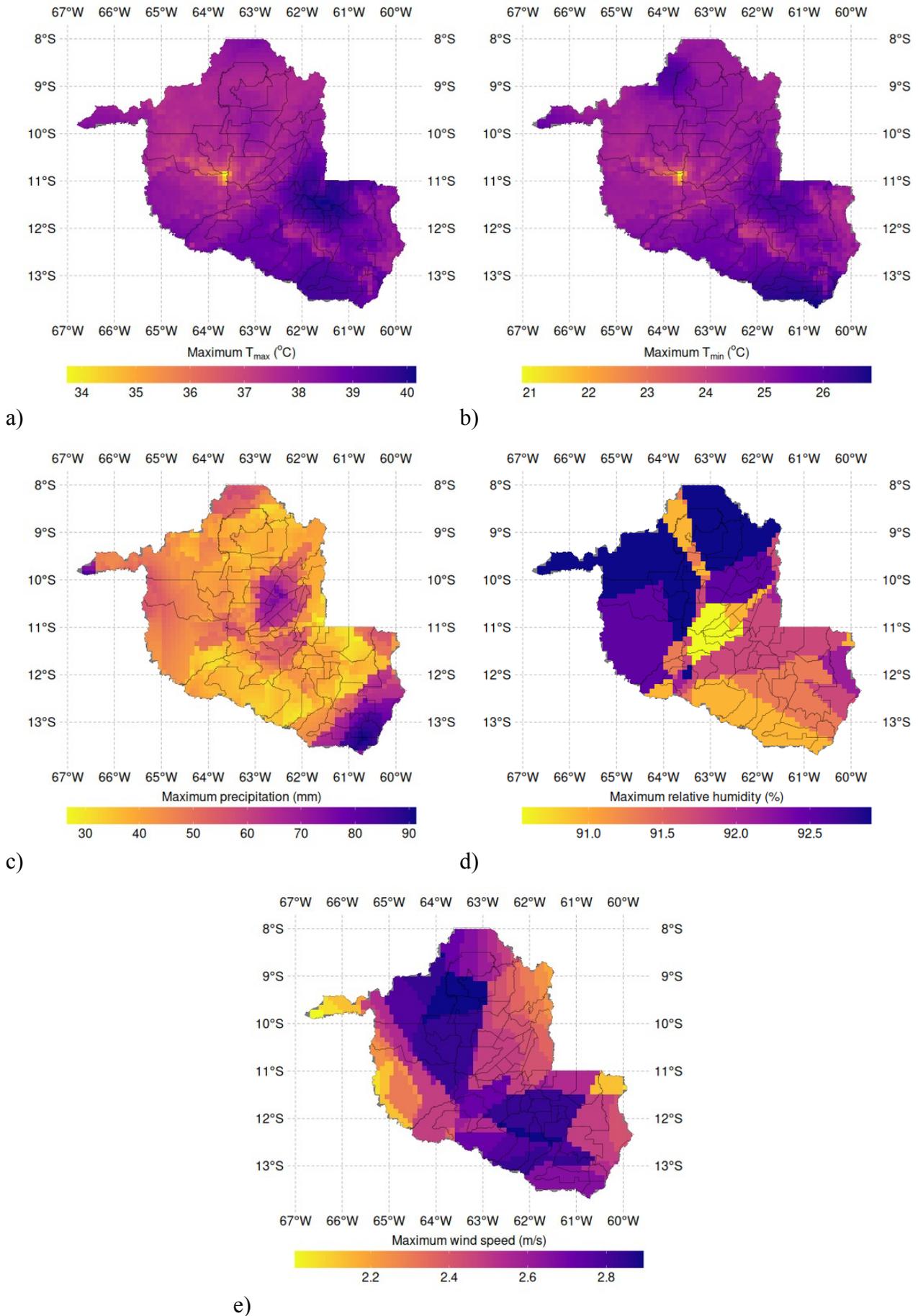

Figure 3S. Map of Rondônia State, showing city divisions and maximum of year 2023 daily maximum temperature (a; ºC), minimum temperature (b; ºC), precipitation (c; mm), relative humidity (d; %), and wind speed at 2 m (e; m/s) data from BR-DWGD (Xavier et al., 2022).

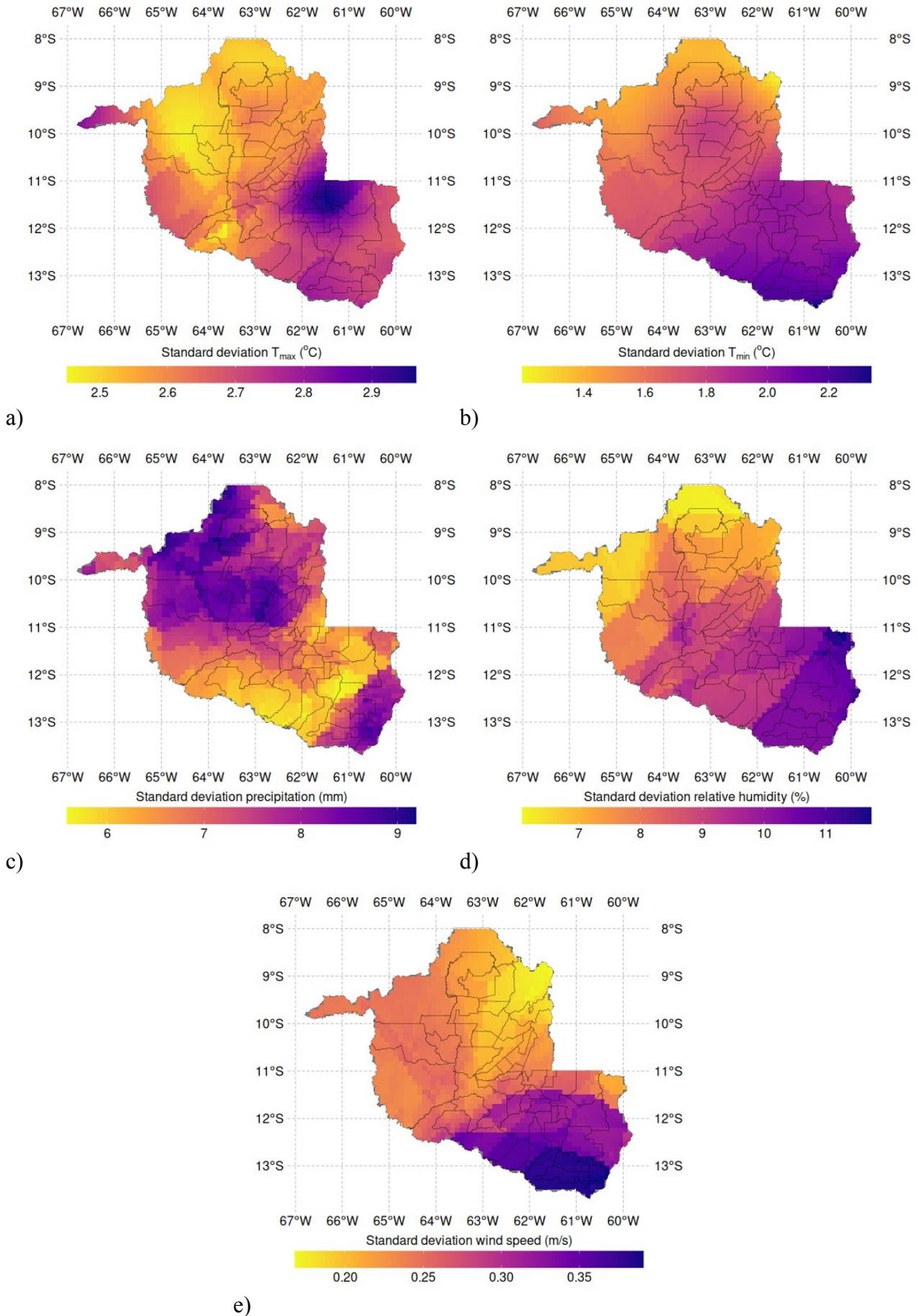

Figure 4S. Map of Rondônia State, showing city divisions and standard deviation of year 2023 daily maximum temperature (a; ºC), minimum temperature (b; ºC), precipitation (c; mm), relative humidity (d; %), and wind speed at 2 m (e; m/s) data from BR-DWGD (Xavier et al., 2022).

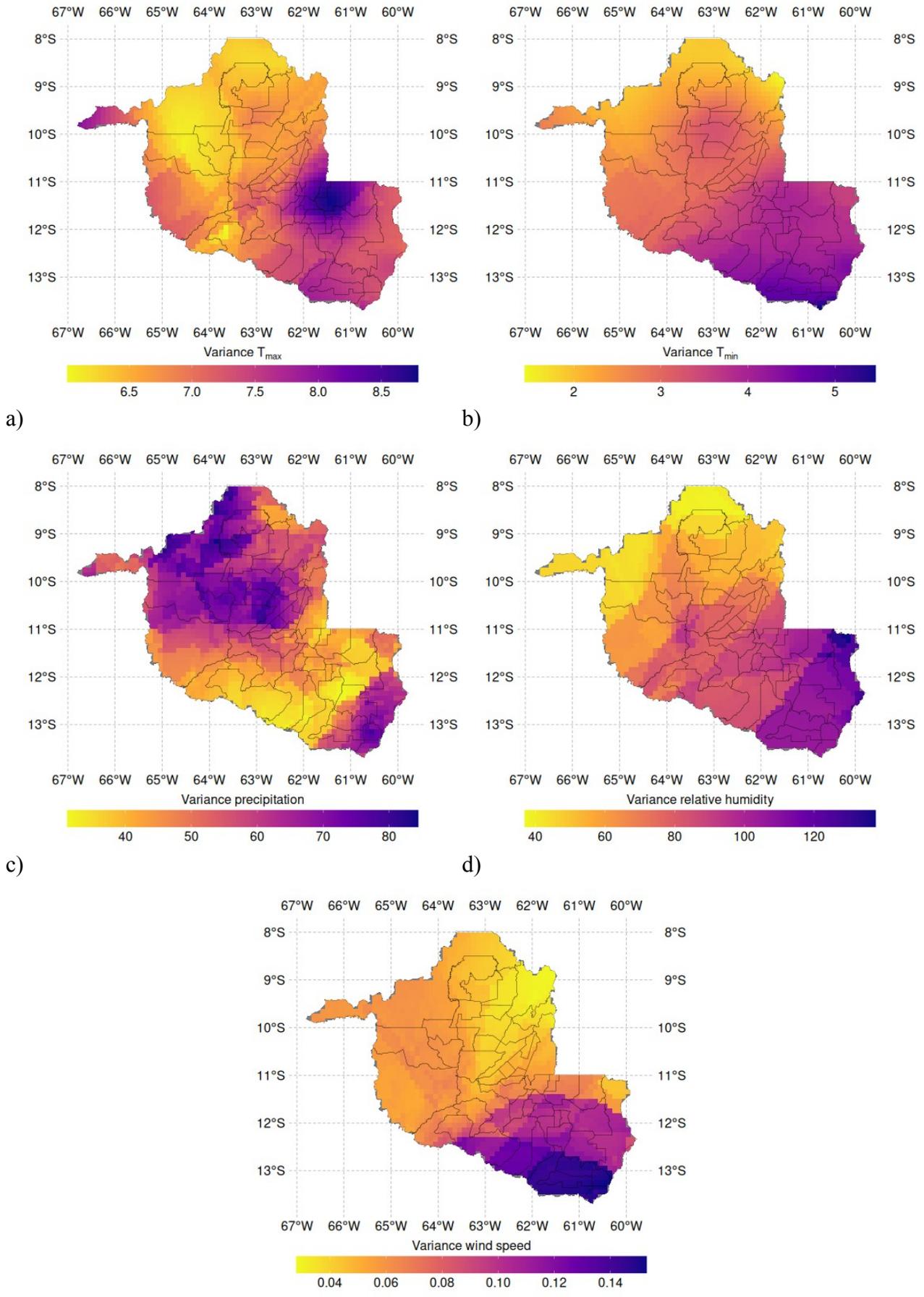

Figure 5S. Map of Rondônia State, showing city divisions and variance of year 2023 daily maximum temperature (a; ºC), minimum temperature (b; ºC), precipitation (c; mm), relative humidity (d; %), and wind speed at 2 m (e; m/s) data from BR-DWGD (Xavier et al., 2022).

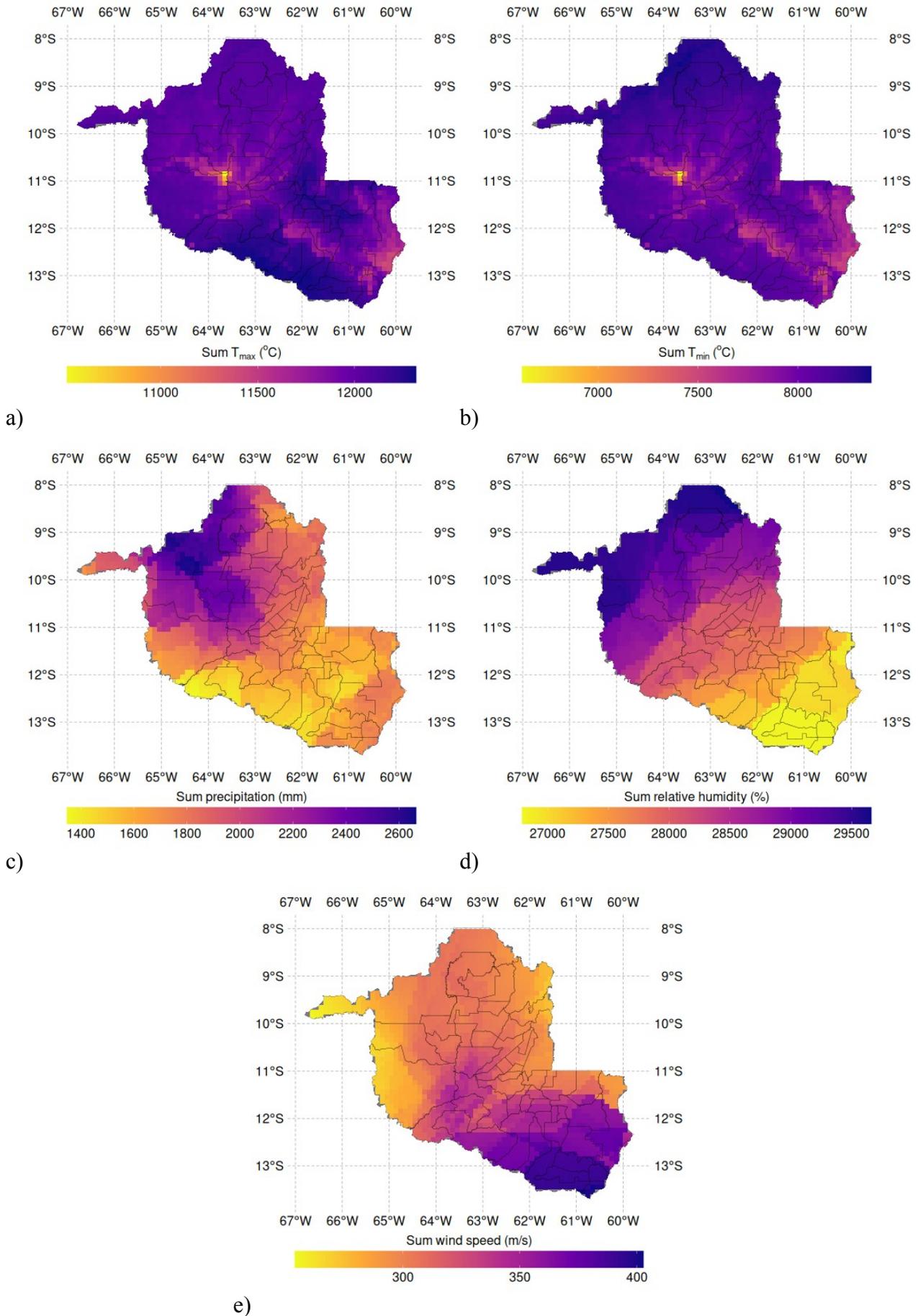

Figure 6S. Map of Rondônia State, showing city divisions and variance of year 2023 daily maximum temperature (a; ºC), minimum temperature (b; ºC), precipitation (c; mm), relative humidity (d; %), and wind speed at 2 m (e; m/s) data from BR-DWGD (Xavier et al., 2022).

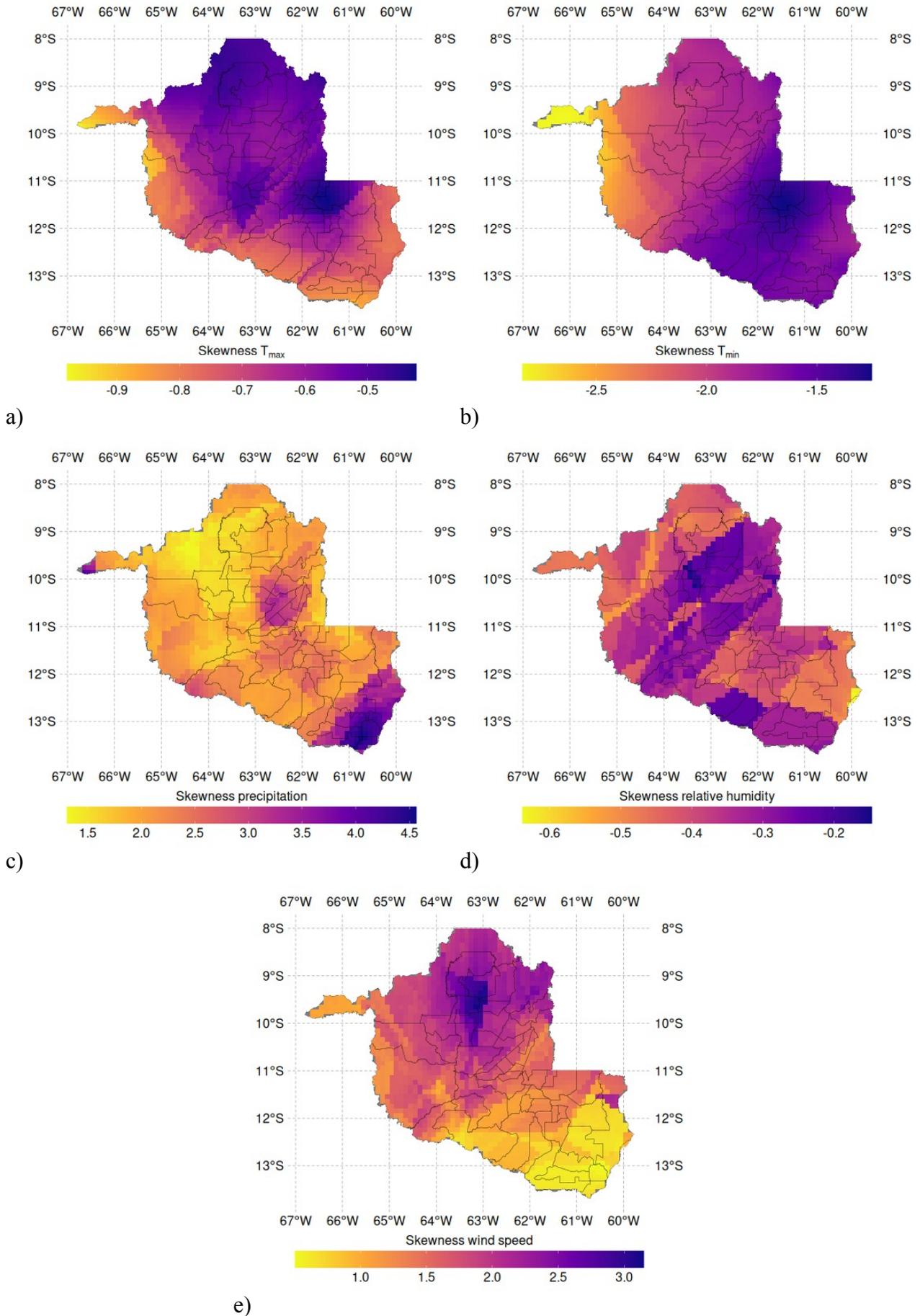

Figure 7S. Map of Rondônia State, showing city divisions and skewness of year 2023 daily maximum temperature (a; ºC), minimum temperature (b; ºC), precipitation (c; mm), relative humidity (d; %), and wind speed at 2 m (e; m/s) data from BR-DWGD (Xavier et al., 2022).

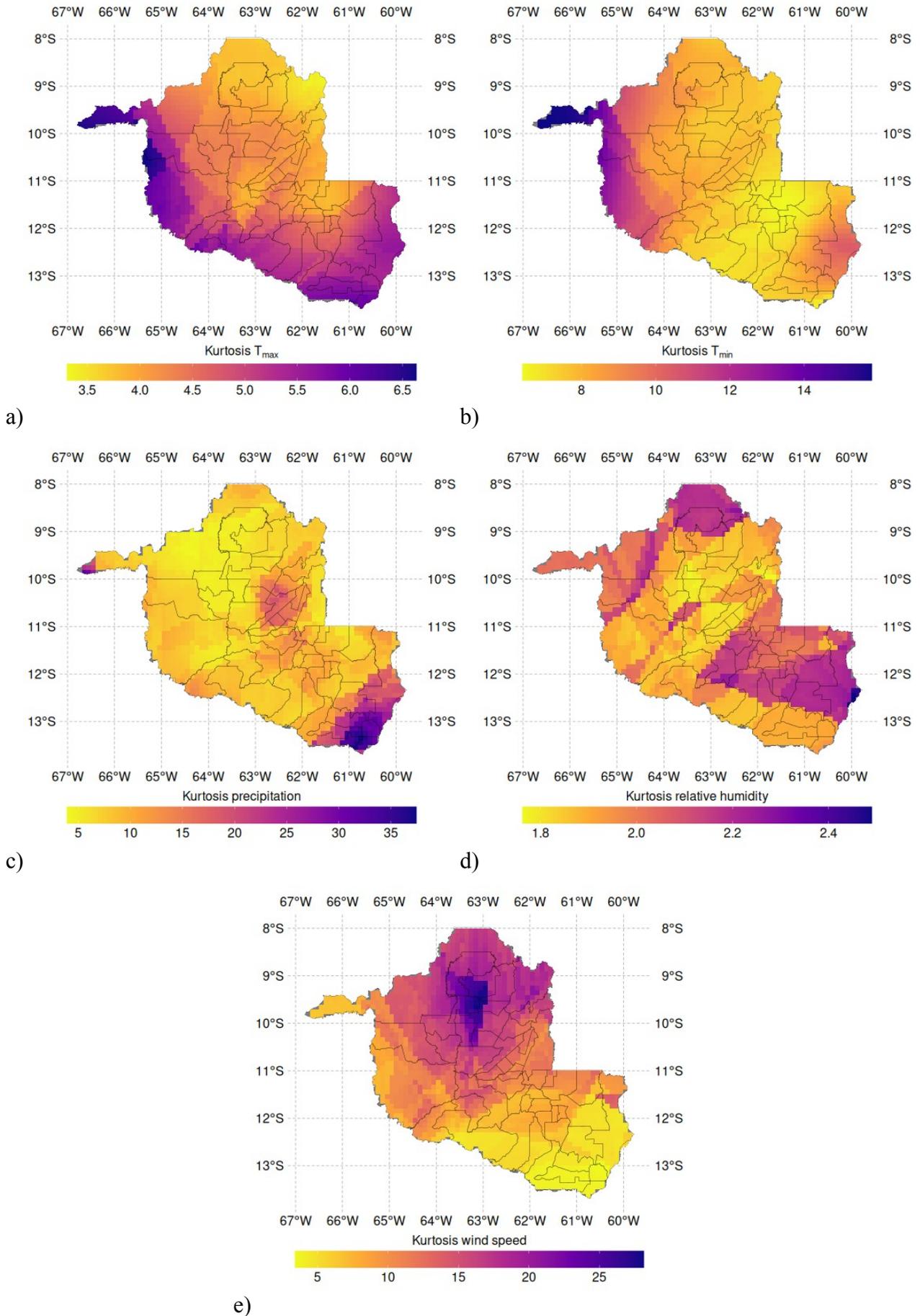

Figure 8S. Map of Rondônia State, showing city divisions and kurtosis of year 2023 daily maximum temperature (a; ºC), minimum temperature (b; ºC), precipitation (c; mm), relative humidity (d; %), and wind speed at 2 m (e; m/s) data from BR-DWGD (Xavier et al., 2022).

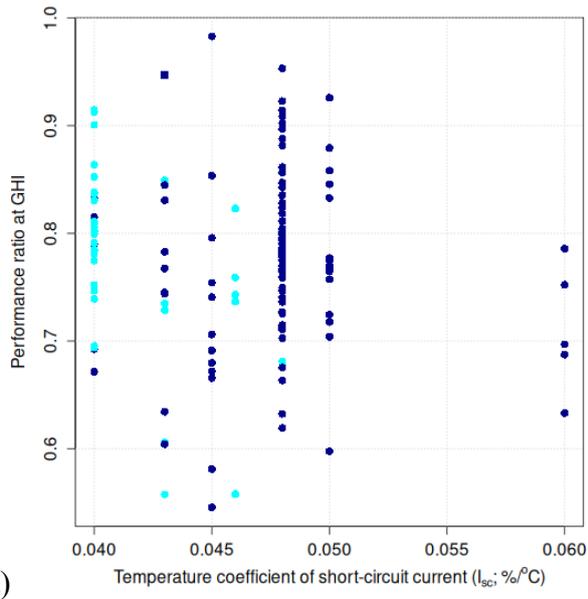
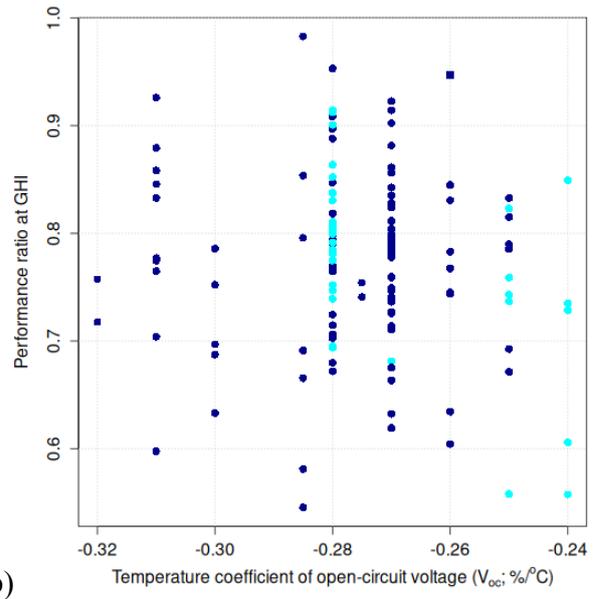
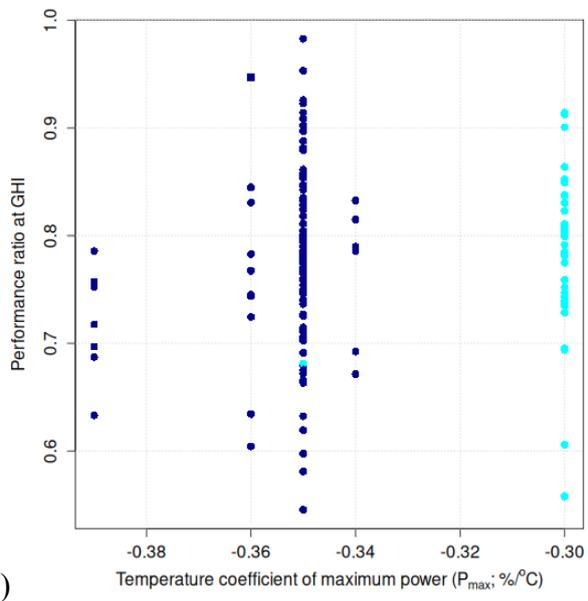
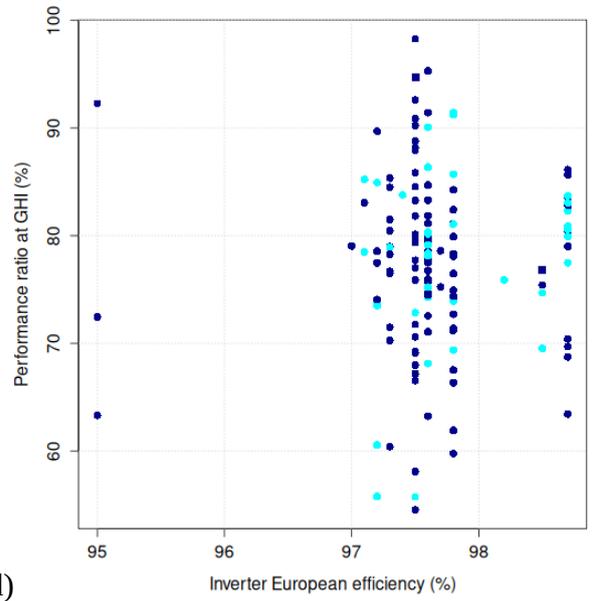
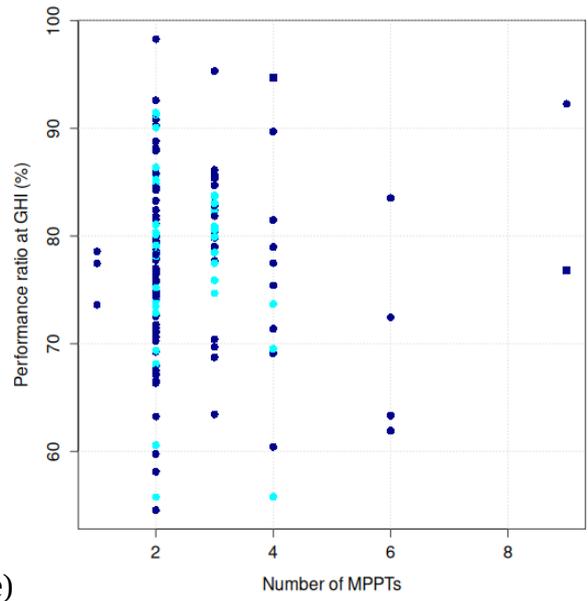
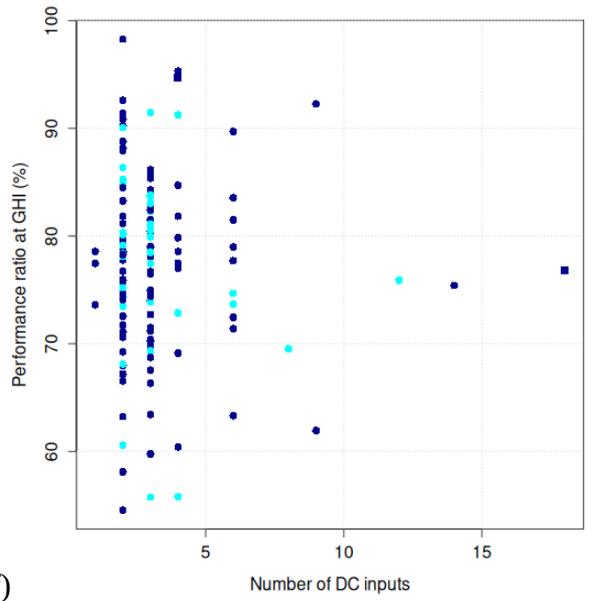

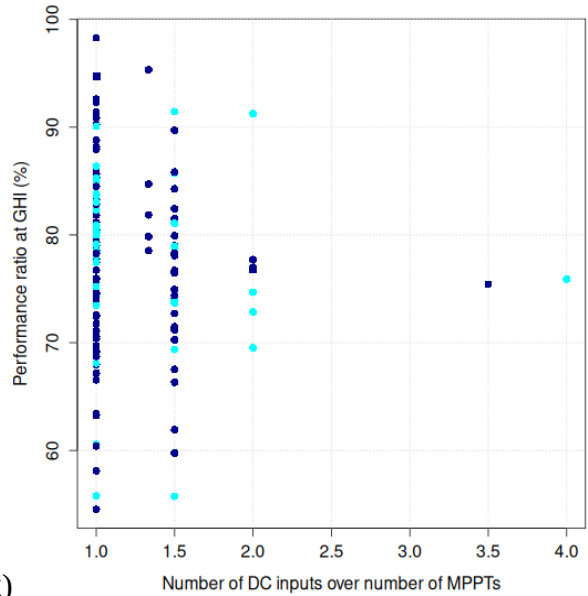

g)

Figure 9S. Relationship between annual performance ratio at global horizontal irradiance ($PR_{GHI}$) and temperature coefficient of modules for short-circuit current (a), temperature coefficient of modules for open-circuit voltage (b), temperature coefficient of modules for maximum power (c), inverter European efficiency (d), number of MPPTs of inverters (e), number of DC inputs of inverters (f), number of DC inputs over number of MPPTs of inverters (g). Blue represents p-type modules (104 samples), while cyan represents n-type modules (38 samples). Circles represent monofacial modules (140 samples), while squares represent bifacial modules (2 samples).

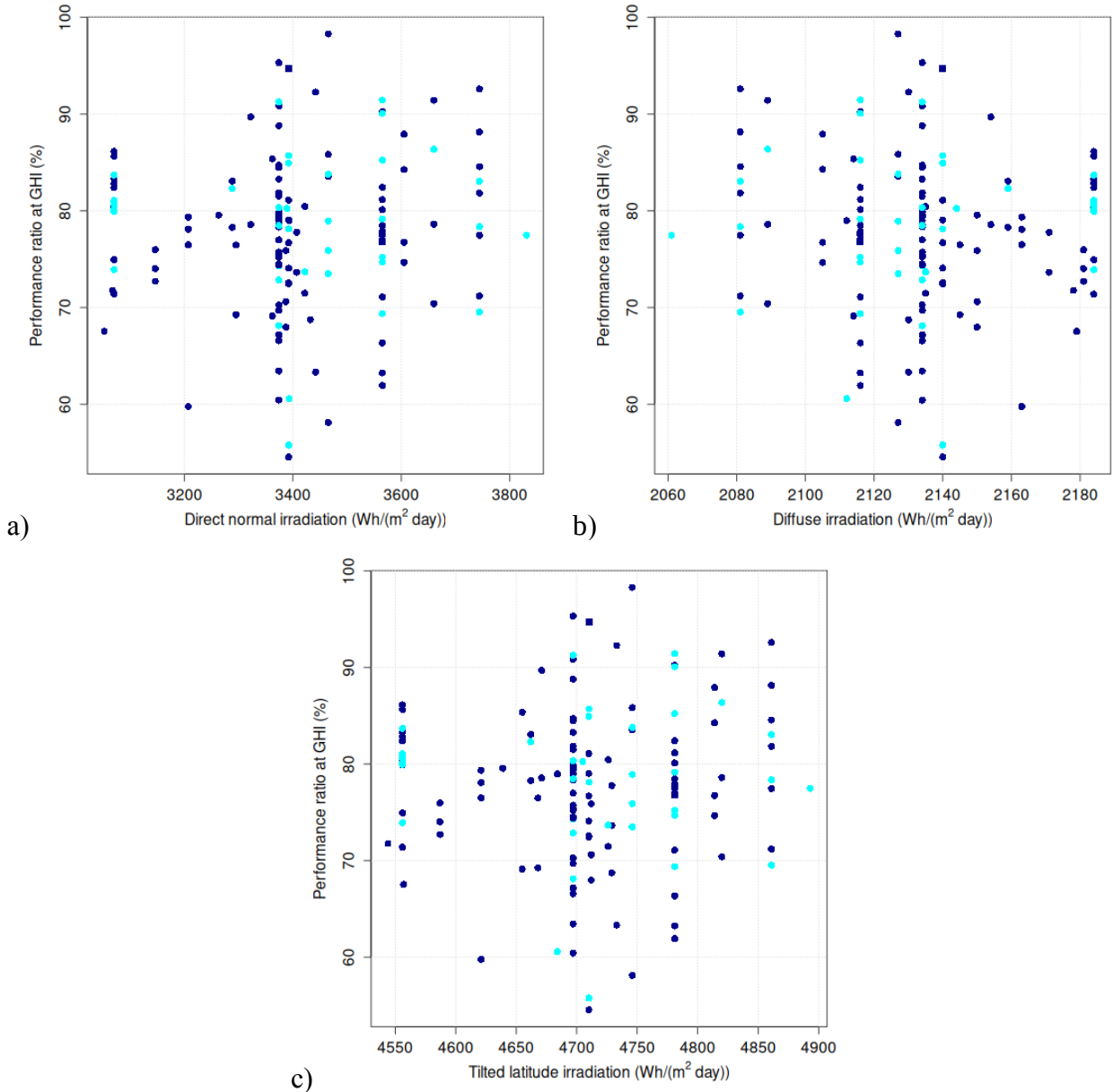

Figure 10S. Relationship between annual performance ratio at global horizontal irradiance ($PR_{GHI}$) and direct normal irradiation (a), diffuse irradiation (b), and tilted latitude irradiation (c). Blue represents p-type modules (104 samples), while cyan represents n-type modules (38 samples). Circles represent monofacial modules (140 samples), while squares represent bifacial modules (2 samples).

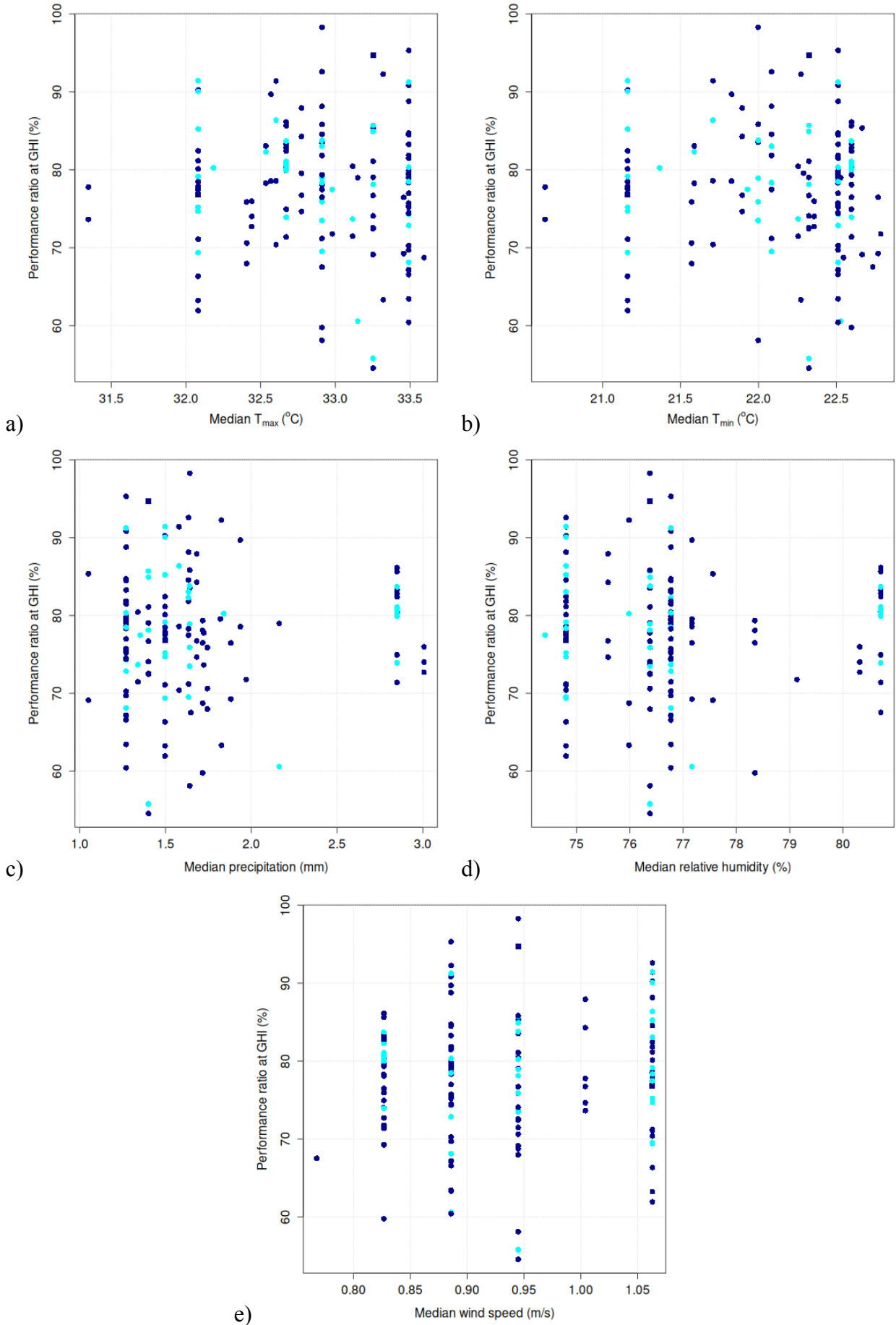

Figure 11S. Relationship between annual performance ratio at global horizontal irradiance ($PR_{GHI}$) and median of year 2023 daily maximum temperature (a; ºC), minimum temperature (b; ºC), precipitation (c; mm), relative humidity (d; %), and wind speed at 2 m (e; m/s) data from BR-

DWGD (Xavier et al., 2022). Blue represents p-type modules (104 samples), while cyan represents n-type modules (38 samples). Circles represent monofacial modules (140 samples), while squares represent bifacial modules (2 samples).

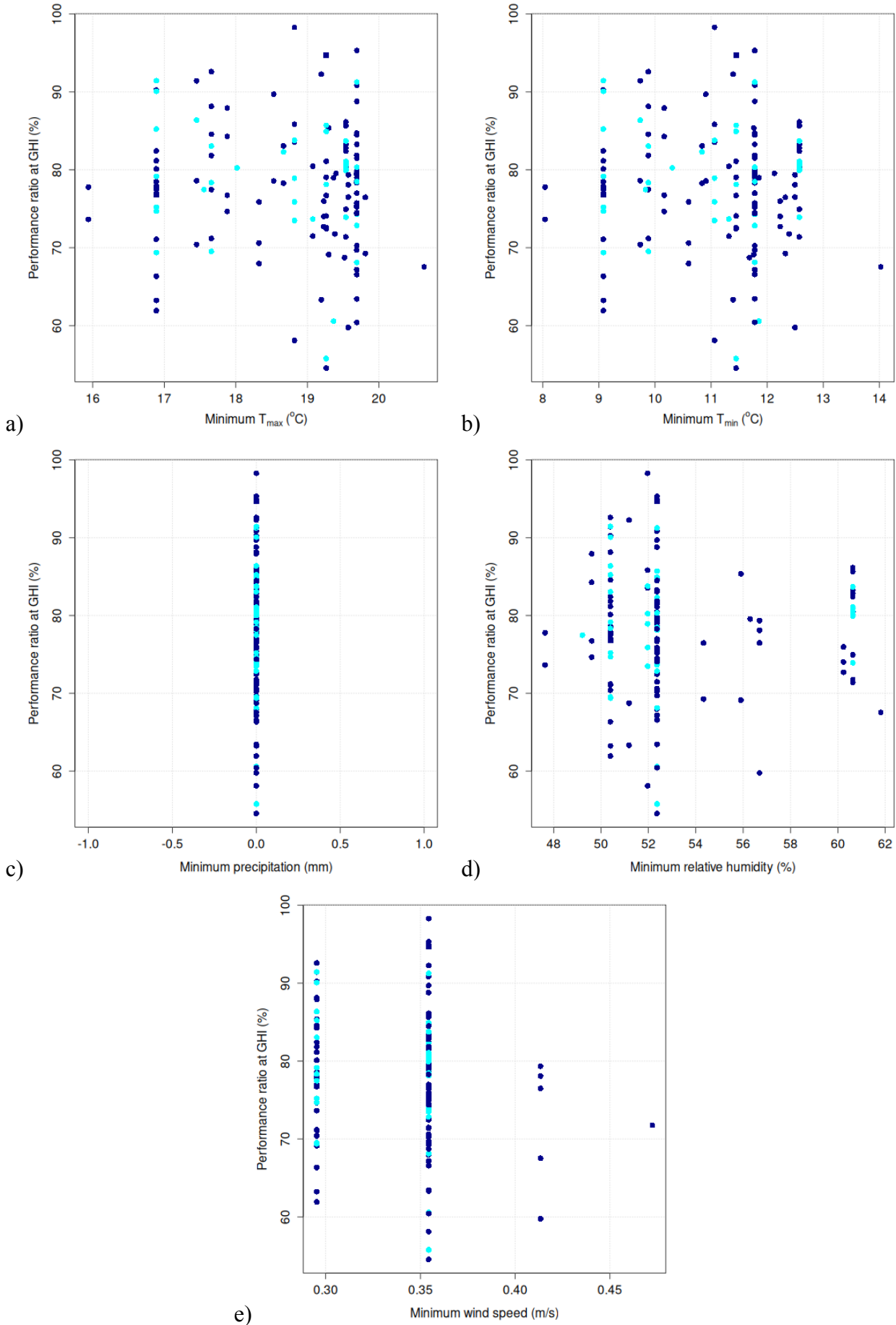

Figure 12S. Relationship between annual performance ratio at global horizontal irradiance ($PR_{GHI}$) and minimum of year 2023 daily maximum temperature (a; ºC), minimum temperature (b; ºC), precipitation (c; mm), relative humidity (d; %), and wind speed at 2 m (e; m/s) data from BR-

DWGD (Xavier et al., 2022). Blue represents p-type modules (104 samples), while cyan represents n-type modules (38 samples). Circles represent monofacial modules (140 samples), while squares represent bifacial modules (2 samples).

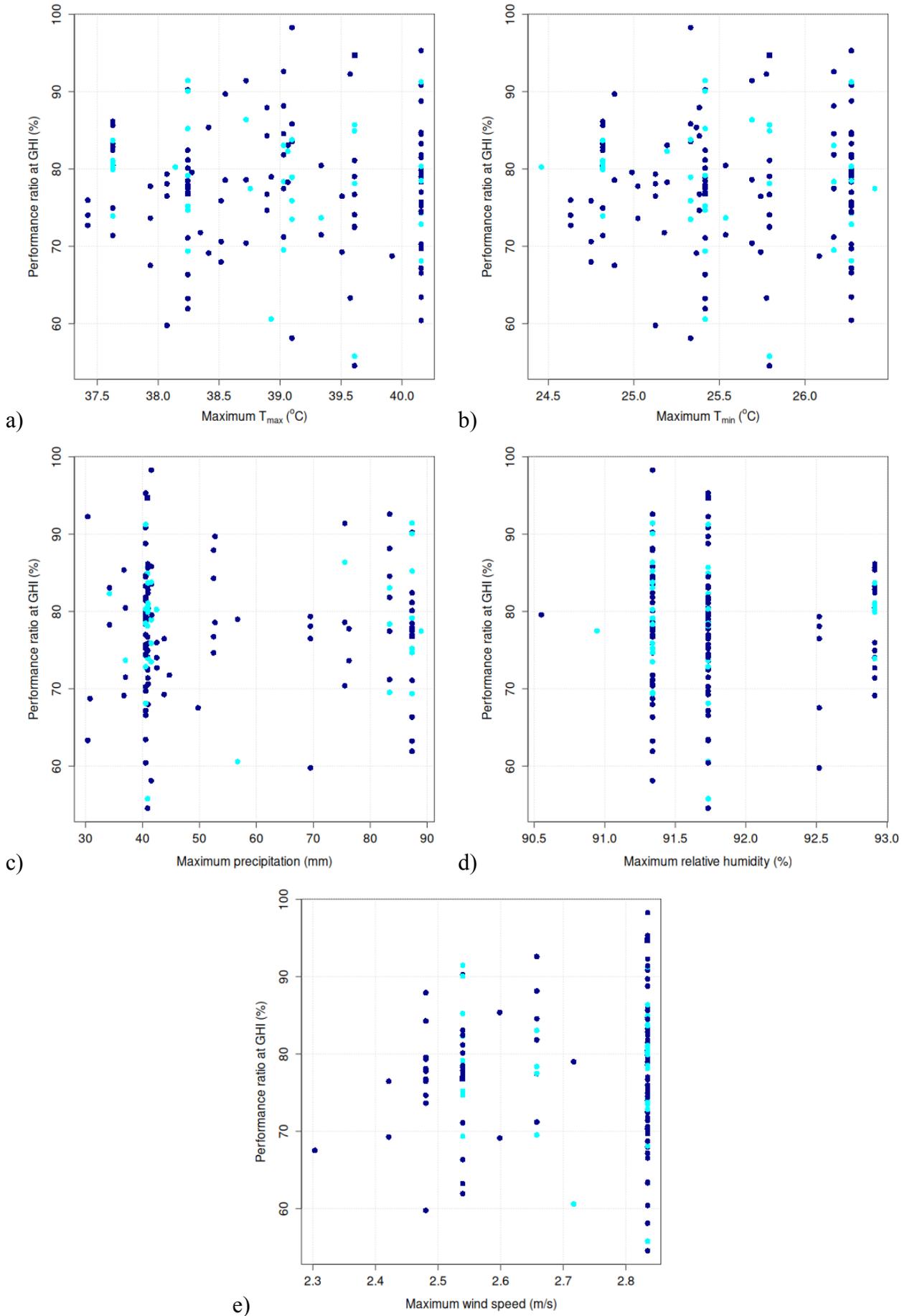

Figure 13S. Relationship between annual performance ratio at global horizontal irradiance ($PR_{GHI}$) and maximum of year 2023 daily maximum temperature (a; °C), minimum temperature (b; °C), precipitation (c; mm), relative humidity (d; %), and wind speed at 2 m (e; m/s) data from BR-

DWGD (Xavier et al., 2022). Blue represents p-type modules (104 samples), while cyan represents n-type modules (38 samples). Circles represent monofacial modules (140 samples), while squares represent bifacial modules (2 samples).

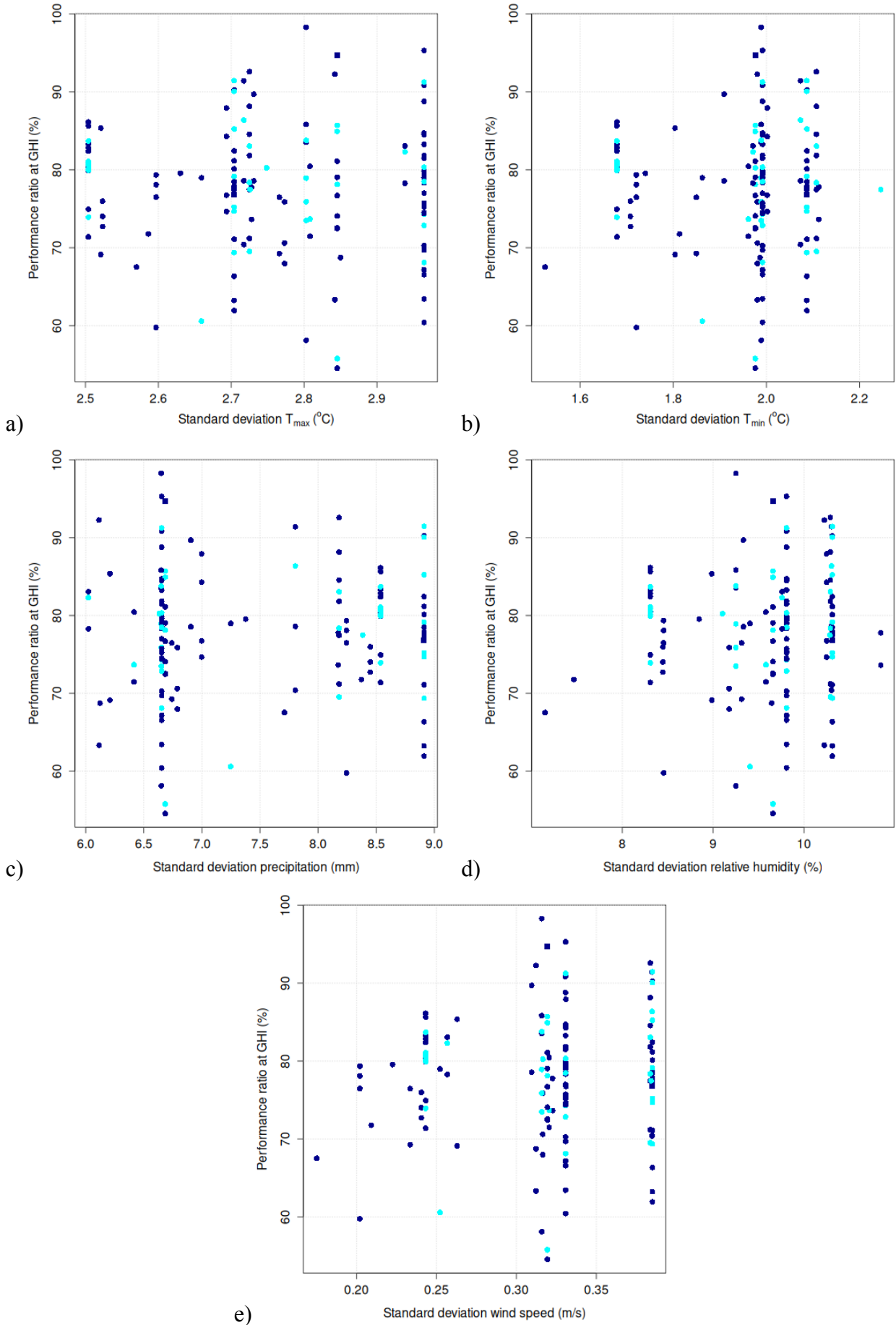

Figure 14S. Relationship between annual performance ratio at global horizontal irradiance ($PR_{GHI}$) and standard deviation of year 2023 daily maximum temperature (a; ºC), minimum temperature (b; ºC), precipitation (c; mm), relative humidity (d; %), and wind speed at 2 m (e; m/s) data from BR-

DWGD (Xavier et al., 2022). Blue represents p-type modules (104 samples), while cyan represents n-type modules (38 samples). Circles represent monofacial modules (140 samples), while squares represent bifacial modules (2 samples).

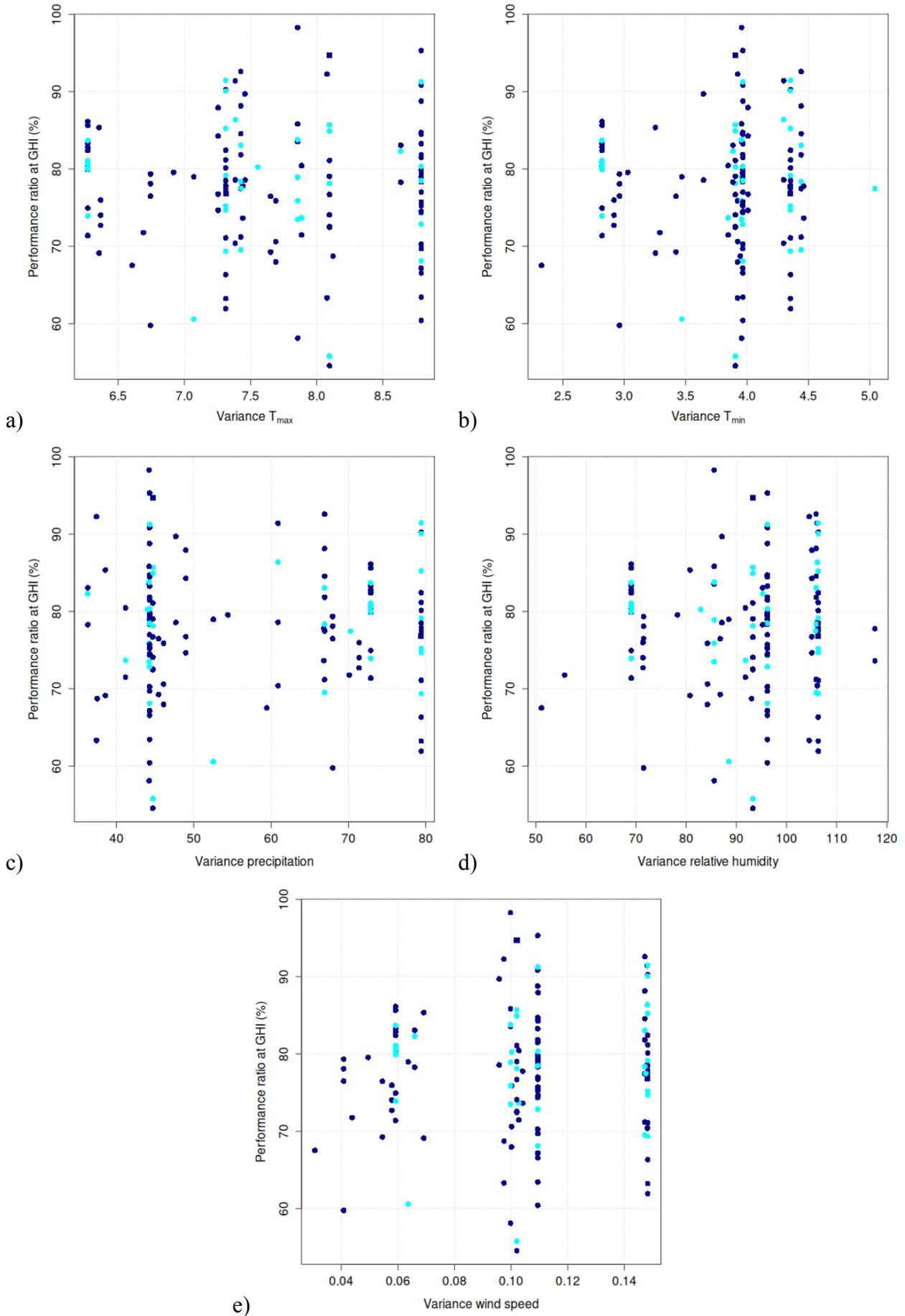

Figure 15S. Relationship between annual performance ratio at global horizontal irradiance ($PR_{GHI}$) and variance of year 2023 daily maximum temperature (a; ºC), minimum temperature (b; ºC), precipitation (c; mm), relative humidity (d; %), and wind speed at 2 m (e; m/s) data from BR-

DWGD (Xavier et al., 2022). Blue represents p-type modules (104 samples), while cyan represents n-type modules (38 samples). Circles represent monofacial modules (140 samples), while squares represent bifacial modules (2 samples).

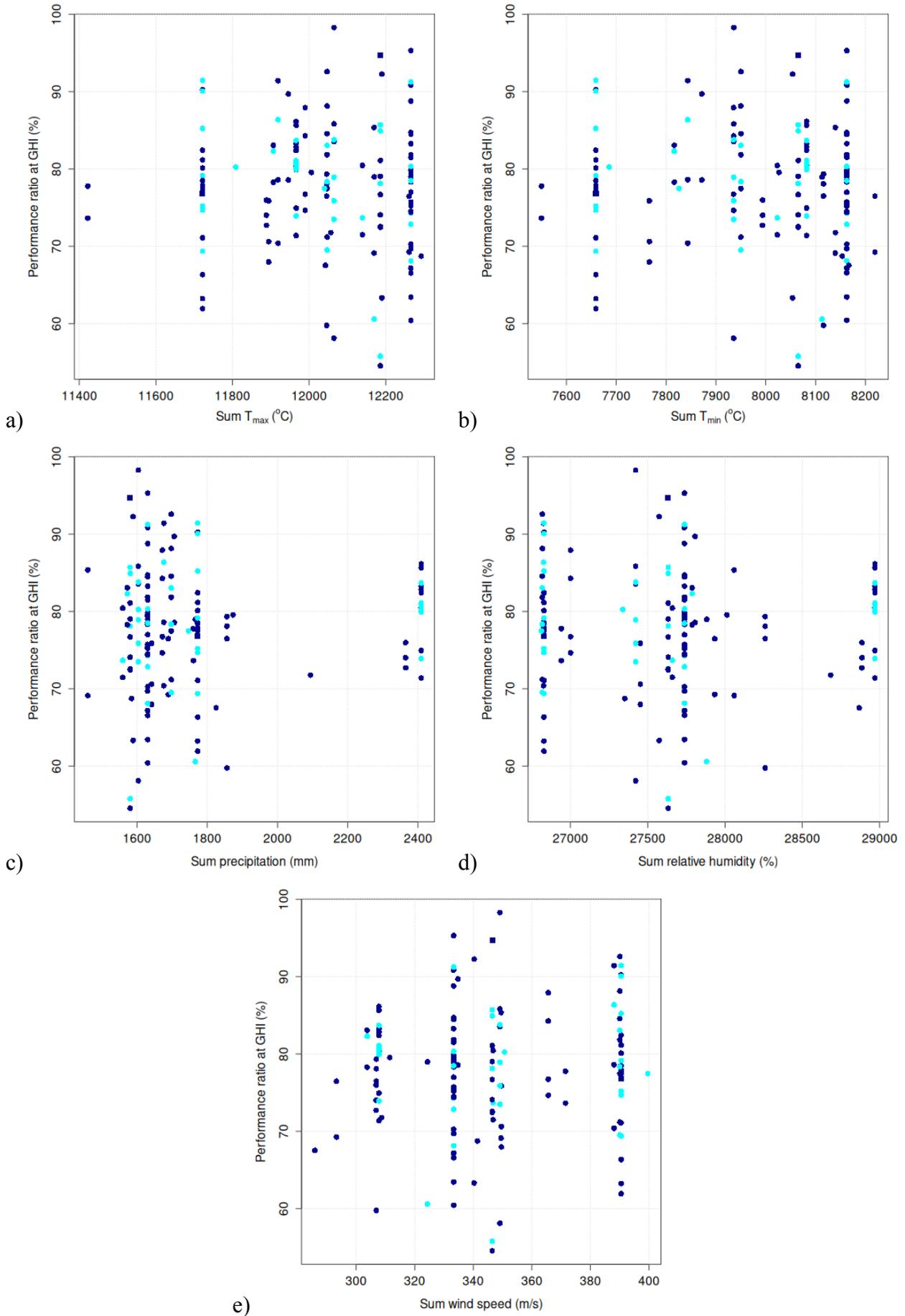

Figure 16S. Relationship between annual performance ratio at global horizontal irradiance ($PR_{GHI}$) and sum of year 2023 daily maximum temperature (a; ºC), minimum temperature (b; ºC), precipitation (c; mm), relative humidity (d; %), and wind speed at 2 m (e; m/s) data from BR-

DWGD (Xavier et al., 2022). Blue represents p-type modules (104 samples), while cyan represents n-type modules (38 samples). Circles represent monofacial modules (140 samples), while squares represent bifacial modules (2 samples).

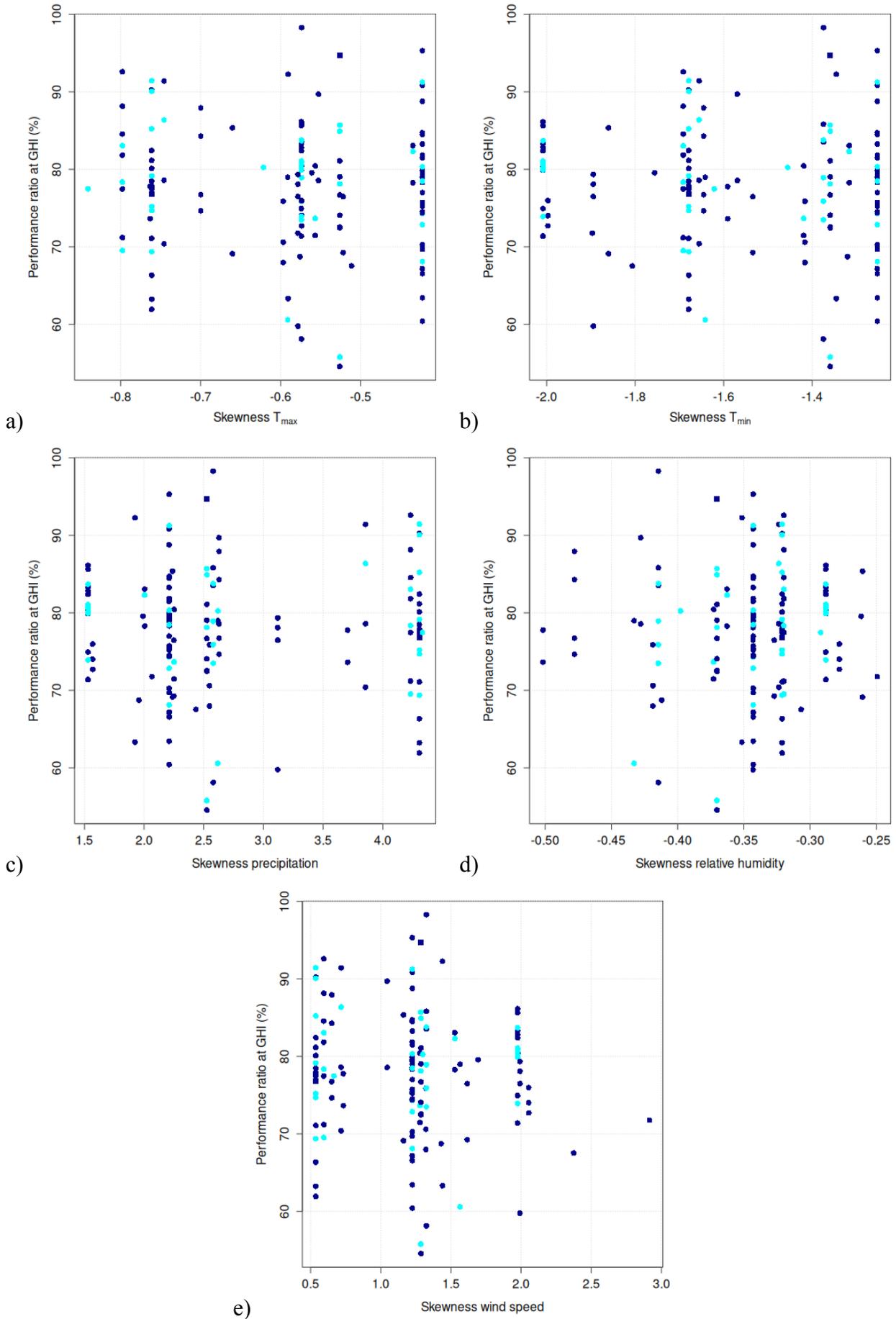

Figure 17S. Relationship between annual performance ratio at global horizontal irradiance ($PR_{GHI}$) and skewness of year 2023 daily maximum temperature (a; ºC), minimum temperature (b; ºC), precipitation (c; mm), relative humidity (d; %), and wind speed at 2 m (e; m/s) data from BR-

DWGD (Xavier et al., 2022). Blue represents p-type modules (104 samples), while cyan represents n-type modules (38 samples). Circles represent monofacial modules (140 samples), while squares represent bifacial modules (2 samples).

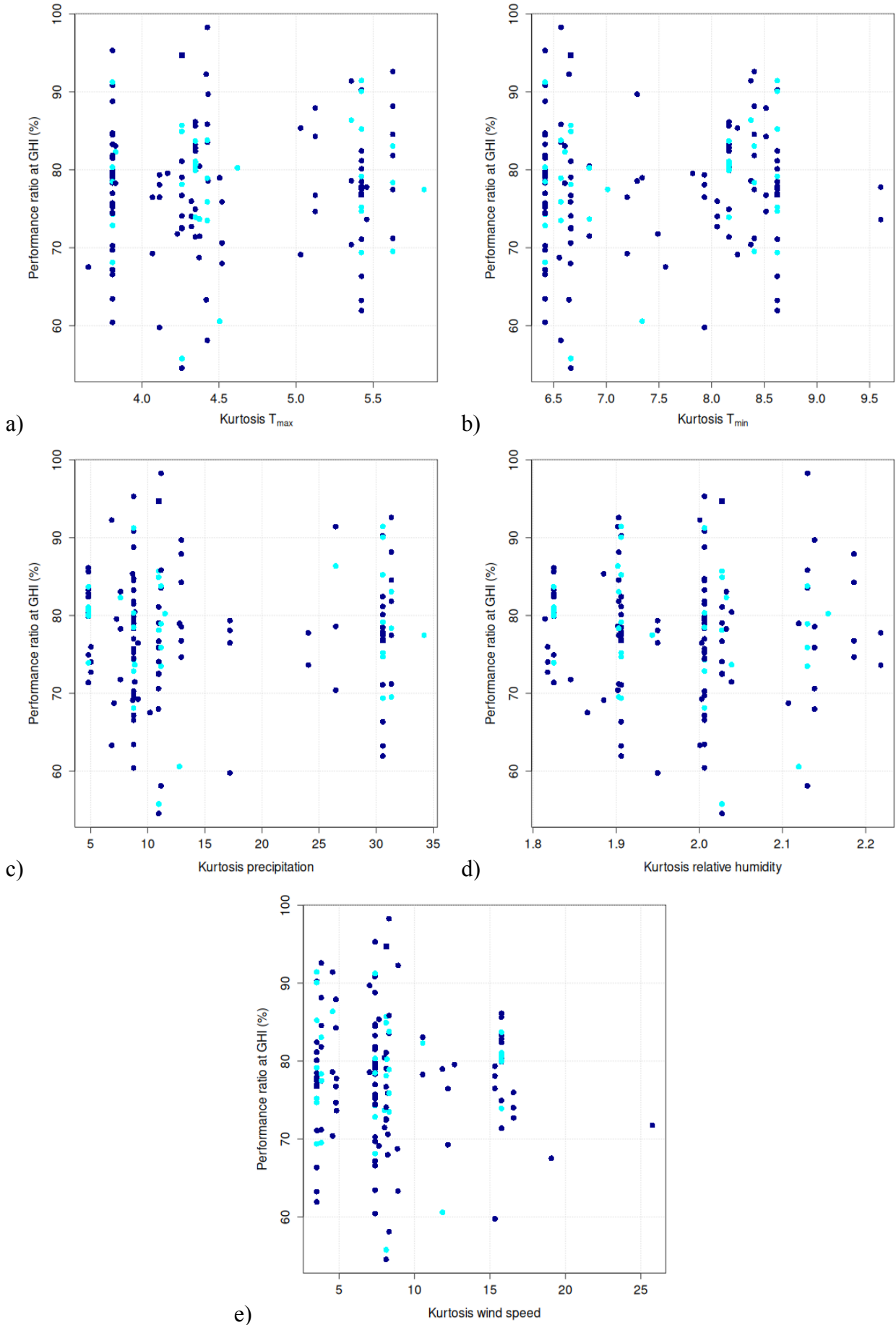

Figure 18S. Relationship between annual performance ratio at global horizontal irradiance ($PR_{GHI}$) and kurtosis of year 2023 daily maximum temperature (a; ºC), minimum temperature (b; ºC), precipitation (c; mm), relative humidity (d; %), and wind speed at 2 m (e; m/s) data from BR-

DWGD (Xavier et al., 2022). Blue represents p-type modules (104 samples), while cyan represents n-type modules (38 samples). Circles represent monofacial modules (140 samples), while squares represent bifacial modules (2 samples).

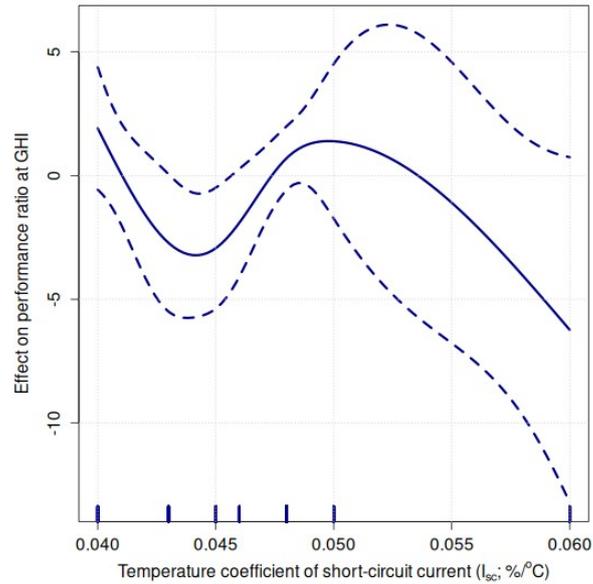

Figure 19S. Non-linear effect of module temperature coefficient of short-circuit current ($I_{SC}$) on annual performance ratio at global horizontal irradiance ($PR_{GHI}$), learned from the data. The learned non-linear function describes that the effect is decreasing, then increasing, and then decreasing again (intercept = 77.52%), which is counter-intuitive, since one could expect only a monotonic effect (i.e., only increasing or decreasing). To enforce a physically correct relationship, a linear function was fitted between the two and was found not statistically significant (p = 0.272 without correction using the False Discovery Rate). This indicates a overfitting to the data. More details can be found in the code (Milan, 2025).

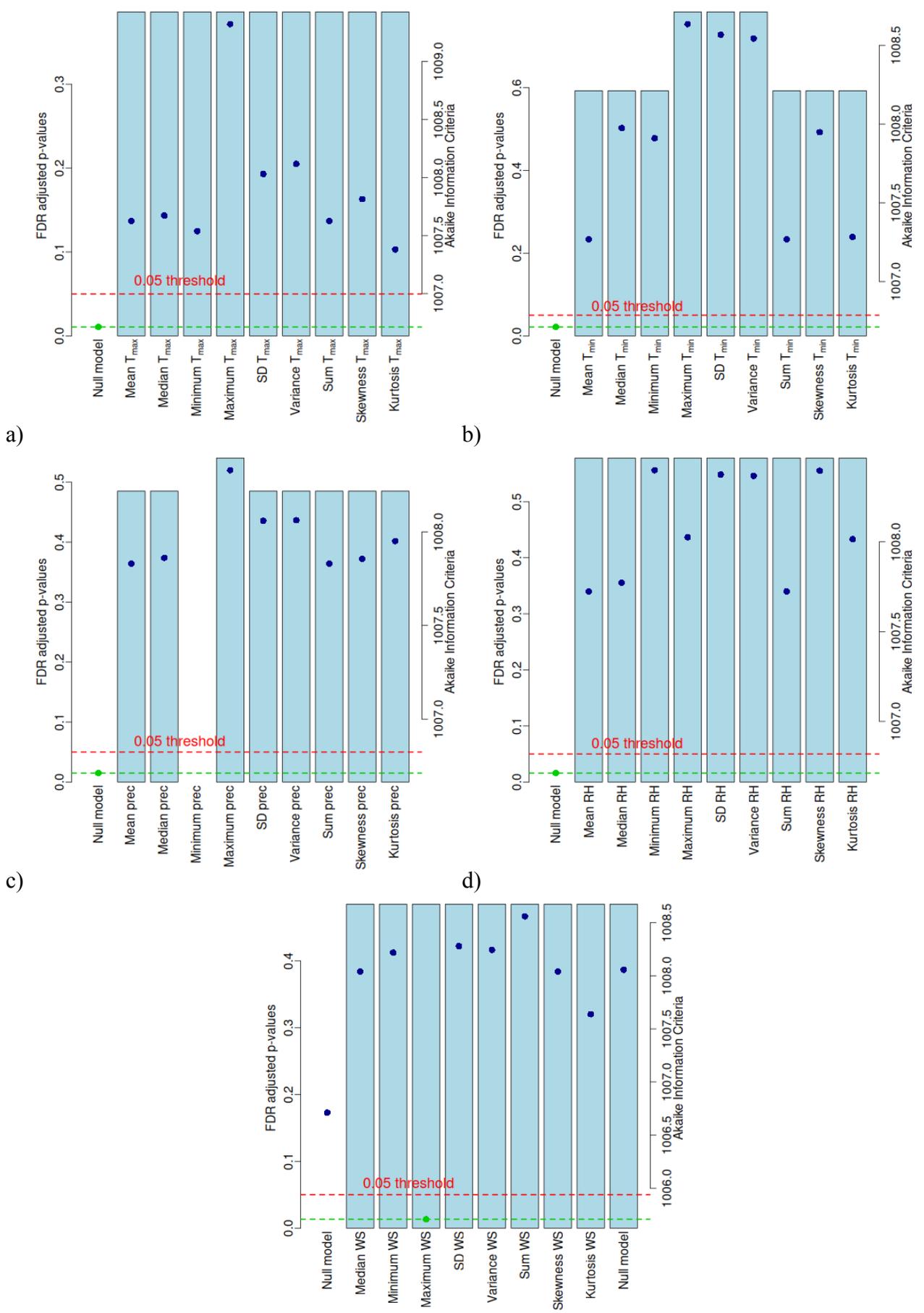

Figure 20S. P-values adjusted using the False Discovery Rate (FDR) methodology and Akaike information criteria (AIC) calculated to statistically compare a Generalize Additive Model (GAM) null model (only the intercept) for predicting annual performance ratio at global horizontal

irradiance ($PR_{GHI}$) vs. GAM models with maximum temperature ($T_{max}$; a), minimum temperature ($T_{min}$, b), precipitation (prec; c), relative humidity (RH, d), and wind speed at 2 m (WS, e). Bars represent FDR adjusted p-values. Red dashed line indicates 0.05 p-value threshold for statistically significant more complex model. Blue circles represent AIC values. Green dashed line indicates lowest AIC value and green circle represents lowest AIC value. All GAM models with AIC lower than the null model were further investigated and found as overfitting or statistical noisy (Milan, 2025).

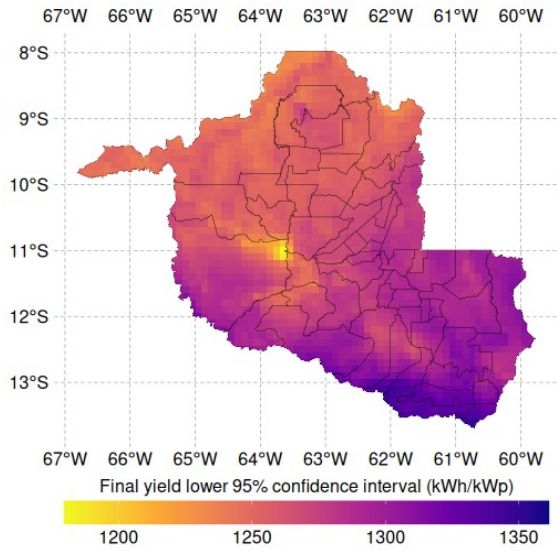
a)
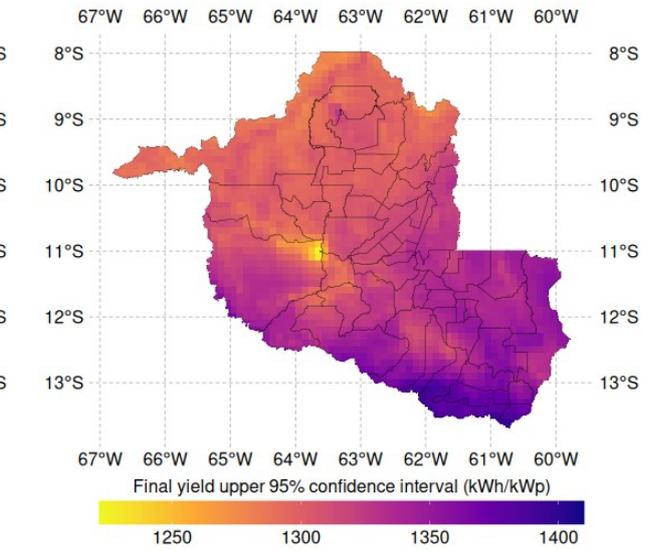
b)
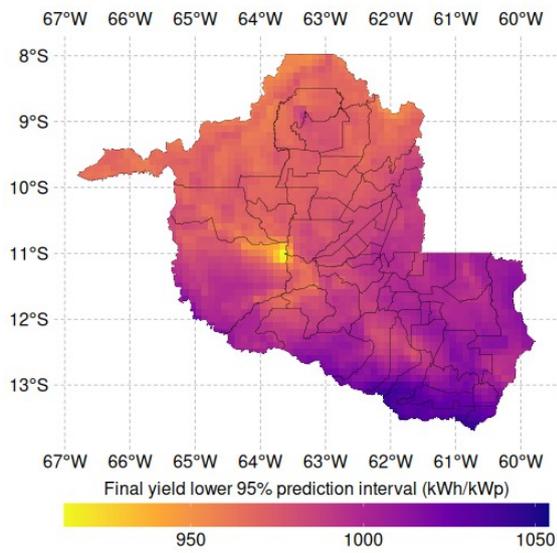
c)
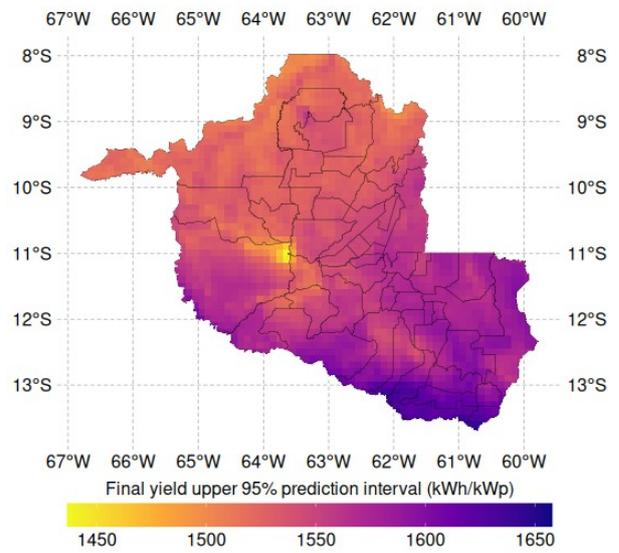
d)

Figure 21S. Map of Rondônia State showing city divisions and lower (a) and upper (b) 95% confidence interval, as well as lower (c) and upper (d) 95% prediction interval.

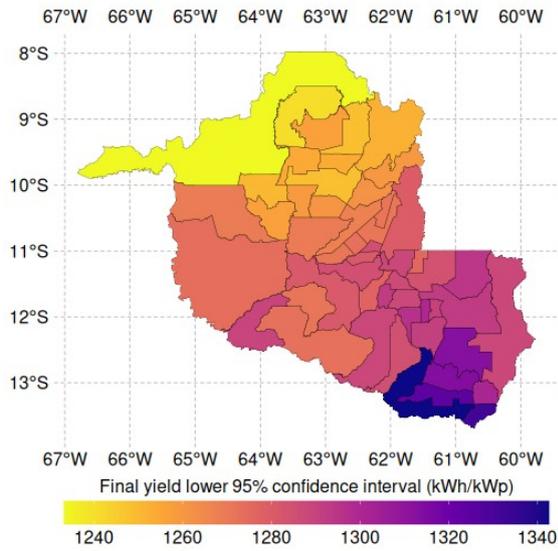
a)
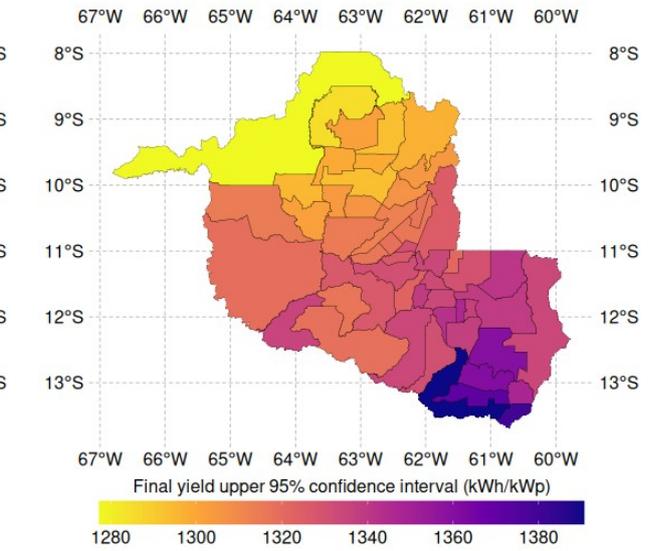
b)

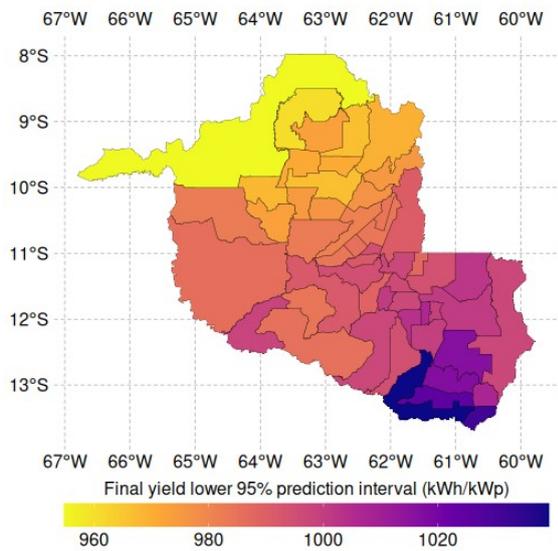
c)
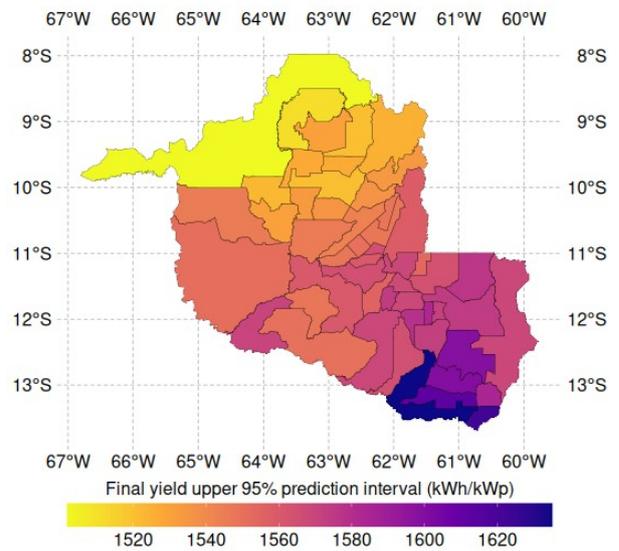
d)

Figure 22S. Map of Rondônia State showing city divisions and lower (a) and upper (b) 95% confidence interval, as well as lower (c) and upper (d) 95% prediction interval, for the urban concentrations in each city.

# References

FUNAI. Brazilian National Indian Foundation [Fundação Nacional do Índio]. Indigenous conservation lands, loaded using function read_indigenous_land() of geobr R-package, 2019.

IBGE. Brazilian Institute of Geography and Statistics [Instituto Brasileiro de Geografia e Estatística]. Urbanized areas in Brazil, loaded using function read_urban_areas() of geobr R-package, 2015.

Milan HFM. Performance_Ration_PV_Systems: Data and analysis for performance ratio of solar energy PV systems in Rondônia, Brazil. Available at https://github.com/hugomilan/performance_ratio_PV_systems, 2025. https://doi.org/10.5281/zenodo.266350

MMA. Brazilian Ministry of Environment and Climate Change [Ministério do Meio Ambiente e Mudança do Clima]. Environmental conservation units, loaded using function read_conservation_units() of geobr R-package, 2019.

Pereira HMF, Gonçalves CN. Geobr: download official spatial data sets of Brazil. R package version 1.9.1, 2024. https://doi.org/10.32614/CRAN.package.geobr

Xavier AC, Scanlon BR, King CW, Alves AI. New improved Brazilian daily weather gridded data (1961–2020). *Int J Climatology* 42(16):8390–8404, 2022. https://doi.org/10.1002/joc.7731